\documentclass[a4paper,12pt]{article}
\pdfoutput=1
\usepackage{epsfig,amsmath,amsfonts,amsthm}
\usepackage[figuresright]{rotating}
\usepackage[left=28mm,right=20mm,top=30mm,bottom=20mm]{geometry}
\usepackage{subfig}
\usepackage{graphicx}
\graphicspath{{RoMnewfigures/}}
\usepackage{rotating}
\usepackage{booktabs}
\usepackage{moreverb}

\DeclareGraphicsExtensions{.pdf,.png,.jpg}
\usepackage{natbib}
\numberwithin{equation}{section}

\def\e{\hbox{E}}
\def\cov{\hbox{Cov}}

\def\var{\hbox{Var}}

\def\max{\hbox{max}}

\def\logit{\hbox{logit}}
\def\expit{\hbox{expit}}

\usepackage[nokeyprefix]{refstyle}
\usepackage{varioref}
\usepackage{xr}
\usepackage{hyperref}

\title{Simulations for a $Q$ statistic with constant weights to assess heterogeneity in meta-analysis of mean difference}
\author{Elena Kulinskaya, David C. Hoaglin, Joseph Newman, and Ilyas Bakbergenuly}

\begin{document}
\maketitle

\abstract{
A variety of problems in random-effects meta-analysis arise from the conventional $Q$ statistic, which uses estimated inverse-variance (IV) weights. In previous work on standardized mean difference and log-odds-ratio, we found superior performance with an estimator of the overall effect whose weights use only group-level sample sizes. The $Q$ statistic with those weights has the form proposed by DerSimonian and Kacker. The distribution of this $Q$ and the $Q$ with IV weights must generally be approximated. We investigate approximations for those distributions, as a basis for testing and estimating the between-study variance ($\tau^2$). Some approximations require the variance and third moment of $Q$, which we derive. We describe the design and results of a simulation study, with mean difference as the effect measure, which provides a framework for assessing accuracy of the approximations, level and power of the tests, and bias in estimating $\tau^2$. Use of $Q$ with sample-size-based weights and its exact distribution (available for mean difference and evaluated by Farebrother's algorithm) provides precise levels even for very small and unbalanced sample sizes. The corresponding estimator of $\tau^2$ is almost unbiased for 10 or more small studies. Under these circumstances this performance compares favorably with the extremely liberal behavior of the standard tests of heterogeneity and the largely biased estimators based on inverse-variance weights.
}

\vspace{1in}


\vspace{1in}
\section{Introduction} \label{sec:Intro}

In meta-analysis, using weights based on estimated variances, without accounting for their sampling variation, is the source of many shortcomings in assessing heterogeneity and estimating an overall effect. Our studies of methods for random-effects meta-analysis of standardized mean difference and log-odds-ratio (\cite{BHK2018SMD, BHK2020LOR}) included an estimator of the overall effect that combines the studies' estimates with weights based only on their groups' sample sizes. That estimator, SSW,  outperformed estimators that use (estimated) inverse-variance-based (IV) weights. The IV weights use estimates of the between-study variance ($\tau^2$) derived from the popular $Q$ statistic discussed by \cite{cochran1954combination}; we denote it by $Q_{IV}$ to show its use of IV weights. Thus, parallel to SSW, we investigate an alternative, $Q_{SW}$, in which the studies' weights are their effective sample sizes. $Q_{SW}$ is an instance of the generalized $Q$ statistics introduced by \cite{dersimonian2007random}, in which the weights are fixed positive constants.

The setting is the following random-effects model (REM): For Study $i$ ($i = 1, \ldots, K$), with sample size $n_i = n_{iT} + n_{iC}$,  the estimate of the effect is $\hat\theta_i \sim G(\theta_i, v_i^2)$, where the effect-measure-specific distribution $G$ has mean $\theta_i$ and variance $v_i^2$, and $\theta_i \sim N(\theta, \tau^2)$. Thus, the $\hat\theta_i$ are unbiased estimators of the true conditional effects $\theta_i$, and the $v_i^2 = \var(\hat\theta_i | \theta_i)$ are the true conditional variances.

The general $Q$ statistic is a weighted sum of squared deviations of the estimated effects $\hat\theta_i$ from their weighted mean $\bar\theta_w = \sum w_i \hat\theta_i / \sum w_i$:
\begin{equation} \label{Q}
Q = \sum w_i (\hat\theta_i - \bar\theta_w)^2.
\end{equation}
In \cite{cochran1954combination} $w_i$ is the reciprocal of the \textit{estimated} variance of $\hat{\theta}_i$, resulting in $Q_{IV}$. In meta-analysis those $w_i$ come from the fixed-effect model. In what follows, we discuss moment-based approximations to the distribution of  $Q$ and estimation of $\tau^2$ when the $w_i$ are arbitrary positive constants. Because it is most tractable, but still instructive, we focus on a single measure of effect, the mean difference (MD). In this favorable situation, the cumulative distribution function of $Q$ is that of a  quadratic form in normally distributed random variables, evaluated by the algorithm of  \cite{Farebrother1984}.   We also consider two- and three-moment  approximations to the distribution of $Q$; we derive the required variance and third moment of $Q$. For comparison we also include some of the popular inverse-variance-based methods of estimating $\tau^2$, approximating the distribution of $Q_{IV}$, and testing for the presence of heterogeneity. A simulation study provides a framework for assessing accuracy of the approximations, level and power of the tests based on $Q_{SW}$ and $Q_{IV}$, and bias in estimating $\tau^2$.

\section{Expected value of $Q$ and estimation of $\tau^2$}
Define $W = \sum w_i$,  $q_i = w_i / W$, and $\Theta_i = \hat\theta_i - \theta$.  In this notation, and expanding $\bar\theta_w$, Equation (\ref{Q}) can be written  as
\begin{equation} \label{Q1}
Q = W \left[ \sum q_i (1 - q_i) \Theta_i^2 - \sum_{i \not = j} q_i q_j \Theta_i \Theta_j \right].
\end{equation}
Under the above REM, it is straightforward to obtain the first moment of $Q$ as
\begin{equation} \label{M1Q}
\e(Q) = W  \sum q_i (1 - q_i) \var(\Theta_i)  = W  \sum q_i (1 - q_i) (\e(v_i^2) + \tau^2).
\end{equation}
This expression is similar to Equation (4) in \cite{dersimonian2007random}. Rearranging the terms gives the moment-based estimator of $\tau^2$
\begin{equation} \label{tau_DSK}
\hat\tau^2_{M} = \max \left( \frac{Q / W - \sum q_i (1 - q_i) \widehat \e(v_i^2)} {\sum q_i (1 - q_i)},\;0 \right).
\end{equation}
This equation is similar to Equation (6) in \cite{dersimonian2007random}; they use the within-study (i.e., conditional) estimate $s_i^2$ instead of $\widehat \e(v_i^2)$, an important distinction because $v_i^2$ is a random variable whose distribution depends on that of $\theta_i$.

\section{Approximations to the distribution of quadratic forms in normal variables}

The $Q$ statistic, Equation~(\ref{Q1}), is a quadratic form in the random variables $\Theta_i$. We can write $Q = \Theta^{T} A \Theta$ for a symmetric matrix $A$ of rank $K - 1$ with the elements $a_{ij} = q_i \delta_{ij}  - q_iq_j$, $1 \leq i,j \leq K$, where $\delta_{ij}$ is the Kronecker delta.  In this section we assume constant weights unless stated otherwise. Unconditionally, the $\Theta_i$ are centered at 0, but they are not, in general, normally distributed.  However, for large sample sizes $n_i$, their distributions are approximately normal. Normality holds exactly for the mean difference (MD). However, the cumulative distribution function of $Q$ needs to be evaluated numerically.  Therefore, we consider available approximations to the distribution of quadratic forms in normal variables. The exact distribution in this case is that of a weighted sum of central chi-square variables.

Quadratic forms in normal variables have an extensive literature. When the vector $\Theta$ has the multivariate normal distribution $N(\mu,\Sigma)$, the exact distribution  of $Q$ is
$\sum_{r = 1}^m \lambda_r \chi^2_{h_r}(\delta_r^2)$,
where the $\lambda_r$ are the eigenvalues of $A \Sigma$, the $h_r$ are their multiplicities, and the $\delta_r^2$ are the non-centrality parameters for the independent chi-square variables $\chi^2_{h_i}(\delta_i^2)$ with $h_i$ degrees of freedom. (The $\delta_r$ are linear combinations of $\mu_1, \ldots, \mu_K$.)

Interest typically centers on the upper-tail probabilities $P(Q>x)$. Moment-based approximations match a particular distribution, often a gamma distribution or, equivalently, a scaled chi-square distribution, to several moments of $Q$. These methods include the well-known Welch-Satterthwaite approximation, which uses $c \chi^2_p$ and matches the first two moments (\cite{Welch1938, Satterthwaite1941}).

\cite{YuanBentler2010} studied, by simulation, the Type I errors of a $Q$ test with the critical values based on the Welch-Satterthwaite approximation. They concluded that this approximation is satisfactory when the eigenvalues do not have too large a coefficient of variation, preferably under $1$. For larger CV, the Type I errors may be larger than nominal.

For the general case of a noncentral quadratic form, the distribution of $Q$ can be approximated by the distribution of $c U^r$, where the distribution of $U$ can depend on one or two parameters. The choice of $c,\;r$, and the parameters of $U$ then permits matching the necessary moments. \cite{Solomon1977} consider three moment-based approximations: a four-moment approximation by a Type III Pearson curve and two three-moment approximations, one with $U \sim N(\mu,\sigma^2)$ and the other with $U \sim \chi^2_p$. They recommend the latter as fitting better in the lower tail, partly because it necessarily starts at zero, whereas the other approximations do not. This approximation matches the constants $c,\;r$, and $p$ to the first three moments of $Q$. For $c (\chi^2_p)^r$ the moments about 0 are $\mu_k' = c^k 2^{kr} \Gamma(kr + p/2)/ \Gamma(p/2)$.

Other, more-complicated methods include the \cite{Sheil_1977_JRSC_92} algorithm, improved by \cite{Farebrother1984}, which represents the value of the c.d.f. for a noncentral quadratic form by an infinite sum of central chi-square probabilities. In particular, when $\mu = 0$, the representation in Farebrother's algorithm is exact; it involves a finite number of chi-square probabilities.

 \cite{Bodenham2016} and \cite{ChenLumly2019} discuss the behavior of various approximations when $K$ is large.

In meta-analysis, approximations to the distribution of $Q$ have usually been sought only for the $Q_{IV}$
version with non-constant inverse-variance weights. Typically, the chi-square distribution with $K - 1$ degrees of freedom is used indiscriminately as the null distribution of $Q_{IV}$. For MD, \cite{Kulin2004contrasts} introduced an improved two-moment approximation to this version of $Q$ based on the \cite{welch1951comparison} test in the heteroscedastic ANOVA. The distribution of this  Welch test for MD is approximated under the null by a rescaled F distribution, and under alternatives by a shifted chi-square distribution.  \cite{biggerstaff2008exact} used the Farebrother approximation to the distribution of a quadratic form in normal variables as the \lq\lq exact" distribution of $Q_{IV}$ with  IV weights. This is not correct when the weights are the reciprocals of estimated variances, but with constant weights it is correct for MD.   When $\tau^2=0$, the Biggerstaff and Jackson approximation to the distribution of $Q_{IV}$ is the $\chi^2_{K-1}$ distribution.

\section{Variance and third moment of $Q$} \label{sec:varQ}
For approximations that require the first two or three moments, we derive the second and the third moments of $Q$ under the REM introduced in Section~\ref{sec:Intro}.

We distinguish between the conditional distribution of $Q$ (given $\theta_i$) and the unconditional distribution, and the respective moments of $\Theta_i$. For instance, the conditional second moment of $\Theta_i$ is $M_{2i}^c = v_i^2$, and the unconditional second moment  is  $M_{2i} = \e(\Theta_i^2) = \var(\hat{\theta}_i) = \e(v_i^2) + \tau^2$.
Similarly,  $M_{4i} = \e(\Theta_i^4)$ is the fourth (unconditional) central moment of $\hat{\theta}_i$. These two moments are required to calculate the variance of $Q$, given by
\begin{equation} \label{M2Q}
W^{-2} \var(Q) = \sum_i q_i^2 (1 - q_i)^2 (M_{4i} - M_{2i}^2) + 2 \sum_{i \not = j} q_i^2 q_j^2 M_{2i} M_{2j}.
\end{equation}
The details are in Appendix~\ref{2m_det}.
When the weights are not related to the effect, these expressions for the mean and variance of $Q$ are the same as in \cite{kulinskaya2011testing}.

For (known) inverse-variance weights $w_i = v_i^{-2}$, and assuming that each $\hat\theta_i$ is normally distributed and $\tau^2 = 0$, so that $M_{2i} = v_i^2$ and $M_{4i} = 3v_i^4$, the first moment of $Q$ is $K - 1$, and the variance is $2(K - 1)$, as it should be for a chi-square distribution with $K - 1$ degrees of freedom.

In general, the unconditional moments $M_{2i}$ and $M_{4i}$ depend on the effect measure (through its second and fourth conditional moments) and on the REM that defines the unconditional moments. Appendix~\ref{sec:NecMoments} gives the details.

In the null distribution, $\tau^2 = 0$, and the unconditional moments of $Q$ coincide with its conditional moments.

The derivation for the unconditional third moment of $Q$
\begin{equation*}
W^{-3} \e(Q^3)  =  \e\{ [\sum q_i (1 - q_i) \Theta_i^2 - \mathop{\sum\sum}_{i\neq j} q_i q_j \Theta_i \Theta_j]^3 \} \\
\end{equation*}
parallels that for the second moment, starting from Equation~(\ref{Q1}). Appendix~\ref{3m_det} gives the details of the derivation. The result is
\begin{equation} \label{M3Q}
\begin{array}{lll}
W^{-3} \e(Q^3) & = & \sum_{i} q_i^3 (1 - q_i)^3 ( M_{6i} - 3 M_{4i} M_{2i} + 2 M_{2i}^3 ) \\
                        &    & + 3 [ \sum_{j} q_j (1 - q_j) M_{2j} ] [ \sum_{i} q_i^2 (1 - q_i)^2 ( M_{4i} - M_{2i}^2 ) ] \\
                        &    & + [ \sum_{i} q_i (1 - q_i) M_{2i} ]^3 \\
                        &    & - 6 \mathop{\sum\sum}_{i\neq j} q_i^2 (1 - q_i) q_j^2 (1 - q_j) M_{3i} M_{3j} \\
                        &    & + 12 \mathop{\sum\sum}_{i\neq j} q_i^3 (1 - q_i) q_j^2 M_{4i} M_{2j} \\
                        &    & + 6 \mathop{\sum\sum\sum}_{i\neq j\neq k} q_i (1 - q_i) q_j^2 q_k^2 M_{2i} M_{2j} M_{2k} \\
                        &    & - 4 \mathop{\sum\sum}_{i\neq j} q_i^3 q_j^3 M_{3i} M_{3j} \\
                        &    & - 8 \mathop{\sum\sum\sum}_{i\neq j\neq k} q_i^2 q_j^2 q_k^2 M_{2i} M_{2j} M_{2k} .
\end{array}
\end{equation}

Importantly, $M_{3i} = \e(\Theta_i^3)$ and  $M_{6i} = \e(\Theta_i^6)$, the third and the sixth unconditional central moments of $\hat{\theta}_i$, are required for this calculation in addition to the second and fourth  central moments used in calculating the second moment of $Q$.

Unconditional central moments of $\hat{\theta}_i$ are linear combinations of expected values of conditional moments, their cross-products, and powers of $\tau^2$. Appendix~\ref{sec:NecMoments} provides the requisite expressions for the first six unconditional central moments for a general effect measure. Calculations of unconditional  moments  are much simpler for the mean difference (MD), as we show next.

\section{Unconditional central moments of $\hat\theta_i$ for mean difference}
Assume that each of $K$ studies consists of two groups whose data are normally distributed with sample sizes $n_{iC}$ and $n_{iT}$ and
means $\mu_{iC}$ and $\mu_{iT} = \mu_{iC} + \Delta_i$, and possibly different variances
$\sigma_{iC}^2$ and $\sigma_{iT}^2$.
Then the mean difference $\Delta_i$ in Study $i$ is estimated by
\begin{equation}
\hat\Delta_i = \bar X_{iT} - \bar X_{iC} ,
\end{equation}
and its (conditional) variance $v_i^2 = \sigma_{iT}^2 / n_{iT} + \sigma_{iC}^2 / n_{iC}$.
The conditional distribution of $\hat{\Delta}_i$ is $N(\Delta_i, v_i^2)$, so its odd central moments are zero, and its even moments are $M_{i,2r}^c = [(2r)! / (2^r r!)] v_i^{2r}$. As the conditional moments do not involve $\Delta_i$, it is easy to write out the unconditional moments: \\
$M_{2i} = v_i^2 + \tau^2$, \\
$M_{4i} = 3 v_i^4 + 6 v_i^2 \tau^2 + 3 \tau^4$,\\
$M_{6i} = 15 v_i^6 + 15 * 3 v_i^4 \tau^2 + 15 v_i^2 * 3 \tau^4 + 15 \tau^6$.

The first three moments of $Q$ can be calculated by substituting these moments into Equations~(\ref{M1Q}), (\ref{M2Q}), and (\ref{M3Q})

\section{Simulation study for mean difference}

For MD as the effect measure, we use simulation of the distribution of $Q$  with constant effective-sample-size weights (SW) $\tilde n_{i}=n_{iC}n_{iT}/(n_{iC}+n_{iT})$ to study three approximations: the Farebrother approximation (F SW), implemented in the R package {\it CompQuadForm} \citep{Duchesne2010};  the  two-moment Welch-Satterthwaite approximation (M2 SW); and the three-moment chi-square approximation (M3 SW) by \cite{Solomon1977}. Details of these two moment-based approximations are given in Appendix~\ref{M23}. We also study the bias of the moment estimator  $\hat \tau^2_M$ in Equation~(\ref{tau_DSK}), denoted by SDL, for this choice of constant weights.

For comparison, we also simulate $Q$ with IV weights, and  study three approximations to its distribution: the standard chi-square approximation   and the approximation based on the Welch test to the null distribution of $Q$,  and the \lq\lq exact" distribution of \cite{biggerstaff2008exact} (BJ) when $\tau^2>0$. To compare the bias of SDL with that of estimators  of $\tau^2$ that use the IV weights, we also consider  \cite{dersimonian1986meta} (DL), \cite{mandel1970interlaboratory} (MP), REML, and a corrected DL estimator (CDL), which uses an improved non-null first moment of $Q$ \citep{BHK2018SMD}.

We varied five parameters: the number of studies $K$, the total (average) sample size of each study $n$ (or $\bar{n}$), the proportion of observations in the Control arm $f$, the between-study variance $\tau^2$, and the within-study variance $\sigma_{T}^2$ (keeping $\sigma_{C}^2 = 1$).  We set the overall true MD $\mu = 0$ because the estimators of $\tau^2$ do not involve $\mu$ and the estimators of $\mu$ are equivariant.

We generate the within-study sample variances $s_{ij}^2$ ($j = T, \;C$) from chi-square distributions $\sigma_{ij}^2 \chi_{n_{ij} - 1}^2 / (n_{ij} - 1)$ and the estimated mean differences $y_i$ from a normal distribution with mean $\mu$ and variance $\sigma_{iT}^2 / n_{iT} + \sigma_{iC}^2 / n_{iC} + \tau^2$. We obtain the estimated within-study variances as $\hat{v}_{i}^2 = s_{iT}^2 / n_{iT} + s_{iC}^2 / n_{iC}$. As would be required in practice, all approximations use these  $\hat{v}_{i}^2$, even though the $\sigma_{iT}^2 / n_{iT} + \sigma_{iC}^2 / n_{iC}$ are available in the simulation.

All simulations use the same numbers of studies  $K = 5, \;10, \;30$ and,  for each combination of parameters, the same vector of total sample sizes $n = (n_{1}, \ldots, n_{K})$ and the same proportions of observations in the Control arm $f_i = .5, \;.75$ for all $i$. The sample sizes in the Treatment and Control arms are $n_{iT}=\lceil{(1 - f_i)n_{i}}\rceil$ and $n_{iC}=n_{i}-n_{iT}$, $i=1,\ldots, K$. The values of $f$ reflect two situations for the two arms of each study: approximately equal (1:1) and quite unbalanced (1:3).

We study equal and unequal study sizes. For equal study sizes $n_i$ is as small as 20, and for unequal study sizes average sample size $\bar{n}$ is as small as 13 (individual $n_i$ are as small as 4), in order to examine how the methods perform for the extremely small sample sizes that arise in some areas of application. In  choosing unequal study sizes, we follow a suggestion of \cite{sanches-2000}.
Table~\ref{tab:altdataMD} gives the details.

\begin{table}[ht]
	\caption{\label{tab:altdataMD} \emph{Data patterns in the simulations}} \label{tab1}
	\begin{footnotesize}
		\begin{center}
			
			\begin{tabular}
				{|l|l|l|}
				\hline
				$Q$ for MD&Equal study sizes& Unequal study sizes \\ 				
				&& \\
				\hline
				$K$ (number of studies)& 5, 10, 30&5, 10, 30\\
				$n$ or $\bar{n}$  (average (individual) study size ---  & 20, 40, 100, 250& 13 (4, 6, 7, 8, 40), \\
				total of the two arms)&&15 (6, 8, 9, 10, 42), \\
            For  $K=10$ and $K=30$,  the same set  	&&30 (12, 16, 18, 20, 84), \\
				of unequal study sizes is used twice or &&60 (24, 32, 36, 40, 168) \\
                six times, respectively&&\\
				
				$f$ (proportion  in the control arm) & 1/2, 3/4&1/2, 3/4\\
				\hline
				$\mu^{}$&0&0  \\
				$\sigma_{C}^2,\sigma_{T}^2$ (within-study variances)&(1,1), (1,2)&(1,1), (1,2)\\
				$\tau^{2}$ (variance of random effects)&0(0.1)1&0(0.1)1   \\
				\hline
			\end{tabular}
		\end{center}
	\end{footnotesize}
\end{table}

We use a total of $10,000$ repetitions for each combination of parameters. Thus, the simulation standard error for an empirical p-value $\hat p$ under the null is roughly $\sqrt{1/(12*10,000)}=0.0029$.
The simulations were programmed in R version 3.6.2 using the University of East Anglia 140-computer-node High Performance Computing (HPC) Cluster, providing a total of 2560 CPU cores, including parallel processing and large memory resources. For each configuration, we divided the 10,000 replications into 10 parallel sets of 1000.

\section{Results}

For each configuration of parameters in the simulation study and for each approximation, we calculated, for each generated value of $Q$, the probability of a larger $Q$: $\tilde{p} = 1 - \hat{F}(Q)$ ($\hat{F}$ denotes the distribution function of the approximation). We recorded empirical p-values $\hat{p} = \# (\tilde{p} < p) / 10000$ at $p$ = .001, .0025, .005, .01, .025, .05, .1, .25, .5 and the complementary values .75, \ldots, .999. The values of $\tau^2$ included both null ($\tau^2 = 0$) and non-null ($\tau^2 > 0$) values (Table~\ref{tab1}). The parameters of the approximations (e.g., moments and IV weights) were based on the value of $\tau^2$ used in the simulation. These data provide the basis for P--P plots (versus the true null distribution) for three  approximations to the distribution of $Q$ with effective-sample-size weights (F SW, M2 SW, and M3 SW) and two approximations to the distribution of $Q$ with IV weights (chi-square/BJ and Welch) and for estimating their null levels, non-null empirical tail areas, and (roughly) their power. We also estimate the bias of five point estimators of $\tau^2$ (SDL, DL, REML, MP, and CDL). The full results are presented, graphically, in Appendix B.

In some instances M3 SW produced anomalous results or no results at all (because numerical problems kept us from obtaining estimates of its parameters).

\subsection{P--P plots}
To compare an approximation for a distribution function of $Q$ against the theoretical distribution function, with no heterogeneity ($\tau^2 = 0$), we use probability--probability (P--P) plots \citep{Wilk_1968_BMTA_1}. Evaluating two distribution functions, $F_1$ and $F_2$, at $x$ yields $p_1 = F_1(x)$ and $p_2 = F_2(x)$. One varies $x$, either continuously or at selected values, and plots the points $( p_1(x), p_2(x) )$ to produce the usual P--P plot of $F_2$ versus $F_1$. If $F_2 = F_1$, the points lie on the line from (0, 0) to (1, 1). If smaller $x$ are more likely under $F_2$, the points will lie above the line, and conversely. (Working with upper tail areas reverses these interpretations.) If $F_2$ is similar to $F_1$,  the points will lie close to the line, and departures will show areas of difference. To make these more visible, we flatten the plot by subtracting the line; that is, we plot $p_2 - p_1$ versus $p_1$.

The simulations offer a shortcut that does not require evaluating the true distribution function of $Q$ (which is unknown for IV weights). If $F$ is the distribution of the random variable $X$, $F(X)$ has the uniform distribution on [0, 1], and so does $1 - F(X)$. Thus, for the values of $p$ listed above, we plot $\hat{p} - p$ versus $p$.

Our P--P plots (illustrated by Figure~\ref{BiasPhatMD0_S1_1q052Sigma2T1_unequal}, Figure~\ref{BiasPhatMD0_S1_1q052Sigma2T1}, Figure~\ref{BiasPhatMD0_S1_1q0752Sigma2T1}, and Figure~\ref{BiasPhatMD0_S2_1q0752Sigma2T1}) show no differences between the M3 and M2  approximations for $Q$ with constant weights. Very minor differences between the Farebrother and the moment approximations are visible, mainly at very small sample sizes. Other comparisons show three distinct patterns.

The chi-square  approximation has strikingly higher empirical tail areas than the true distribution of $Q$ with IV weights over the whole domain. This pattern is especially noticeable for $K=30$ and small unequal sample sizes, though it persists for equal sample sizes as large as 100. It indicates that the approximating chi-square distribution produces values that are systematically too large,

The Welch test provides a much better fit that is especially good for balanced sample sizes, equal variances,  and small $K$. When sample sizes are small and vary among studies or are unbalanced between arms, however, its fit is worse. It produces values of $Q$ that are systematically too small when $K = 5$; produces more small values and, to a lesser extent, more large values when $K = 10$; and produces more large values and, to a lesser extent, more small values when $K = 30$.

The three approximations to $Q$ with constant weights provide reasonably good fits, which appear to be similar to the fit of the Welch test to $Q$ with IV weights.

\subsection{Empirical levels when $\tau^2 = 0$}
To better visualize the quality of the approximations as the basis for a test for heterogeneity at the .05 level, we plotted their empirical levels under the null $\tau^2=0$ versus sample size. Figure~\ref{PlotOfPvaluesAgainstN_005level} and Figure~\ref{PlotOfPvaluesAgainstN_005level_unequal} present typical results for small sample sizes at the .05 level.

For equal variances, the empirical levels depend on the sample size.  The  chi-square test is very liberal up to $n = 100$, especially for unbalanced arms, and the problem becomes worse as $K$ increases.  It  is considerably better than the chi-square, but is still noticeably liberal when the arms are unbalanced. Tests based on $Q$ with constant weights are generally less liberal, though they may have level up to .07 for $n = 20$, for unbalanced arms and small $K$. The M3 approximation breaks down and results in very liberal levels for unequal sample sizes and unbalanced arms and large $K$. The Farebrother and M2 approximations perform better for larger $K$, and overall are the best choice. They also hold the level well at smaller nominal levels. The Welch test is rather unstable for very low levels such as $\alpha = .001$ (where in our simulations with 10,000 replicates, the empirical values are  based  just on 10 observations), but improves from  $\alpha =  .005$.

\subsection{Empirical levels when $\tau^2 > 0$}
To understand how the approximations behave as $\tau^2$ increases, we plotted the empirical p-values ($\hat{p}$) vs $\tau^2$ for the nominal levels .05 and .01 (Figure~\ref{PlotOfPhatAt001Sigma2T2andq05MD_underH1}, Figure~\ref{PlotOfPhatAt001Sigma2T2andq075MD_underH1}, Figure~\ref{PlotOfPhatAt001Sigma2T2andq05MD_underH1_unequal}, and Figure~\ref{PlotOfPhatAt005Sigma2T2andq075MD_underH1_unequal}). For unequal sample sizes, the Farebrother and the 3-moment approximation differ slightly at the .01 level, but those differences disappear at the .05 level and for equal sample sizes. When $K = 30$,  M3 sometimes fails; and when it does not, it breaks down for small and large values of $\tau^2$.   The 2-moment approximation is almost indistinguishable from the Farebrother approximation.

Overall, the Farebrother approximation performs superbly across all $\tau^2$ values. This is as it should be, as it is practically an exact distribution in the case of MD. The M2 approximation is reasonably good at the .05 level. The BJ approximation is much too liberal, especially at smaller values of $\tau^2$ and for larger $K$.

\subsection{Power of tests for heterogeneity}
\lq\lq Power" is a reasonable term as a heading, but not as an accurate description for most of the results. Although discussions of simulation results in meta-analysis do not always do this, comparisons of power among tests that are intended to have a specified level (i.e., rate of Type I error) are not valid unless the tests' estimated levels are equal or nearly so. This complication is evident in Figure~\ref{PlotOfPhatAt005Sigma2T2andq05MD_underH0}, Figure~\ref{PlotOfPhatAt005Sigma2T2andq075MD_underH0}, Figure~\ref{PlotOfPhatAt005Sigma2T2andq05MD_underH0_unequal}, and Figure~\ref{PlotOfPhatAt005Sigma2T2andq075MD_underH0_unequal}, which depict the power of tests of heterogeneity at the .05 level for $n = 20$ and equal and unequal sample sizes.

The chi-square test appears to be more powerful, and the Welch test  slightly less powerful, than the tests based on $Q$ with constant weights.  These differences are much smaller when $n = 40$ (not shown) and disappear when $n$ is larger.  But even for $n=20$, these appearances are misleading. For $n=20$, Figure~\ref{PlotOfPvaluesAgainstN_005level} and Figure~\ref{PlotOfPvaluesAgainstN_005level_unequal} show that for balanced arms, the level of the chi-square test is $.08$ for $K = 5$, $.1$ for $K = 10$, and considerably higher than $.1$ for $K = 30$. For unbalanced arms, the level of the chi-square test substantially exceeds $.1$ for all $K$. Thus, our results do not show that the chi-square test has higher power, and its power may actually be lower. It is not clear how to modify the chi-square test so that it has the correct level in a broad range of situations.

The Welch test has levels similar to those of the tests based on $Q$ with constant weights when $K = 5$ or 10. But for $K = 30$ and $f = .75$, its level is approximately $.09$. This may mean that it does have somewhat lower power.

At $n = 100$ (or at $\bar{n} = 60$ for unequal sample sizes), visible differences among the traces for the tests disappear. Given higher levels of the chi-square test, this means that its power is the same or even lower than that of the tests based on $Q$ with constant weights.

\subsection{Bias in estimation of $\tau^2$}
Here we compare the SDL estimator of $\tau^2$ with the well-known estimators DL, MP, and REML and the recently suggested CDL. Figure~\ref{BiasTauMD0_S1_1q052Sigma2T1_unequal}, Figure ~\ref{BiasTauMD0_S1_1q052Sigma2T1}, Figure~\ref{BiasTauMD0_S1_1q0752Sigma2T1}, and Figure~\ref{BiasTauMD0_S2_1q0752Sigma2T1} depict the biases of the five estimators for small sample sizes.

All five estimators have positive bias at $\tau^2 = 0$, because of truncation at zero. The bias across all values of $\tau^2$  is quite substantial,  and it increases for unequal variances and/or sample sizes. Among the standard estimators, DL has the most bias and MP the least.  SDL and CDL generally have similar bias, considerably less than the standard estimators. The relation of their bias to $K$ when $\bar{n} = 13$ is interesting, but atypical. As $K$ increases, the trace for SDL flattens toward 0, demonstrating no bias at all for larger values of $\tau^2$, whereas the trace for CDL rises toward the other three.  By $n = 100$ or $\bar{n} = 60$, the differences among the five estimators of $\tau^2$ are quite small.

\section{Discussion}
It is well known that a large number of small studies is the worst-case scenario for the statistical properties  of meta-analysis \citep{BHK2018SMD}.  This situation may not be very widespread in medical meta-analyses, but it is very common in the social sciences and in ecology \citep{sanches-2010, Hamman2018}.

We considered  a heterogeneity test based on a $Q$ statistic with constant sample-size-based weights. We derived its first three moments for a general effect measure and, in a simulation study, compared the properties of the test for heterogeneity for MD with its IV-weights-based counterparts. We also studied, by simulation, bias of the resulting estimator SDL of the heterogeneity variance $\tau^2$ in comparison with several popular standard estimators.

Overall, the proposed  test for heterogeneity for MD, combined with its exact distribution as obtained by the \cite{Farebrother1984} algorithm or, alternatively, with  the two-moment approximation,  provides very precise control of the significance level, even in the case of small and unbalanced sample sizes, in contrast to the extremely liberal behavior of the standard tests, especially for a large number of studies. (These results suggest that the null distribution of $Q_{IV}$ is more difficult to approximate than the null distribution of $Q$ with constant weights.) Similarly, the proposed SDL estimator is almost unbiased for $K \geq 10$, even in the case of extremely small sample sizes.

It is enlightening to observe that, for the non-null distribution of $Q_{IV}$, the approximation of \cite{biggerstaff2008exact} (using Farebrother's algorithm) is no better than the standard chi-square approximation to the null distribution. The problem here evidently lies with the IV weights.

We found that, even though both moment approximations performed well overall, the three-moment approximation sometimes fails, and it breaks down in the case of very small and unbalanced sample sizes and a large number of studies.  Therefore, for MD we recommend the \cite{Farebrother1984} approximation to the distribution of $Q$ with constant weights.

In further work we intend to develop tests for heterogeneity in other effect measures based on $Q$ with constant weights. Even though we derived general expressions for the moments of $Q$, application of these expressions to such effect measures as SMD and the odds ratio involves a lot of tedious algebra.
The moment approximations are less precise than the exact distribution or the approximation by  \cite{Farebrother1984} for the case of normal variables in the quadratic form, but they are much faster and  may be a better option when the distribution is only asymptotically normal.

\bibliography{Qfixed}

\begin{thebibliography}{24}
\providecommand{\natexlab}[1]{#1}
\providecommand{\url}[1]{\texttt{#1}}
\expandafter\ifx\csname urlstyle\endcsname\relax
  \providecommand{\doi}[1]{doi: #1}\else
  \providecommand{\doi}{doi: \begingroup \urlstyle{rm}\Url}\fi

\bibitem[Bakbergenuly et~al.(2020{\natexlab{a}})Bakbergenuly, Hoaglin, and
  Kulinskaya]{BHK2018SMD}
Ilyas Bakbergenuly, David~C. Hoaglin, and Elena Kulinskaya.
\newblock Estimation in meta-analyses of mean difference and standardized mean
  difference.
\newblock \emph{Statistics in Medicine}, 39\penalty0 (2):\penalty0 171--191,
  2020{\natexlab{a}}.

\bibitem[Bakbergenuly et~al.(2020{\natexlab{b}})Bakbergenuly, Hoaglin, and
  Kulinskaya]{BHK2020LOR}
Ilyas Bakbergenuly, David~C. Hoaglin, and Elena Kulinskaya.
\newblock Methods for estimating between-study variance and overall effect in
  meta-analyses of odds-ratios.
\newblock \emph{Research Synthesis Methods}, 11:\penalty0 426--442,
  2020{\natexlab{b}}.
\newblock \doi{10.1002/jrsm.1404}.

\bibitem[Biggerstaff and Jackson(2008)]{biggerstaff2008exact}
Brad~J. Biggerstaff and Dan Jackson.
\newblock The exact distribution of {C}ochran's heterogeneity statistic in
  one-way random effects meta-analysis.
\newblock \emph{Statistics in Medicine}, 27\penalty0 (29):\penalty0 6093--6110,
  2008.

\bibitem[Bodenham and Adams(2016)]{Bodenham2016}
D.A. Bodenham and N.M. Adams.
\newblock A comparison of efficient approximations for a weighted sum of
  chi-squared random variables.
\newblock \emph{Statistics and Computing}, 26:\penalty0 917--928, 2016.
\newblock \doi{10.1007/s11222-015-9583-4}.

\bibitem[Chen and Lumley(2019)]{ChenLumly2019}
Tong Chen and Thomas Lumley.
\newblock Numerical evaluation of methods approximating the distribution of a
  large quadratic form in normal variables.
\newblock \emph{Computational Statistics \& Data Analysis}, 139:\penalty0
  75--81, 2019.
\newblock \doi{10.1016/j.csda.2019.05.002}.

\bibitem[Cochran(1954)]{cochran1954combination}
William~G. Cochran.
\newblock The combination of estimates from different experiments.
\newblock \emph{Biometrics}, 10\penalty0 (1):\penalty0 101--129, 1954.

\bibitem[DerSimonian and Kacker(2007)]{dersimonian2007random}
Rebecca DerSimonian and Raghu Kacker.
\newblock Random-effects model for meta-analysis of clinical trials: an update.
\newblock \emph{Contemporary Clinical Trials}, 28\penalty0 (2):\penalty0
  105--114, 2007.

\bibitem[DerSimonian and Laird(1986)]{dersimonian1986meta}
Rebecca DerSimonian and Nan Laird.
\newblock Meta-analysis in clinical trials.
\newblock \emph{Controlled Cinical Trials}, 7\penalty0 (3):\penalty0 177--188,
  1986.

\bibitem[Duchesne and {Lafaye De Micheaux}(2010)]{Duchesne2010}
Pierre Duchesne and Pierre {Lafaye De Micheaux}.
\newblock Computing the distribution of quadratic forms: Further comparisons
  between the {L}iu–{T}ang–{Z}hang approximation and exact methods.
\newblock \emph{Computational Statistics \& Data Analysis}, 54\penalty0
  (4):\penalty0 858--862, 2010.

\bibitem[Farebrother(1984)]{Farebrother1984}
R.~W. Farebrother.
\newblock Algorithm {AS} 204: The distribution of a positive linear combination
  of $\chi^2$ random variables.
\newblock \emph{Journal of the Royal Statistical Society, {S}eries {C}},
  33\penalty0 (3):\penalty0 332--339, 1984.

\bibitem[Hamman et~al.(2018)Hamman, Pappalardo, Bence, Peacor, and
  Osenberg]{Hamman2018}
Elizabeth~A. Hamman, Paula Pappalardo, James~R. Bence, Scott~D. Peacor, and
  Craig~W. Osenberg.
\newblock Bias in meta-analyses using {H}edges\rq {} d.
\newblock \emph{Ecosphere}, 9\penalty0 (9):\penalty0 e02419, 2018.
\newblock \doi{10.1002/ecs2.2419}.
\newblock URL
  \url{https://esajournals.onlinelibrary.wiley.com/doi/abs/10.1002/ecs2.2419}.

\bibitem[Kulinskaya et~al.(2004)Kulinskaya, Dollinger, Knight, and
  Gao]{Kulin2004contrasts}
E.~Kulinskaya, M.~B. Dollinger, E.~Knight, and H.~Gao.
\newblock A {W}elch-type test for homogeneity of contrasts under
  heteroscedasticity with application to meta-analysis.
\newblock \emph{Statistics in Medicine}, 23\penalty0 (23):\penalty0 3655--3670,
  2004.
\newblock \doi{10.1002/sim.1929}.

\bibitem[Kulinskaya et~al.(2011)Kulinskaya, Dollinger, and
  Bj{\o}rkest{\o}l]{kulinskaya2011testing}
Elena Kulinskaya, Michael~B. Dollinger, and Kirsten Bj{\o}rkest{\o}l.
\newblock Testing for homogeneity in meta-analysis {I}. {T}he one-parameter
  case: standardized mean difference.
\newblock \emph{Biometrics}, 67\penalty0 (1):\penalty0 203--212, 2011.

\bibitem[Mandel and Paule(1970)]{mandel1970interlaboratory}
John Mandel and Robert~C Paule.
\newblock Interlaboratory evaluation of a material with unequal numbers of
  replicates.
\newblock \emph{Analytical Chemistry}, 42\penalty0 (11):\penalty0 1194--1197,
  1970.

\bibitem[S{\'a}nchez-Meca and Mar{\'\i}n-Mart{\'\i}nez(2000)]{sanches-2000}
Julio S{\'a}nchez-Meca and Fulgencio Mar{\'\i}n-Mart{\'\i}nez.
\newblock Testing the significance of a common risk difference in
  meta-analysis.
\newblock \emph{Computational Statistics \& Data Analysis}, 33\penalty0
  (3):\penalty0 299--313, 2000.

\bibitem[S{\'a}nchez-Meca and Mar{\'\i}n-Mart{\'\i}nez(2010)]{sanches-2010}
Julio S{\'a}nchez-Meca and Fulgencio Mar{\'\i}n-Mart{\'\i}nez.
\newblock Meta-analysis in psychological research.
\newblock \emph{International Journal of Psychological Research}, 3\penalty0
  (1):\penalty0 150--162, 2010.

\bibitem[Satterthwaite(1941)]{Satterthwaite1941}
Franklin~E. Satterthwaite.
\newblock Synthesis of variance.
\newblock \emph{Psychometrika}, 6\penalty0 (5):\penalty0 309--316, 1941.

\bibitem[Sheil and O'Muircheartaigh(1977)]{Sheil_1977_JRSC_92}
J.~Sheil and I.~O'Muircheartaigh.
\newblock Algorithm {AS} 106. {T}he distribution of non-negative quadratic
  forms in normal variables.
\newblock \emph{Applied Statistics}, 26:\penalty0 92--98, 1977.

\bibitem[Soetaert et~al.(2020)Soetaert, Hindmarsh, Eisenstat, Moler, Dongarra,
  and Saad]{rootsolve}
Karline Soetaert, Alan~C. Hindmarsh, S.C. Eisenstat, Cleve Moler, Jack
  Dongarra, and Youcef Saad.
\newblock Nonlinear root finding, equilibrium and steady-state analysis of
  ordinary differential equations.
\newblock https://cran.r-project.org/web/packages/root{S}olve/root{S}olve.pdf,
  April 2020.
\newblock Version 1.8.2.1.

\bibitem[Solomon and Stephens(1977)]{Solomon1977}
Herbert Solomon and Michael~A. Stephens.
\newblock Distribution of a sum of weighted chi-square variables.
\newblock \emph{Journal of the American Statistical Association}, 72\penalty0
  (360):\penalty0 881--885, 1977.

\bibitem[Welch(1938)]{Welch1938}
B.~L. Welch.
\newblock The significance of the difference between two means when the
  population variances are unequal.
\newblock \emph{Biometrika}, 29\penalty0 (3/4):\penalty0 350--362, 1938.

\bibitem[Welch(1951)]{welch1951comparison}
B.~L. Welch.
\newblock On the comparison of several mean values: an alternative approach.
\newblock \emph{Biometrika}, 38\penalty0 (3/4):\penalty0 330--336, 1951.

\bibitem[Wilk and Gnanadesikan(1968)]{Wilk_1968_BMTA_1}
M.~B. Wilk and R.~Gnanadesikan.
\newblock Probability plotting methods for the analysis of data.
\newblock \emph{Biometrika}, 53\penalty0 (1):\penalty0 1--17, 1968.

\bibitem[Yuan and Bentler(2010)]{YuanBentler2010}
Ke-Hai Yuan and Peter~M. Bentler.
\newblock Two simple approximations to the distributions of quadratic forms.
\newblock \emph{British Journal of Mathematical and Statistical Psychology},
  63\penalty0 (2):\penalty0 273--291, 2010.

\end{thebibliography}
\renewcommand{\thefigure}{A.\arabic{figure}}
\renewcommand{\thesection}{A.\arabic{section}}
\renewcommand{\thetable}{A.\arabic{table}}
\setcounter{figure}{0}
\setcounter{section}{0}
\setcounter{table}{0}
\clearpage

\text{\LARGE{\bf{Appendix A}}}
\section{Details for calculation of the second moment of $Q$}\label{2m_det}
The second moment of $Q$ (times $W^{-2}$) is
\begin{eqnarray*}
W^{-2} \e(Q^2) & = & \e \left[ \sum q_i (1 - q_i) \Theta_i^2 \right]^2 \\
                        &    & -2 \e \left( \left[ \sum q_k (1 - q_k) \Theta_k^2 \right] \left[ \sum_{i \neq j} q_i q_j \Theta_i \Theta_j \right] \right) \\
                        &    & + \e \left[ \sum_{i \neq j} q_i q_j \Theta_i \Theta_j \right]^2 \\
                        & = & A - 2B + C.
\end{eqnarray*}

The first term,
\begin{eqnarray*}
A & = & \e \left( \sum_{i,j} q_i (1 - q_i) q_j (1 - q_j) \Theta_i^2 \Theta_j^2 \right) \\
   & = & \e \left( \sum_i q_i^2 (1-q_i)^2 \Theta_i^4 \right) + \e \left( \sum_{i \neq j} q_i (1 - q_i) q_j (1 - q_j) \Theta_i^2 \Theta_j^2 \right) \\
   & = & \sum_i q_i^2 (1 - q_i)^2 (M_{4i} - M_{2i}^2) + \left[ \sum q_i (1 - q_i) M_{2i} \right]^2,
\end{eqnarray*}
where $M_{2i} = \e(\Theta_i^2) = \var(\Theta_i) = \e(v_i^2) + \tau^2$ is the variance, and $M_{4i} = \e(\Theta_i^4)$ is the fourth central moment of $\hat\theta_i$.

The second term, $B = 0$ because its terms $\e(\Theta_k^2 \Theta_i \Theta_j)$, with $i \not = j$,  always include a first-order moment of $\Theta_i$ for some $i$.

In the third term, $C = \e[\sum_{i \not = j} \sum_{k \not = l} q_i q_j q_k q_l \Theta_i \Theta_j \Theta_k \Theta_l]$, the only nonzero terms have $i = k$ and $j = l$ or $i = l$ and $j = k$, so
$C = 2 \sum_{i \not = j} q_i^2 q_j^2 M_{2i} M_{2j}$.

To obtain $W^{-2}$ times the variance of $Q$, we subtract the square of its mean, given by Equation~(\ref{M1Q}), which is exactly the second term of $A$:
\begin{equation*}
W^{-2} \var(Q) = \sum_i q_i^2 (1-q_i)^2 (M_{4i} - M_{2i}^2) + 2 \sum_{i \not = j} q_i^2 q_j^2 M_{2i} M_{2j}.
\end{equation*}

\section{Details for calculation of the third moment of $Q$}\label{3m_det}

To support Section~\ref{sec:varQ}, we record selected steps in the calculation of the third moment of $Q$. We have

\begin{eqnarray*}
W^{-3} \e(Q^3) & = & \e\{ [\sum q_i (1 - q_i) \Theta_i^2 - \mathop{\sum\sum}_{i\neq j} q_i q_j \Theta_i \Theta_j]^3 \} \\
                         & = & \e \{ [ \sum q_i (1 - q_i) \Theta_i^2 ]^3 \} \\
                         &    & - 3 \e \{ [ \sum q_i (1 - q_i) \Theta_i^2]^2 [\mathop{\sum\sum}_{i\neq j} q_i q_j \Theta_i \Theta_j ] \} \\
                         &    & + 3 \e \{ [ \sum q_i (1 - q_i) \Theta_i^2 ] [ \mathop{\sum\sum}_{i\neq j} q_i q_j \Theta_i \Theta_j ]^2 \} \\
                         &    & - \e \{ [ \mathop{\sum\sum}_{i\neq j}q_i q_j \Theta_i \Theta_j ]^3 \} \\
                         & = & A - 3B + 3C - D
\end{eqnarray*}

The terms $A$, $B$, $C$, and $D$  are obtained below. 
\begin{equation*}
\begin{array}{ll}
A  &=  \e [ \sum_{i} \sum_{j} \sum_{k} q_i (1 - q_i) q_j (1 - q_j) q_k (1 - q_k) \Theta_i^2 \Theta_j^2 \Theta_k^2
] \\
   & =  \sum q_i^3 (1 - q_i)^3 M_{6i} \\
       & + 3 \mathop{\sum\sum}_{i\neq j} q_i^2 (1 - q_i)^2 q_j (1 - q_j) M_{4i} M_{2j} \\
       & + \mathop{\sum\sum\sum}_{i\neq j\neq k} q_i (1 - q_i) q_j (1 - q_j) q_k (1 - q_k) M_{2i} M_{2j} M_{2k} \\
   & =  \sum q_i^3 (1 - q_i)^3 M_{6i} \\
       & + 3 \{ [ \sum_{i} q_i^2 (1 - q_i)^2 M_{4i} ] [ \sum_{j} q_j (1 - q_j) M_{2j} ] - \sum q_i^3 (1 - q_i)^3 M_{4i} M_{2i} \} \\
       & + \{ [ \sum_{i} q_i (1 - q_i) M_{2i} ]^3 - 3 [ \sum_{i} q_i^2 (1 - q_i)^2 M_{2i}^2 ] [ \sum_{j} q_j (1 - q_j) M_{2j} ] + 2 \sum_{i} q_i^3 (1 - q_i)^3 M_{2i}^3 \}\\
   &=  \sum q_i^3 (1 - q_i)^3 [ M_{6i} - 3 M_{4i} M_{2i} + 2 M_{2i}^3 ] \\
       & + 3 [ \sum_{j} q_j (1 - q_j) M_{2j} ] [ \sum_{i} q_i^2 (1 - q_i)^2 (M_{4i} - M_{2i}^2) ] \\
       & + [ \sum_{i} q_i (1- q_i) M_{2i} ]^3
\end{array}
\end{equation*}

\begin{eqnarray*}
B & = & \e \{ [ \sum_{i} \sum_{j} q_i (1 - q_i) q_j (1 - q_j) \Theta_i^2 \Theta_j^2 ] [ \mathop{\sum\sum}_{i\neq j} q_i q_j \Theta_i \Theta_j ] \} \\
   & = & \e \{ [ \sum_{i} q_i^2 (1 - q_i)^2 \Theta_i^4 + \mathop{\sum\sum}_{i\neq j} q_i (1 - q_i) q_j (1 - q_j) \Theta_i^2 \Theta_j^2 ] [\mathop{\sum\sum}_{i\neq j} q_i q_j \Theta_i \Theta_j ] \} \\
   & = & \e \{ 2 \mathop{\sum\sum}_{i\neq j} q_i^3 (1 - q_i)^2 q_j \Theta_i^5 \Theta_j + \mathop{\sum\sum\sum}_{i \neq j \neq k} q_i ^2 (1 - q_i)^2 q_j q_k \Theta_i^4 \Theta_j \Theta_k \\
   &+& \mathop{\sum\sum}_{i\neq j} \mathop{\sum\sum}_{l \neq k} q_i (1 - q_i) q_j (1 - q_j) q_k q_l \Theta_i^2 \Theta_j^2 \Theta_k \Theta_l \} \\
   & = & 2 \mathop{\sum\sum}_{i\neq j} q_i^3 (1 - q_i)^2 q_j \e(\Theta_i^5) \e(\Theta_j) + \mathop{\sum\sum\sum}_{i \neq j \neq k} q_i^2 (1 - q_i)^2 q_j q_k  \e(\Theta_i^4) \e(\Theta_j) \e(\Theta_k) \\
   &+& \mathop{\sum\sum}_{i\neq j} \mathop{\sum\sum}_{l \neq k} q_i (1 - q_i) q_j (1 - q_j) q_k q_l  \e(\Theta_i^2 \Theta_j^2 \Theta_k \Theta_l )
\end{eqnarray*}
The first two summations are zero because $\e(\Theta_j) = 0$. In the third summation, however, some terms have (for example) $i = k$ and $j = l$, yielding $\e(\Theta_i^3) \e(\Theta_j^3)$. It is straightforward, but somewhat tedious, to identify those terms, The result is
\[ B = 2 \mathop{\sum\sum}_{i\neq j} q_i^2 (1 - q_i) q_j^2 (1 - q_j) M_{3i} M_{3j} \].
\begin{eqnarray*}
C & = & \e \{ [ \sum_{i} q_i (1 - q_i) \Theta_i^2 ] [ \mathop{\sum\sum}_{k \neq j} \mathop{\sum\sum}_{m \neq l} q_j q_k q_l q_m \Theta_j \Theta_k \Theta_l \Theta_m ] \} \\
    & = & \sum_{i} \mathop{\sum\sum}_{k \neq j} \mathop{\sum\sum}_{m \neq l} q_i (1 - q_i) q_j q_k q_l q_m \e( \Theta_i^2  \Theta_j \Theta_k \Theta_l \Theta_m )
\end{eqnarray*}
As in $B$, this summation contains some terms that do not vanish. Identifying those yields
\begin{eqnarray*}
C & = & 4 \mathop{\sum\sum}_{i\neq j} q_i^3 (1 - q_i) q_j^2 M_{4i} M_{2j} \\
    &    & + 2 \mathop{\sum\sum\sum}_{i\neq j\neq k}  q_i (1 - q_i) q_j^2 q_k^2 M_{2i} M_{2j} M_{2k}
\end{eqnarray*}

\[
D =  \e [ \mathop{\sum\sum}_{i\neq j} \mathop{\sum\sum}_{k \neq l} \mathop{\sum\sum}_{m \neq n} q_i q_j q_k q_l q_m q_n \Theta_i  \Theta_j \Theta_k \Theta_l \Theta_m \Theta_n ] \\
\]
As above, removing the terms that vanish leaves
\begin{eqnarray*}
D & = & 4 \mathop{\sum\sum}_{i\neq j} q_i^3 q_j^3 M_{3i} M_{3j} \\
   &     & + 8 \mathop{\sum\sum\sum}_{i\neq j\neq k} q_i^2 q_j^2 q_k^2 M_{2i} M_{2j} M_{2k} .
\end{eqnarray*}

Finally, assembling the four parts (with some simplification) yields
\begin{equation*}
\begin{array}{lll}
W^{-3} \e(Q^3) & = & \sum_{i} q_i^3 (1 - q_i)^3 ( M_{6i} - 3 M_{4i} M_{2i} + 2 M_{2i}^3 ) \\
                        &    & + 3 [ \sum_{j} q_j (1 - q_j) M_{2j} ] [ \sum_{i} q_i^2 (1 - q_i)^2 ( M_{4i} - M_{2i}^2 ) ] \\
                        &    & + [ \sum_{i} q_i (1 - q_i) M_{2i} ]^3 \\
                        &    & - 6 \mathop{\sum\sum}_{i\neq j} q_i^2 (1 - q_i) q_j^2 (1 - q_j) M_{3i} M_{3j} \\
                        &    & + 12 \mathop{\sum\sum}_{i\neq j} q_i^3 (1 - q_i) q_j^2 M_{4i} M_{2j} \\
                        &    & + 6 \mathop{\sum\sum\sum}_{i\neq j\neq k} q_i (1 - q_i) q_j^2 q_k^2 M_{2i} M_{2j} M_{2k} \\
                        &    & - 4 \mathop{\sum\sum}_{i\neq j} q_i^3 q_j^3 M_{3i} M_{3j} \\
                        &    & - 8 \mathop{\sum\sum\sum}_{i\neq j\neq k} q_i^2 q_j^2 q_k^2 M_{2i} M_{2j} M_{2k} .
\end{array}
\end{equation*}

\section{Unconditional moments of $\Theta_i$} \label{sec:NecMoments}
The unconditional moments of $\Theta_i$  for $\theta_i \sim N(\theta, \tau^2)$ are given by
\begin{equation} \label{eq:Moments_unc}
M_{ri} = \e [ (\hat\theta_i - \theta)^r ] = \sum_{j=0}^r {r \choose j} \e [ (\hat\theta_i - \theta_i)^j (\theta_i -\theta)^{r-j} ] = \sum_{j=0}^r {r \choose j} \e [M_{ji}^c(\theta_i - \theta)^{r-j} ],
\end{equation}
for conditional central  moments $M_{ji}^c = \e [ (\hat\theta_i - \theta_i)^j | \theta_i ]$ with $M_{0i}^c = 1$ and $M_{2i}^c = v_i^2$. For unbiased estimators $\hat\theta_i$, \\
$M_{1i} = M_{1i}^c = 0$, \\
$M_{2i} = \e(v_i^2) + \tau^2$, \\
$M_{3i} = \e(M_{3i}^c) + 3 \e(v_i^2 (\theta_i - \theta))$, \\
$M_{4i} = \e(M_{4i}^c) + 4 \e(M_{3i}^c (\theta_i - \theta)) + 6 \e(v_i^2 (\theta_i - \theta)^2) + 3 \tau^4$, \\
$M_{5i} = \e(M_{5i}^c) + 5 \e(M_{4i}^c (\theta_i - \theta)) + 10 \e(M_{3i}^c (\theta_i - \theta)^2) + 10 \e(v_i^2 (\theta_i - \theta)^3)$, \\
$M_{6i} = \e(M_{6i}^c) + 6 \e(M_{5i}^c (\theta_i - \theta)) + 15 \e(M_{4i}^c (\theta_i - \theta)^2) + 20 \e(M_{3i}^c (\theta_i - \theta)^3) + \\
15 \e(v_i^2 (\theta_i - \theta)^4) + 15\tau^6$.

\section{Two- and three-moment approximations to the distribution of $Q$} \label{M23}
The two- and three-moment  approximations to the distribution of $Q$ use the distribution of a transformed chi-square random variable $c(\chi^2_{p})^r$. The parameters $c$, $r$, and $p$ are found by matching the first two or three moments.


The $k$th moment about zero for $c(\chi^2_{p})^r$ is
\begin{equation}
    \mu'_{k} = \frac{c^k 2^{k r}\Gamma(k r + p/2)}{\Gamma(p/2)}.
    \nonumber
\end{equation}

\subsection{Two-moment approximation} \label{2m}
The two-moment approximation by \cite{Satterthwaite1941} and \cite{Welch1938} sets $r = 1$, so $Q \sim c(\chi^2_{p})$. Matching the first moment $\mu'_{1}$ to $\e(Q)$,  we obtain
\begin{equation}
    \frac{2c\Gamma(1 + p/2)}{\Gamma(p/2)} = E[Q].
    \nonumber
\end{equation}
Since  $\Gamma(x + 1) = x \Gamma(x)$, the above equation reduces  to
\begin{equation} \label{cp}
    cp = E[Q].
\end{equation}
For the second moment $\mu'_{2}$,
\begin{equation}
    \frac{4c^2\Gamma(2 + p/2)}{\Gamma(p/2)} = E[Q^2],
    \nonumber
\end{equation}
which reduces to
\begin{equation} \label{c^2p(p+2)}
   c^2p(p+2) = E[Q^2].
\end{equation}
Solving for $c$ in equation (\ref{cp}) and substituting the result into (\ref{c^2p(p+2)}) yield
\begin{equation}
    c = {E[Q]}/{p}, \quad p = 2 \left[ \frac{E[Q^2]} {E[Q]^2} - 1 \right]^{-1} .
    \nonumber
\end{equation}

\subsection{Three-moment approximations} \label{3m}
For the three-moment approximations we have $Q \sim c(\chi^2_{p})^r$. Similar to the two-moment case, we set $k = 1, 2, 3$ to obtain the following system of equations
\begin{equation}
    (\mu'_{1}): \quad \frac{2^r c\Gamma(r + p/2)}{\Gamma(p/2)} = E[Q];
    \nonumber
\end{equation}
\begin{equation}
    (\mu'_{2}): \quad \frac{2^{2r} c^2\Gamma(2r + p/2)}{\Gamma(p/2)} = E[Q^2];
    \nonumber
\end{equation}
\begin{equation}
    (\mu'_{3}): \quad \frac{2^{3r} c^3\Gamma(3r + p/2)}{\Gamma(p/2)} = E[Q^3].
    \nonumber
\end{equation}
Dividing $\mu'_{2}$ by $\mu'_{1}$, we obtain the following expression for $c$:
\begin{equation}
    c = \frac{E[Q^2]\Gamma(r + p/2)}{2^r E[Q]\Gamma(2r + p/2)}.
    \nonumber
\end{equation}
To eliminate $c$,  define $A = \mu'_{2}/(\mu'_{1})^2$ and $B = \mu'_{3}/(\mu'_{1})^3$. Then we have the following two nonlinear equations:
\begin{equation}
    A = \frac{\Gamma(2r + p/2)\Gamma(p/2)}{\Gamma^2(r+p/2)}, \quad B = \frac{\Gamma(3r + p/2)\Gamma^2(p/2)}{\Gamma^3(r+p/2)}.
    \nonumber
\end{equation}
We solve this system for $p$ and $r$ by using the function {\it 'multiroot'} in the R package {\it rootSolve} \citep{rootsolve} with the starting values $r = 1$ and $c$ and $p$ from the two-moment approximation.

\renewcommand{\thefigure}{B.\arabic{figure}}
\renewcommand{\thesection}{B.\arabic{section}}
\renewcommand{\thetable}{B.\arabic{table}}
\setcounter{figure}{0}
\setcounter{section}{0}
\setcounter{table}{0}
\clearpage

\text{\LARGE{\bf{Appendix B}}}

\section*{Simulation results}

The appendices that follow present, graphically, results for equal sample sizes (i.e., $n_i$ is the same in all $K$ studies) and unequal sample sizes (i.e., $n_i$ varies among the studies). Specifically, $n$ = 20, 40, 100, and 250 (and also 640 and 1000 in Appendix B2) and $\bar{n}$ = 13, 15, 30, and 60 or 15, 30, 60, and 100 (and $\bar{n}$ = 13, 15, 30, 60, 100, and 160 in Appendix B2). When the sample sizes are unequal, the $n_i$ for $K = 5$ are \\
\begin{tabular} {ll}
$n$  & $n_i$, $i$ = 1, \ldots, 5 \\
 13   & 4, 6, 7, 8, 40           \\
 15   & 6, 8, 9, 10, 42         \\
 30   & 12, 16, 18, 20, 84   \\
 60   & 24, 32, 36, 40, 168 \\
100  & 64, 72, 76, 80, 208 \\
160  & 124, 132, 136, 140, 268
\end{tabular} \\
(one study is substantially larger than the others). For $K = 10$ and $K = 30$, the set of five $n_i$ is used twice or six times, respectively.

We used two sets of unequal sample sizes because $\bar{n} = 13$ combined with $f$ (the fraction of each study's sample size in the Control arm) = .75 produces studies whose sample size in the Treatment arm is too small. From the study-level sample size $n_i$ we obtain $n_{iT}$ and $n_{iC}$ according to $n_{iT} = \lceil{(1 - f)n_{i}}\rceil$ and $n_{iC} = n_{i} - n_{iT}$. Thus, with $\bar{n} = 13$ and $f = .75$ the $n_{iT}$ (for $K = 5$) would be 1, 2, 2, 2, and 10.

\begin{itemize}
	\item Appendix B1. Probability (P--P) plots of approximations to the null distribution of $Q$
	\item Appendix B2. Empirical levels of the approximations to the distribution of $Q$ vs sample size
	\item Appendix B3: Empirical p-values of approximations to the distribution of $Q$ vs $\tau^2$
	\item Appendix B4. Power of tests for heterogeneity ($\tau^2 = 0$ versus $\tau^2 > 0$) based on approximations to the distribution of $Q$
	\item Appendix B5. Bias in estimation of $\tau^2$
\end{itemize}

\clearpage
\setcounter{section}{0}
\setcounter{figure}{0}
\renewcommand{\thesection}{B1.\arabic{section}}
\section*{B1. Probability (P--P) plots of approximations to the null distribution of $Q$}
Each figure corresponds to a value of $\sigma_T^2$ (= 1, 2), a value of $f$ (= .5, .75), and a pattern of sample sizes (equal or unequal). (For all figures, $\tau^2 = 0$ and $\sigma_C^2 = 1$.) \\
For each combination of a value of $n$ (= 20, 40, 100, 250) or $\bar{n}$ (= 13, 15, 30, 60 or 15, 30, 60, 100)  and a value of $K$ (= 5, 10, 30), a panel plots the error, $\hat{P}(Q > t) - p$, for $p$ = .001, .0025, .005, .01, .025, .05, .1, .25, .5, .75, .9, .95, .975, .99, .995, .9975, .999.\\
The approximations to the distribution of $Q$ are
\begin{itemize}
	\item F SW (Farebrother approximation, effective-sample-size weights)
	\item M3 SW (Three-moment approximation, effective-sample-size weights)
	\item M2 SW (Two-moment approximation, effective-sample-size weights)
	\item $\chi_{K - 1}^2$ (Chi-square, IV weights)
	\item Welch (Welch approximation, IV weights)
\end{itemize}

\clearpage
\renewcommand{\thefigure}{B1.\arabic{figure}}
\begin{figure}[t]
	\centering
	\includegraphics[scale=0.33]{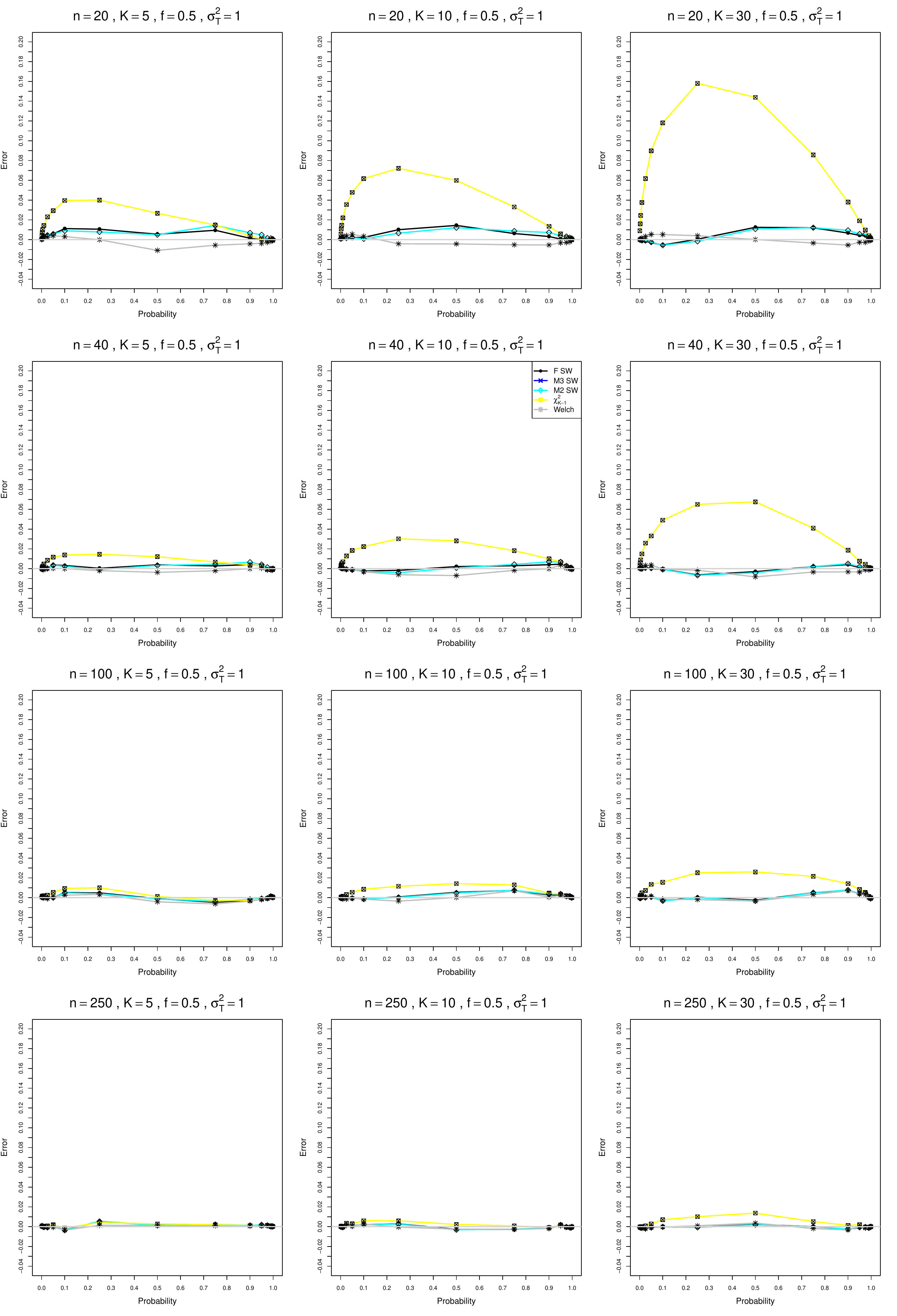}
	\caption{Approximation error for $\sigma_T^2 = 1$, $f = .5$, and equal sample sizes $n$ = 20, 40, 100, 250
		\label{BiasPhatMD0_S1_1q052Sigma2T1}}
\end{figure}

\begin{figure}[t]
	\centering
	\includegraphics[scale=0.33]{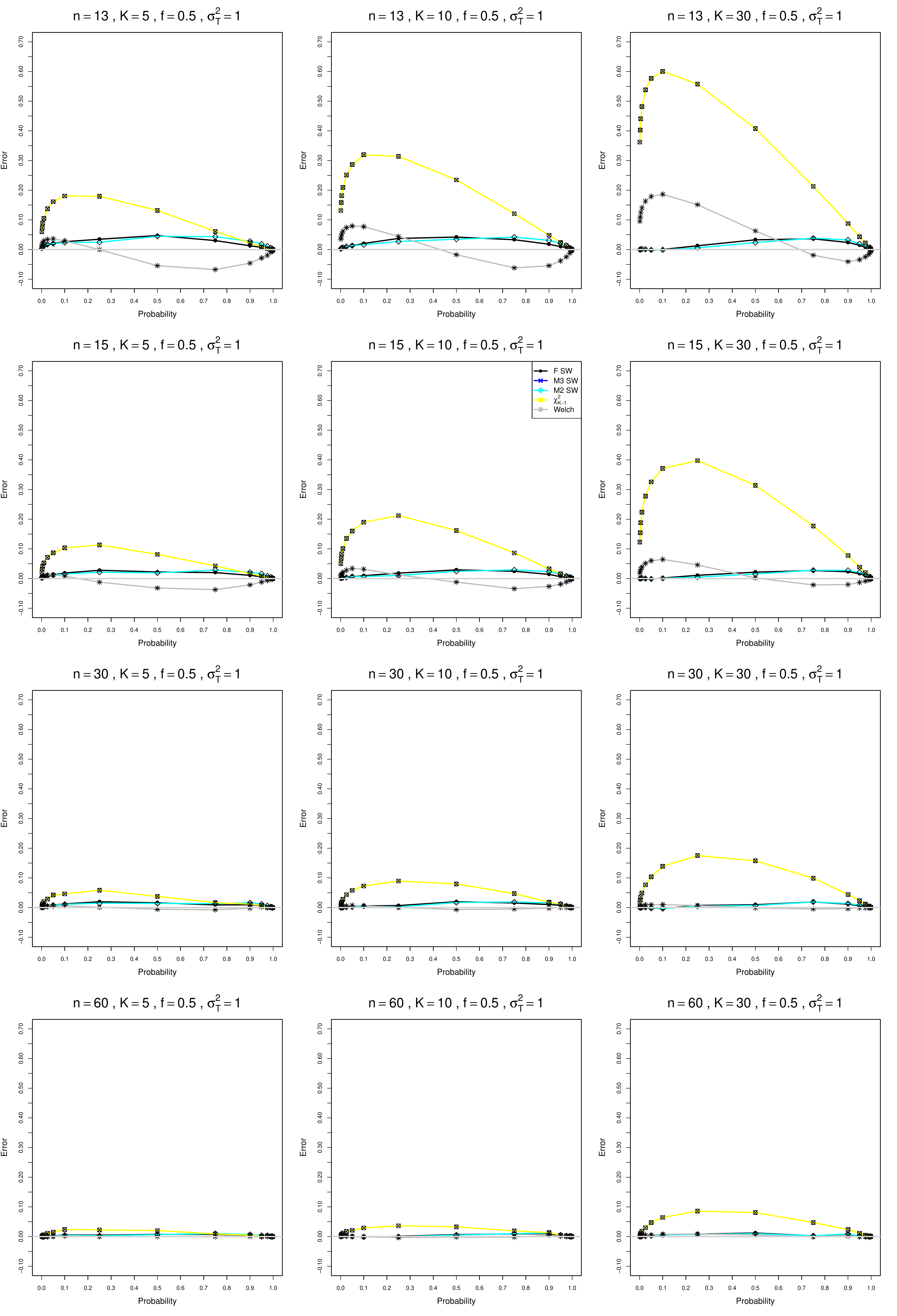}
	\caption{Approximation error for $\sigma_T^2 = 1$, $f = .5$, and unequal sample sizes $\bar{n}$ = 13, 15, 30, 60
		\label{BiasPhatMD0_S1_1q052Sigma2T1_unequal}}
\end{figure}

\begin{figure}[t]
	\centering
	\includegraphics[scale=0.33]{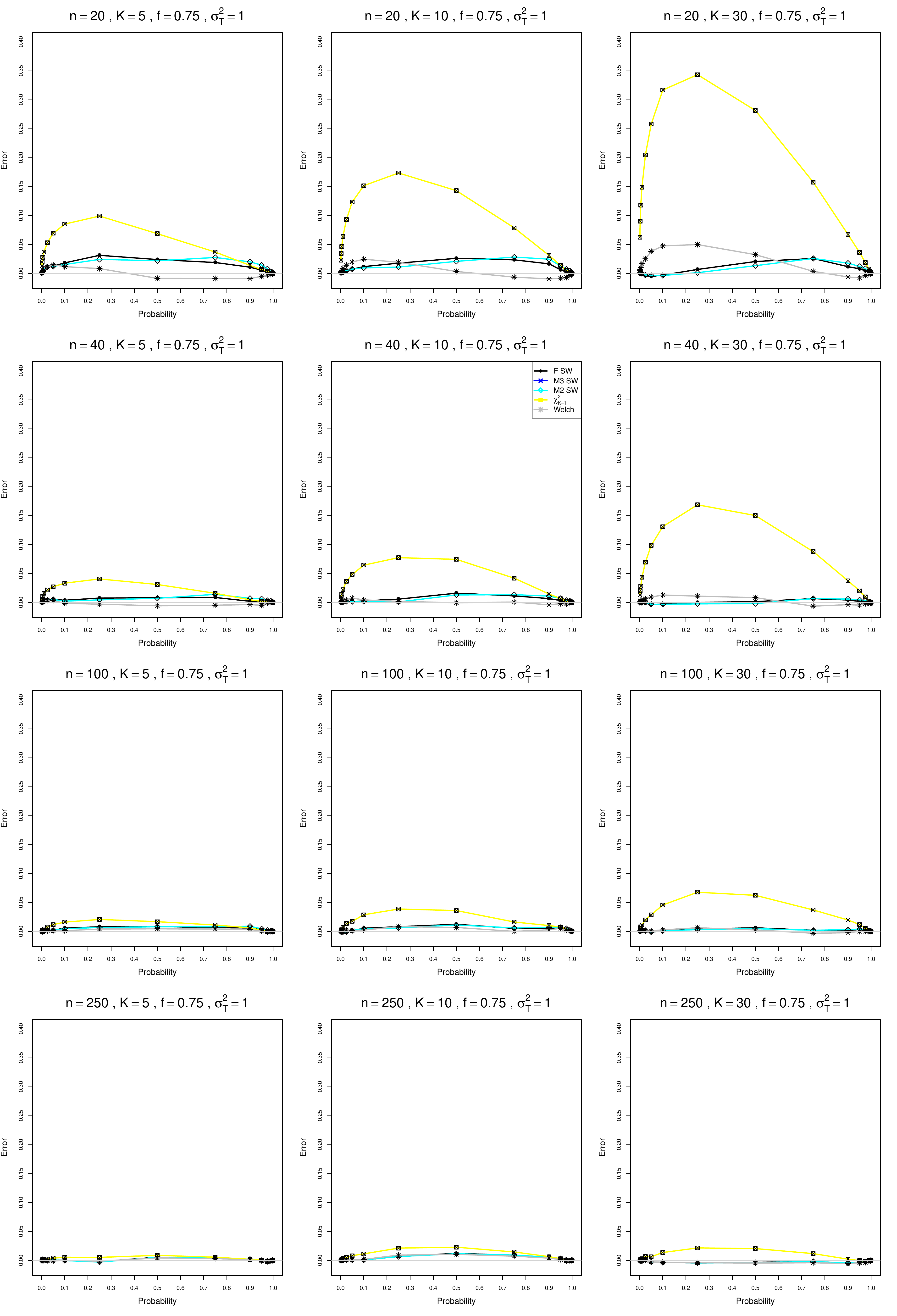}
	\caption{Approximation error for $\sigma_T^2 = 1$, $f = .75$, and equal sample sizes $n$ = 20, 40, 100, 250
		\label{BiasPhatMD0_S1_1q0752Sigma2T1}}
\end{figure}

\begin{figure}[t]
	\centering
	\includegraphics[scale=0.33]{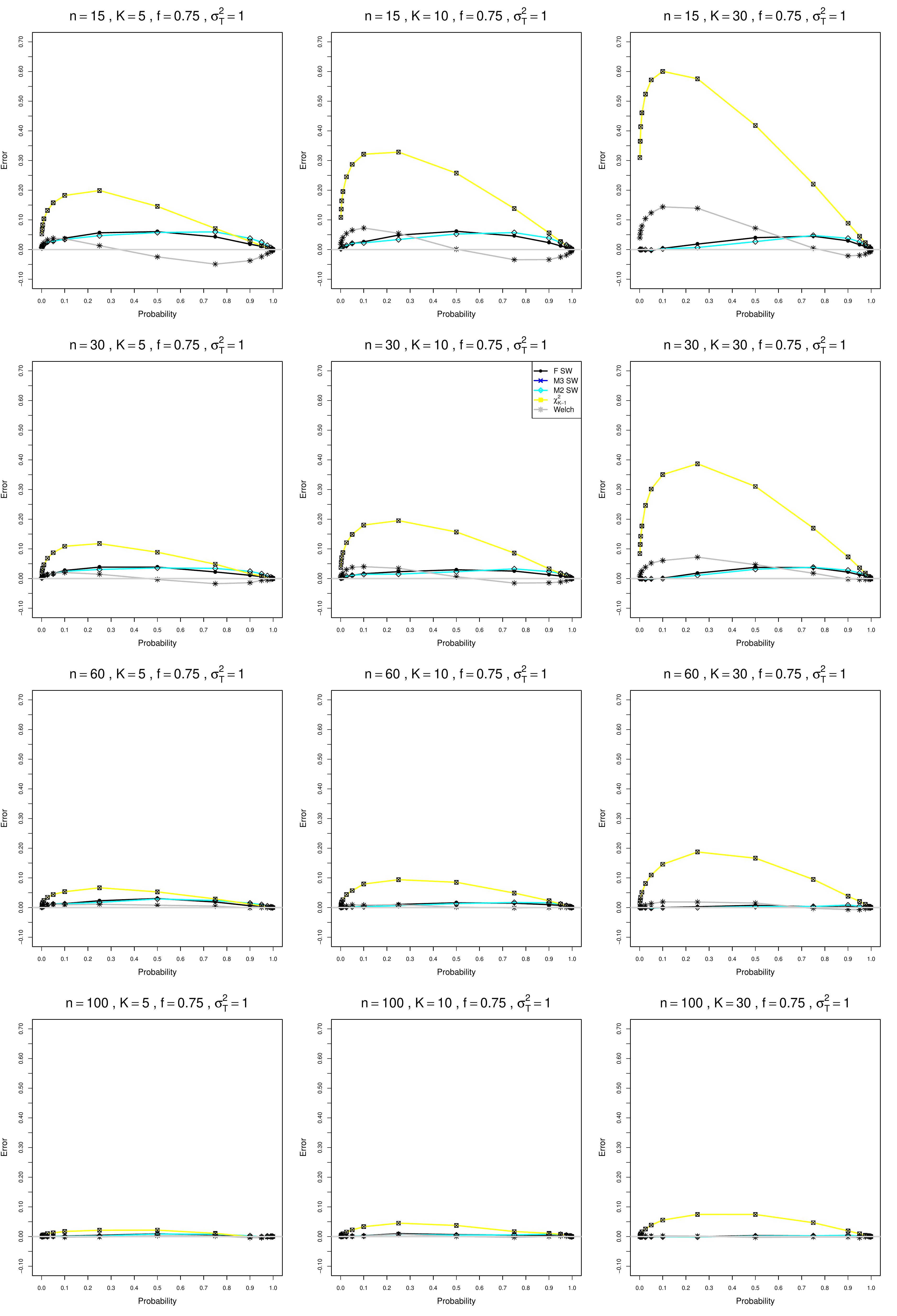}
	\caption{Approximation error for $\sigma_T^2 = 1$, $f = .75$, and unequal sample sizes $\bar{n}$ = 15, 30, 60, 100
		\label{BiasPhatMD0_S1_1q0752Sigma2T1_unequal}}
\end{figure}

\begin{figure}[t]
	\centering
	\includegraphics[scale=0.33]{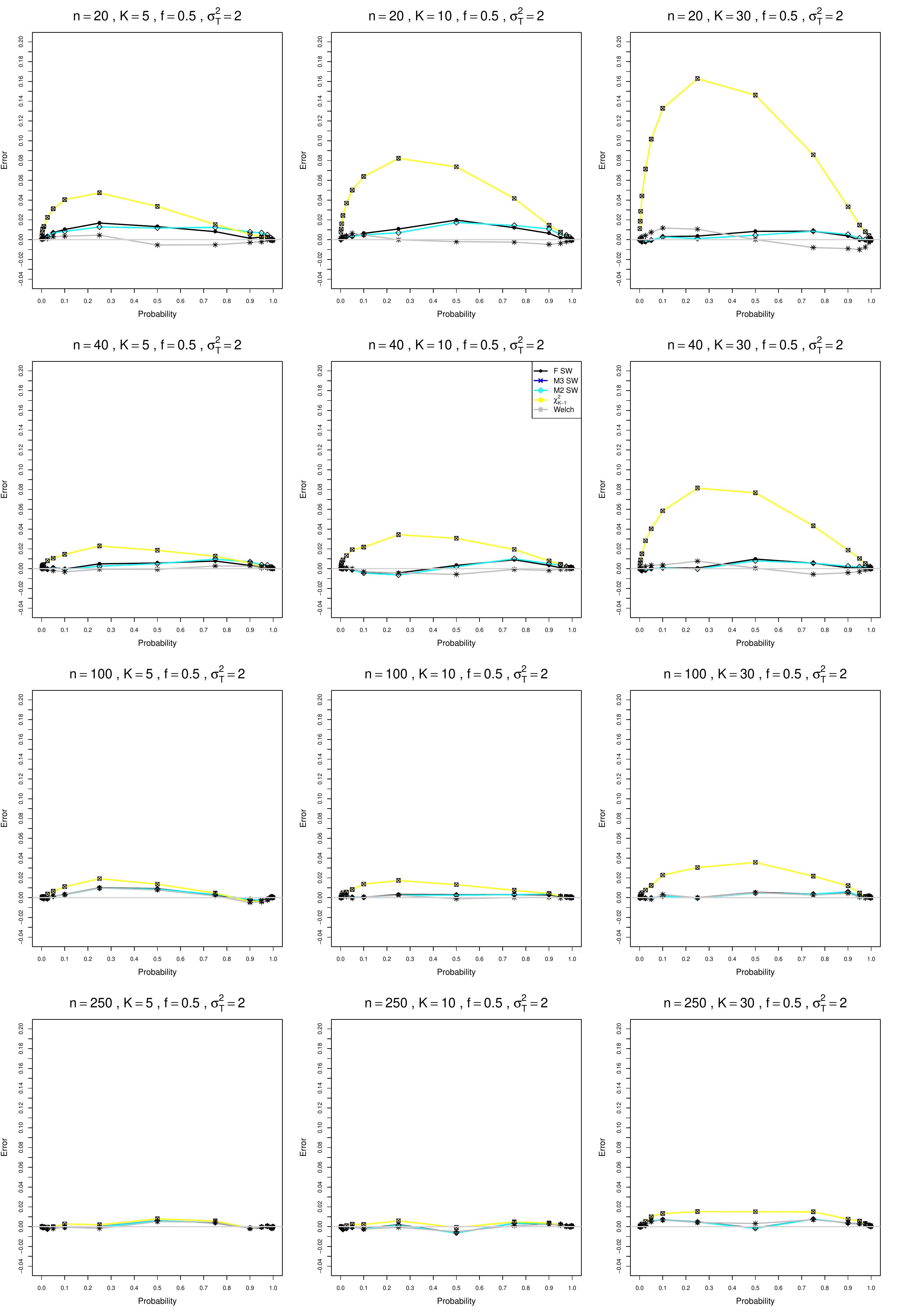}
	\caption{Approximation error for $\sigma_T^2 = 2$, $f = .5$, and equal sample sizes $n$ = 20, 40, 100, 250
		\label{BiasPhatMD0_S2_1q052Sigma2T1}}
\end{figure}

\begin{figure}[t]
	\centering
	\includegraphics[scale=0.33]{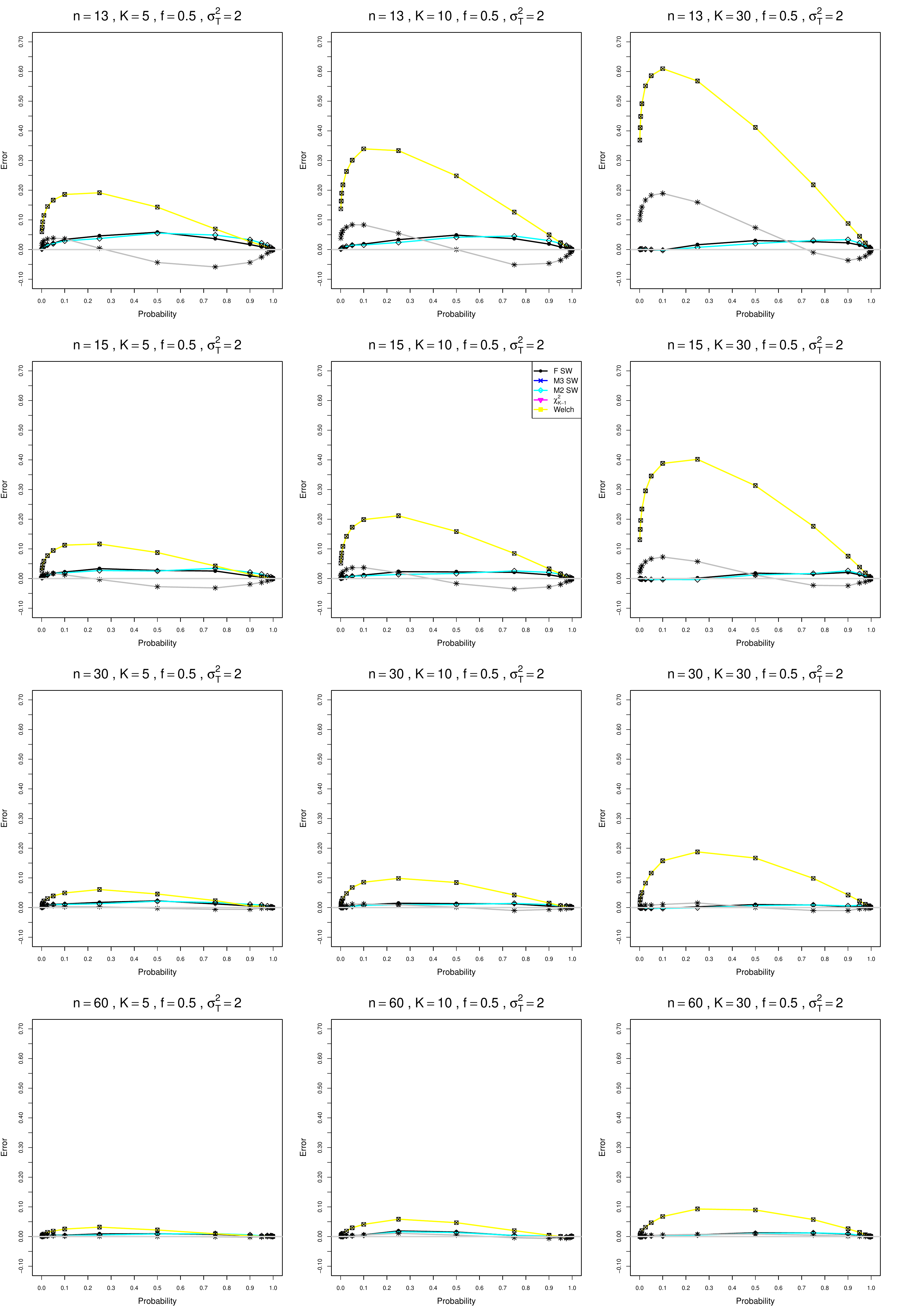}
	\caption{Approximation error for $\sigma_T^2 = 2$, $f = .5$, and unequal sample sizes $\bar{n}$ = 13, 15, 30, 60
		\label{BiasPhatMD0_S2_1q052Sigma2T1_unequal}}
\end{figure}

\begin{figure}[t]
	\centering
	\includegraphics[scale=0.33]{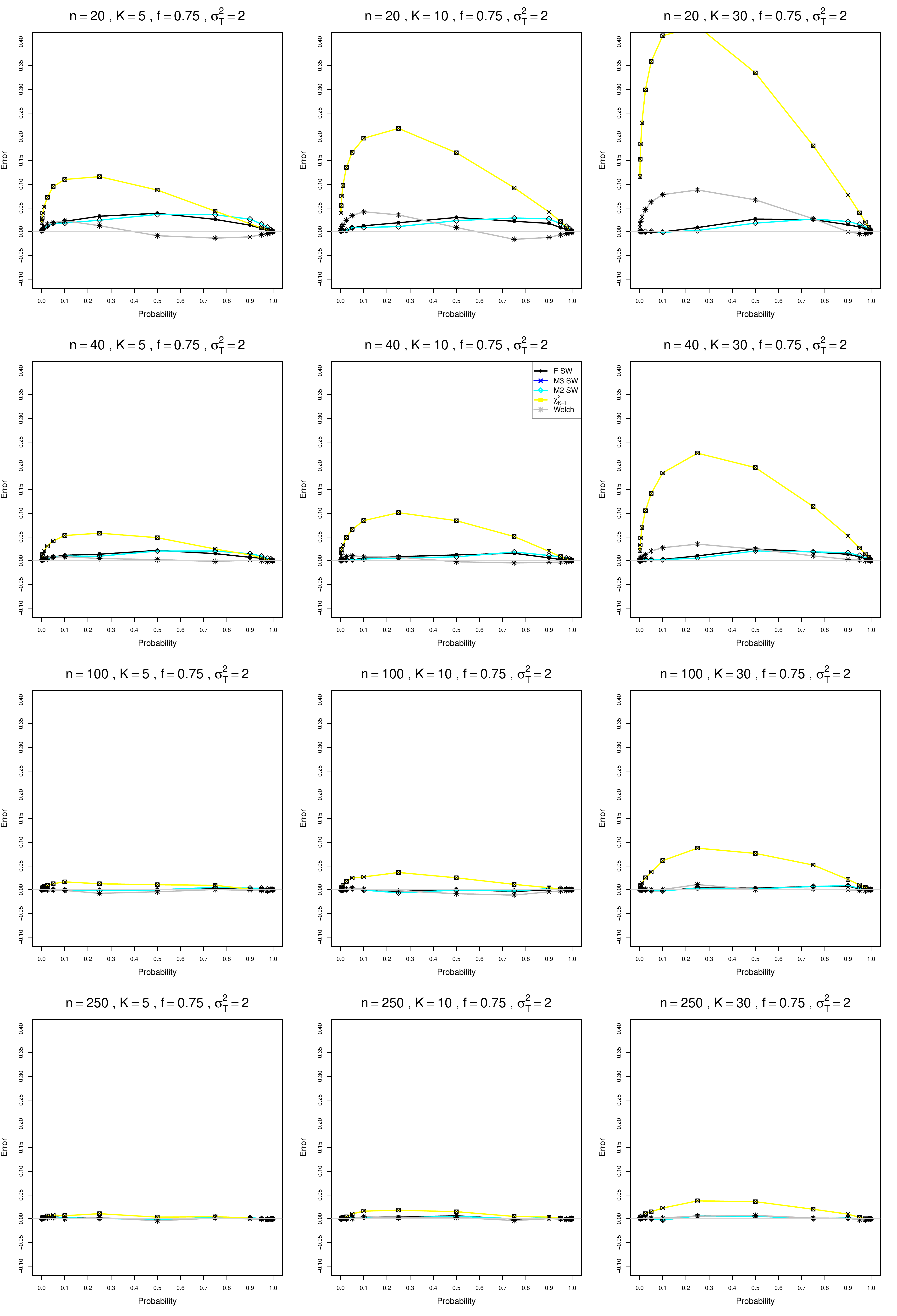}
	\caption{Approximation error for $\sigma_T^2 = 2$, $f = .75$, and equal sample sizes $n$ = 20, 40, 100, 250
		\label{BiasPhatMD0_S2_1q0752Sigma2T1}}
\end{figure}

\begin{figure}[t]
	\centering
	\includegraphics[scale=0.33]{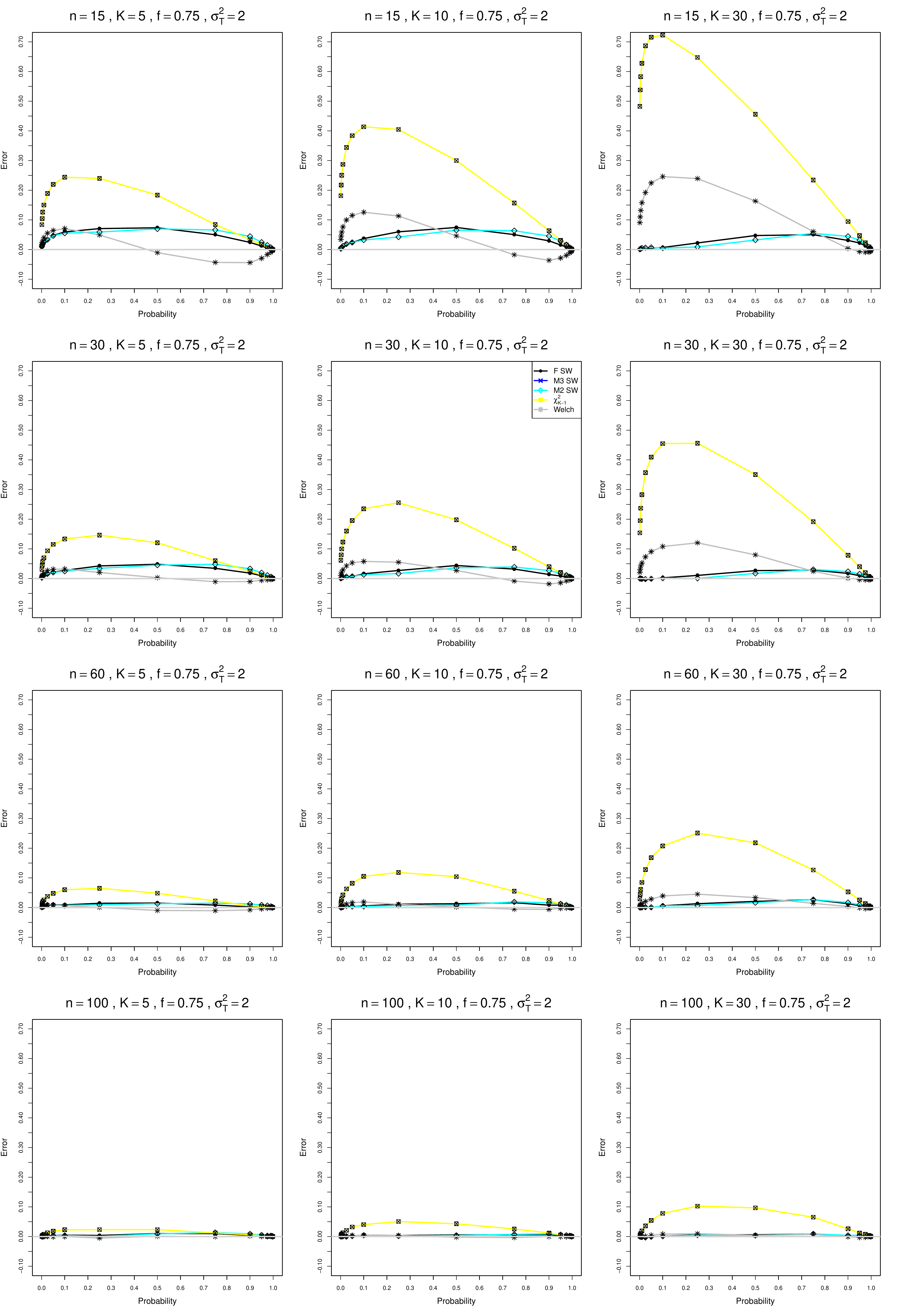}
	\caption{Approximation error for $\sigma_T^2 = 2$, $f = .75$, and unequal sample sizes $\bar{n}$ = 15, 30, 60, 100
		\label{BiasPhatMD0_S2_1q0752Sigma2T1_unequal}}
\end{figure}

\clearpage
\setcounter{figure}{0}
\setcounter{section}{0}
\renewcommand{\thesection}{B2.\arabic{section}}

\section*{B2: Empirical levels, at nominal level $\alpha$, of the approximations to the distribution of $Q$ vs sample size}
Each figure corresponds to a value of $\alpha$ (= .001, .005, .01, .05) and a pattern of sample sizes (equal or unequal). (For all figures, $\sigma_C^2 = 1$ and $\tau^2 = 0$.) \\
For each combination of a value of $f$ (= .5, .75), a value of $\sigma_T^2$ (= 1, 2), a value of $\sigma_C^2$ (= 1, 2), and a value of $K$ (= 5, 10, 30), a panel plots the empirical level versus $n$ (= 20, 40, 100, 250, 640, 1000) or $\bar{n}$ (= 13, 15, 30, 60, 100, 160).\\
The approximations for the distribution of $Q$ are
\begin{itemize}
	\item F SW (Farebrother approximation, effective-sample-size weights)
	\item M3 (Three-moment approximation, effective-sample-size weights)
	\item M2 (Two-moment approximation, effective-sample-size weights)
	\item Welch (Welch approximation, IV weights)
\end{itemize}

\clearpage
\renewcommand{\thefigure}{B2.\arabic{figure}}
\begin{figure}[t]
	\centering
	\includegraphics[scale=0.33]{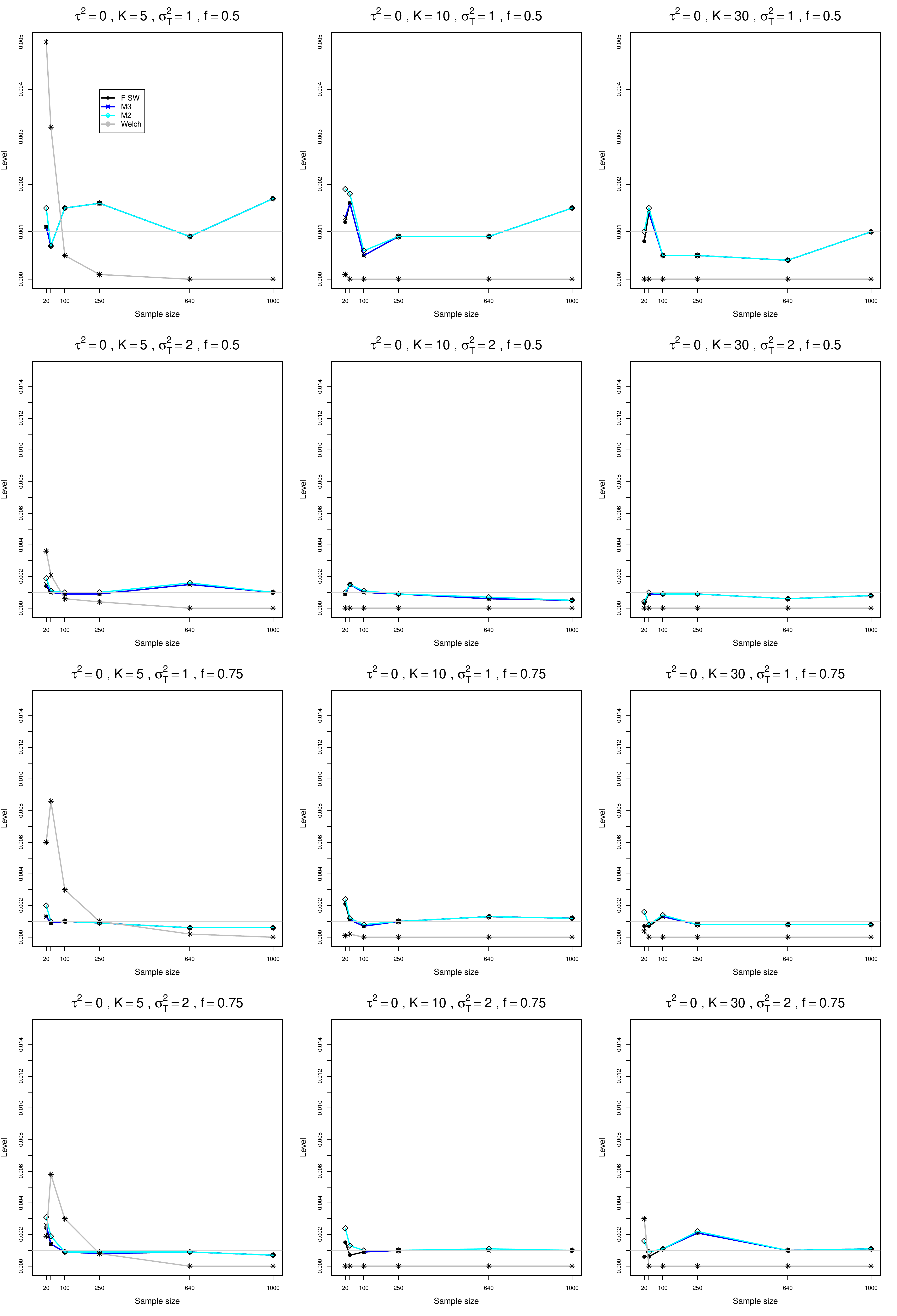}
	\caption{Level at $\alpha = .001$ and equal sample sizes $n$ = 20, 40, 100, 250, 640, 1000.
		First row: $\sigma_T^2 = 1$ and $f = .5$.
		Second row: $\sigma_T^2 = 2$ and $f = .5$.
		Third row: $\sigma_T^2 = 1$ and  $f = .75$.
		Fourth row: $\sigma_T^2 = 2$ and $f = .75$.
		\label{PlotOfPvaluesAgainstN_0001level}}
\end{figure}

\begin{figure}[t]
	\centering
	\includegraphics[scale=0.33]{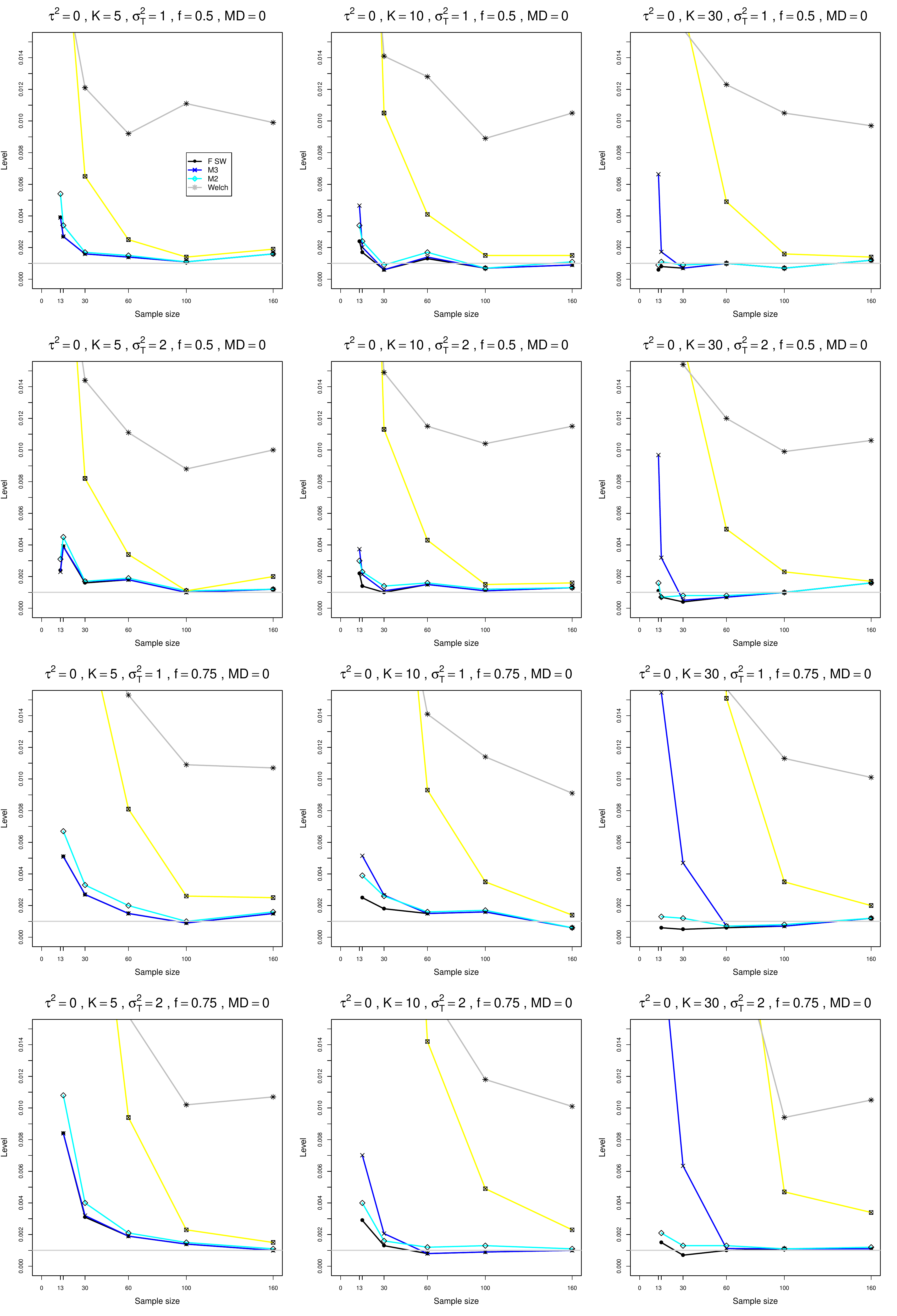}
	\caption{Level at $\alpha = .001$ and unequal sample sizes $\bar{n}$ = 13, 15, 30, 60, 100, 160.
		First row:   $f = .5$, $\sigma_C^2 = 1$, $\sigma_T^2 = 1$.
		Second row: $f = .5$, $\sigma_C^2 = 2$, $\sigma_T^2 = 1$.
		Third row: $f = .75$, $\sigma_C^2 = 1$, $\sigma_T^2 = 1$.
		Fourth row: $f = .75$, $\sigma_C^2 = 2$, $\sigma_T^2 = 1$.
		\label{PlotOfPvaluesAgainstN_0001level_unequal}}
\end{figure}

\begin{figure}[t]
	\centering
	\includegraphics[scale=0.33]{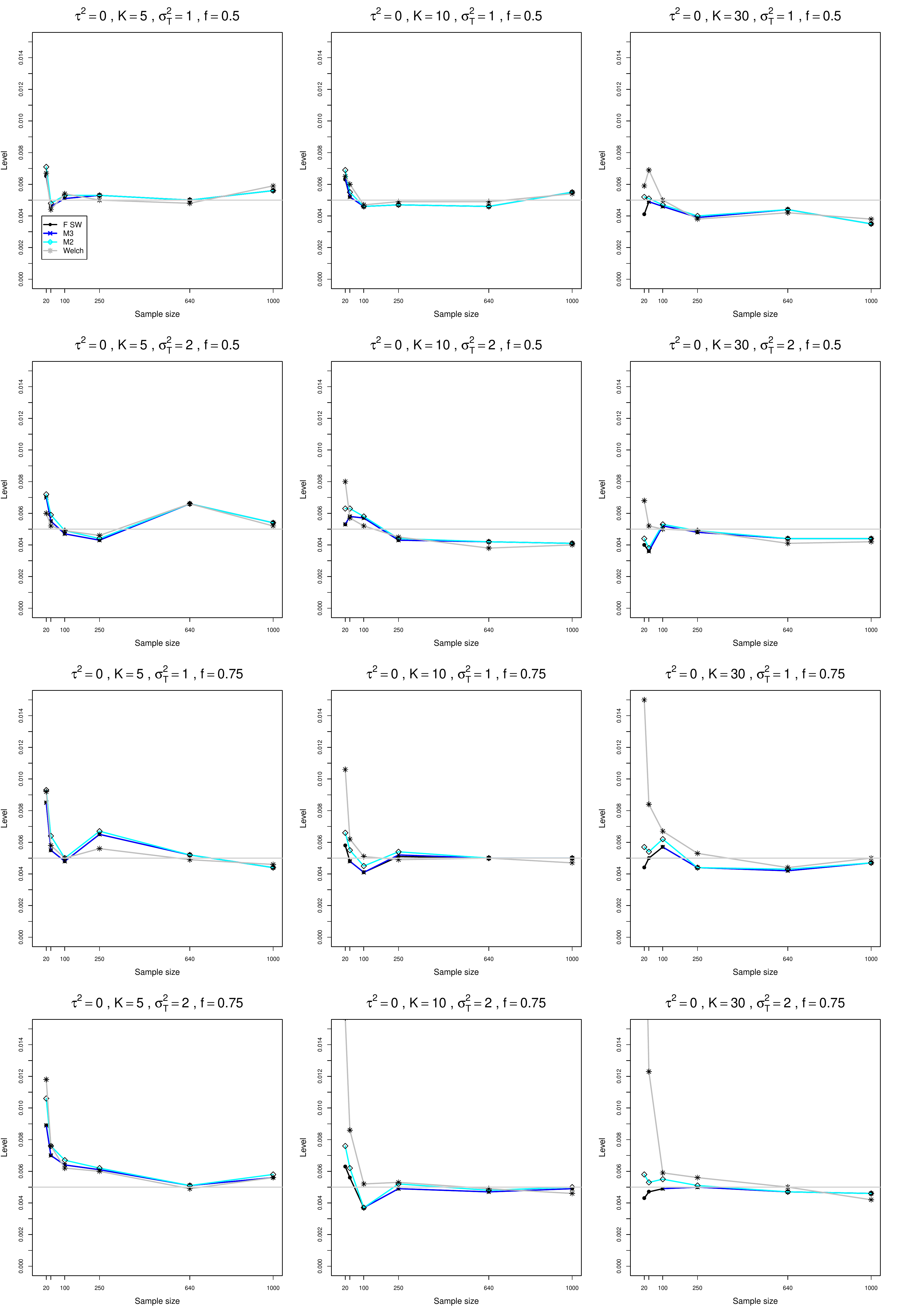}
	\caption{Level at $\alpha = .005$ and equal sample sizes $n$ = 20, 40, 100, 250, 640, 1000.
		First row: $f = .5$ and $\sigma_T^2 = 1$.
		Second row: $f = .5$ and $\sigma_T^2 = 2$.
		Third row: $f = .75$ and $\sigma_T^2 = 1$.
		Fourth row: $f = .75$ and $\sigma_T^2 = 2$.
		\label{PlotOfPvaluesAgainstN_0005level}}
\end{figure}

\begin{figure}[t]
	\centering
	\includegraphics[scale=0.33]{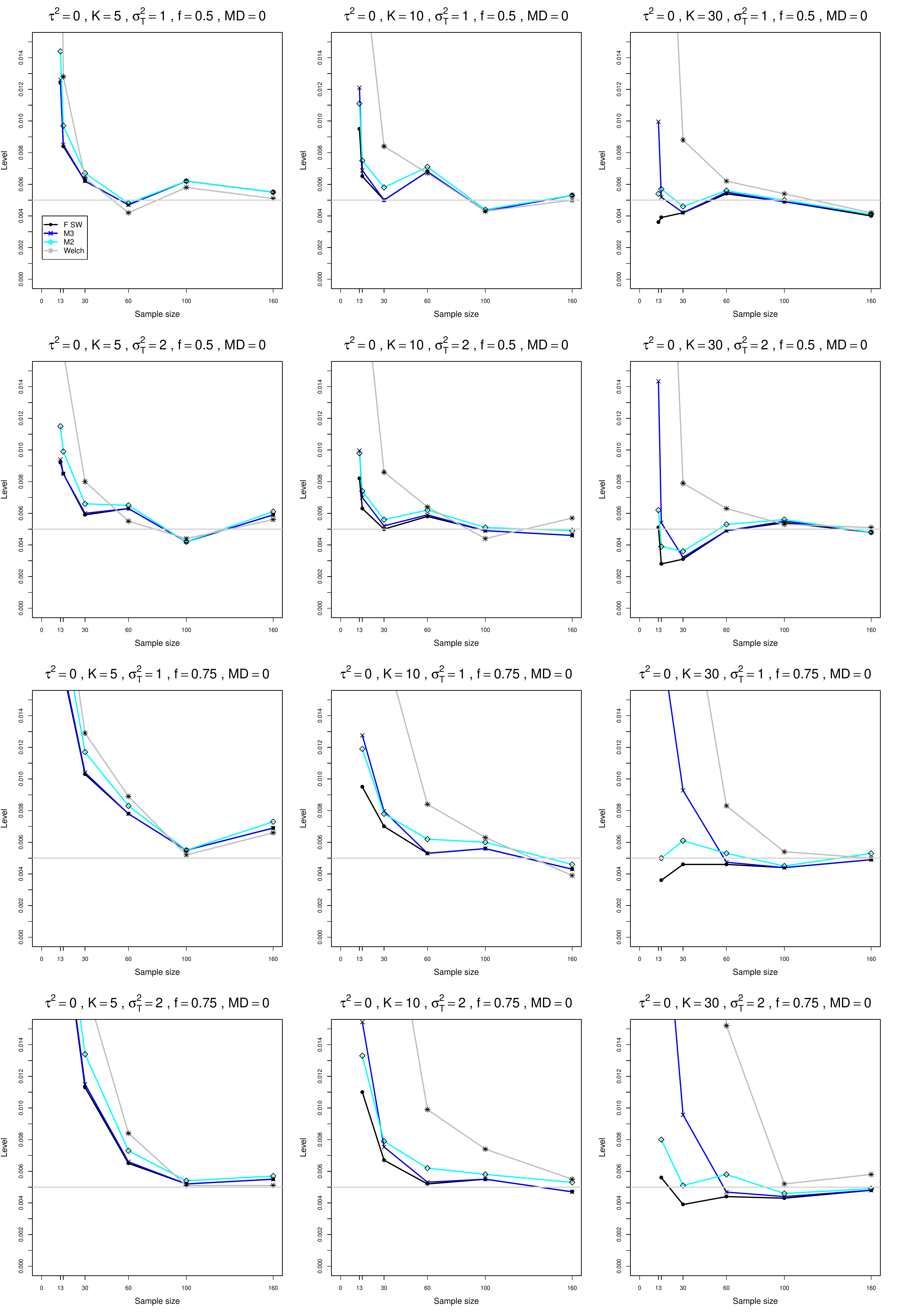}
	\caption{Level at $\alpha=0.005$ and unequal sample sizes $\bar{n}$ = 13, 15, 30, 60, 100, 160.
		First row: $f = .5$, $\sigma_C^2 = 1$, $\sigma_T^2 = 1$.
		Second row:  $f = .5$, $\sigma_C^2 = 2$, $\sigma_T^2 = 1$.
		Third row: $f = .75$, $\sigma_C^2 = 1$, $\sigma_T^2 = 1$.
		Fourth row: $f = .75$, $\sigma_C^2 = 2$, $\sigma_T^2  = 1$.										\label{PlotOfPvaluesAgainstN_0005level_unequal}}
\end{figure}

\begin{figure}[t]
	\centering
	\includegraphics[scale=0.33]{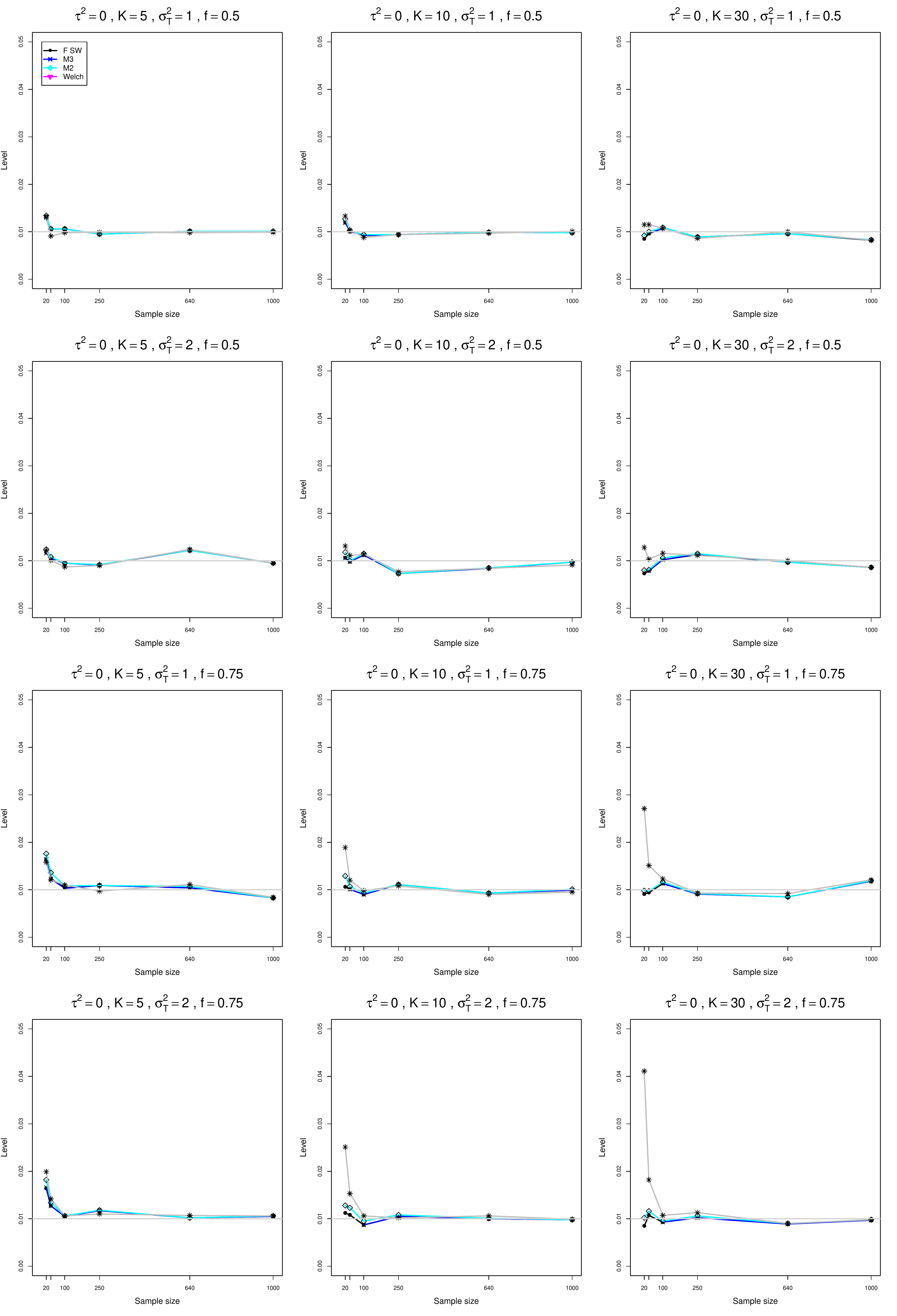}
	\caption{Level at $\alpha = .01$ and equal sample sizes $n$ = 20, 40, 100, 250, 640, 1000.
		First row: $f = .5$ and $\sigma_T^2 = 1$.
		Second row: $f = .5$ and $\sigma_T^2 = 2$.
		Third row: $f = .75$ and $\sigma_T^2 = 1$.
		Fourth row: $f = .75$ and $\sigma_T^2 = 2$.
		\label{PlotOfPvaluesAgainstN_001level}}
\end{figure}

\begin{figure}[t]
	\centering
	\includegraphics[scale=0.33]{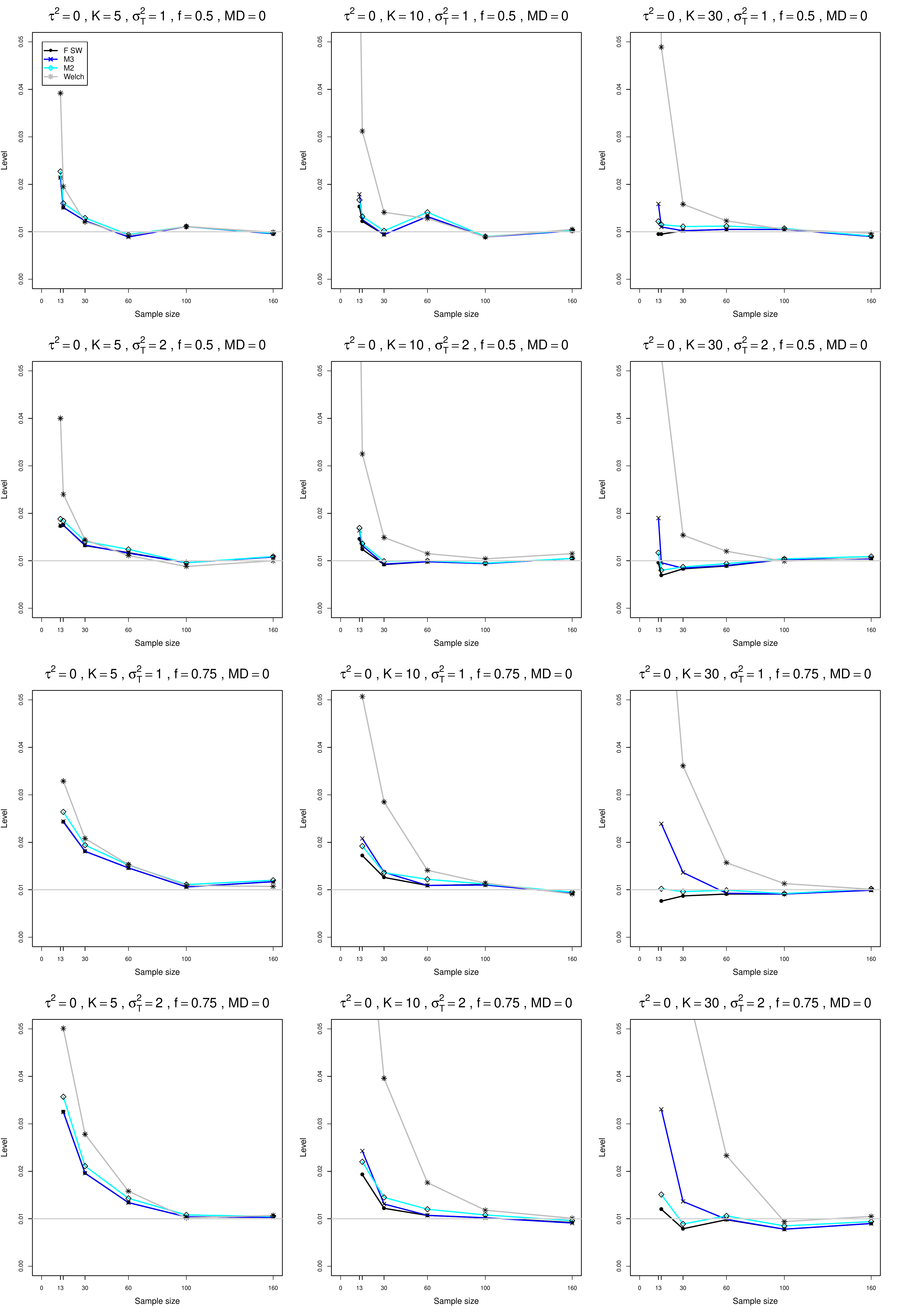}
	\caption{Level at $\alpha = .01$ and unequal sample sizes $\bar{n}$ = 13, 15, 30, 60, 100, 160.
		First row: $f = .5$, $\sigma_C^2 = 1$, $\sigma_T^2 = 1$.
		Second row: $f = .5$, $\sigma_C^2 = 2$, $\sigma_T^2 = 1$.
		Third row: $f  = .75$, $\sigma_C^2=1$, $\sigma_T^2=1$.
		Fourth row: $f = .75$, $\sigma_C^2=2$, $\sigma_T^2=1$
		\label{PlotOfPvaluesAgainstN_001level_unequal}}
\end{figure}

\begin{figure}[t]
	\centering
	\includegraphics[scale=0.33]{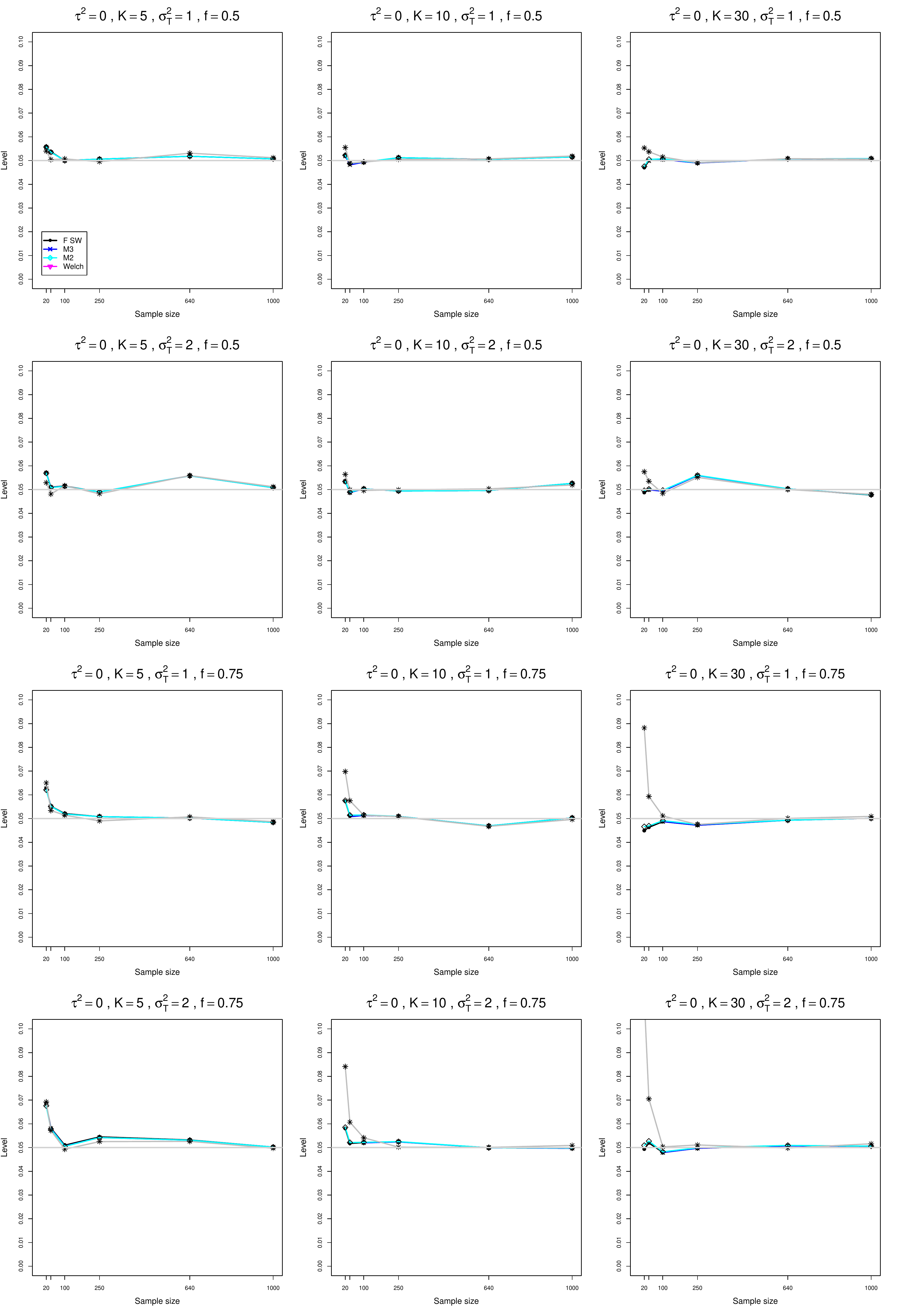}
	\caption{Level at $\alpha = .05$ and equal sample sizes $n$ = 20, 40, 100, 250, 640, 1000.
		First row: $f = .5$, $\sigma_T^2 = 1$.
		Second row: $f = .5$, $\sigma_T^2 = 2$ .
		Third row: $f = .75$, $\sigma_T^2 = 1$.
		Fourth row: $f = .75$, $\sigma_T^2 = 2$.
		\label{PlotOfPvaluesAgainstN_005level}}
\end{figure}

\begin{figure}[t]
	\centering
	\includegraphics[scale=0.33]{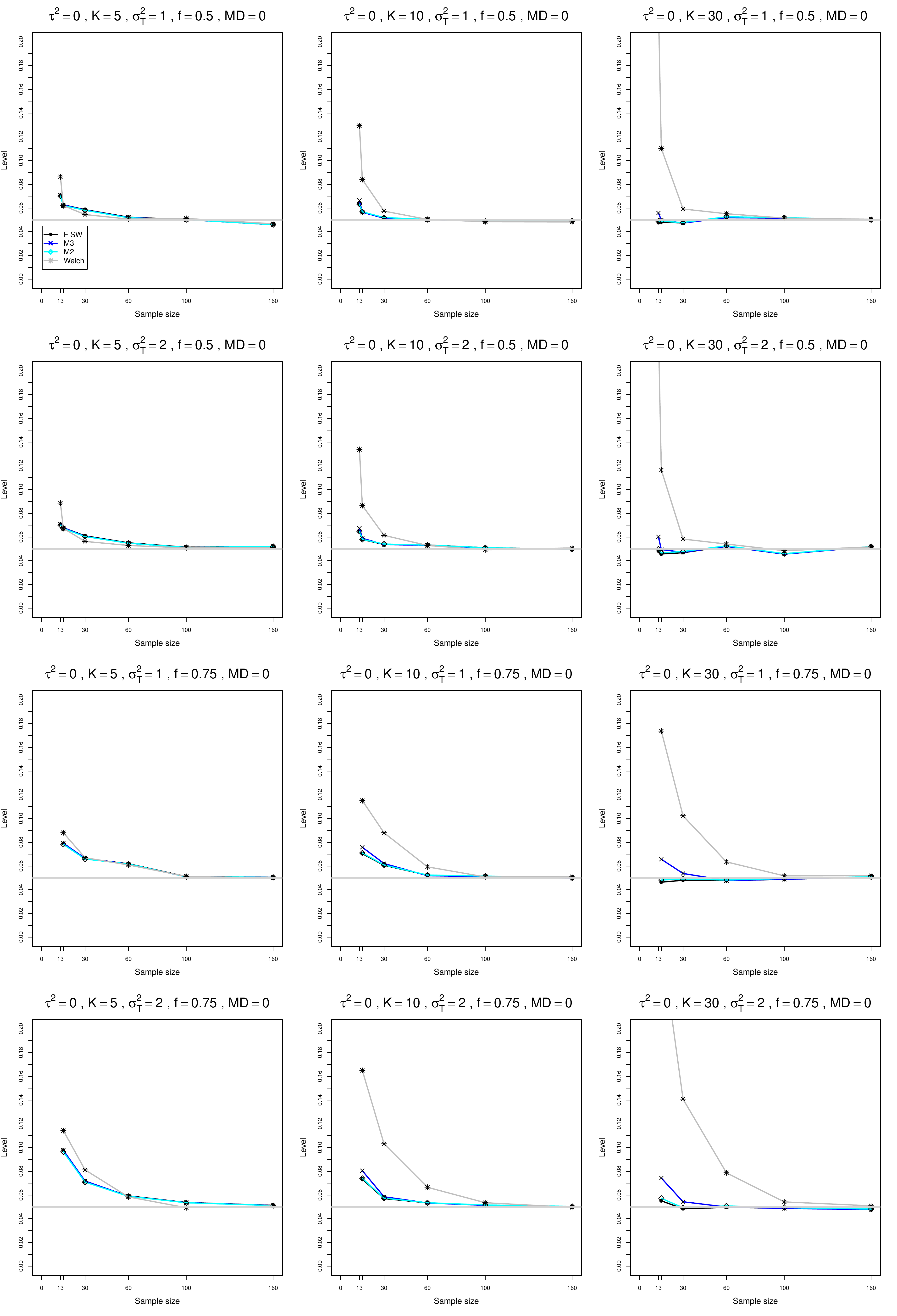}
	\caption{Level at $\alpha = .05$ and unequal sample sizes $\bar{n}$ = 13, 15, 30, 60, 100, 160.
		First row: $f = .5$, $\sigma_T^2 = 1$.
		Second row: $f = .5$, $\sigma_T^2 = 2$.
		Third row: $f = .75$, $\sigma_T^2=1$.
		Fourth row: $f = .75$, $\sigma_T^2=2$.
		\label{PlotOfPvaluesAgainstN_005level_unequal}}
\end{figure}

\clearpage
\setcounter{section}{0}
\setcounter{figure}{0}
\renewcommand{\thesection}{B3.\arabic{section}}

\section*{B3: Empirical p-values of approximations to the distribution of $Q$ vs $\tau^2$}

Each figure corresponds to a value of $\alpha$ (= .01, .05), a value of $\sigma_T^2$ (= 1, 2), a value of $f$ (= .5, .75), and a pattern of sample sizes (equal or unequal). (For all figures, $\sigma_C^2 = 1$.)\\
For each combination of a value of $n$ (= 20, 40, 100, 250) or $\bar{n}$ (= 13, 15, 30, 60 or 15, 30, 60, 100) and a value of $K$ (= 5, 10, 30), a panel plots empirical p-values versus $\tau^2$ = 0.0(0.1)1.0.\\
The approximations to the distribution of $Q$ are
\begin{itemize}
	\item F SW (Farebrother approximation, effective-sample-size weights)
	\item M3 (Three-moment approximation, effective-sample-size weights)
	\item M2 (Two-moment approximation, effective-sample-size weights)
	\item BJ (Biggerstaff and Jackson, IV weights)
\end{itemize}

\clearpage
\renewcommand{\thefigure}{B3.\arabic{figure}}

\begin{figure}[t]
	\centering
	\includegraphics[scale=0.33]{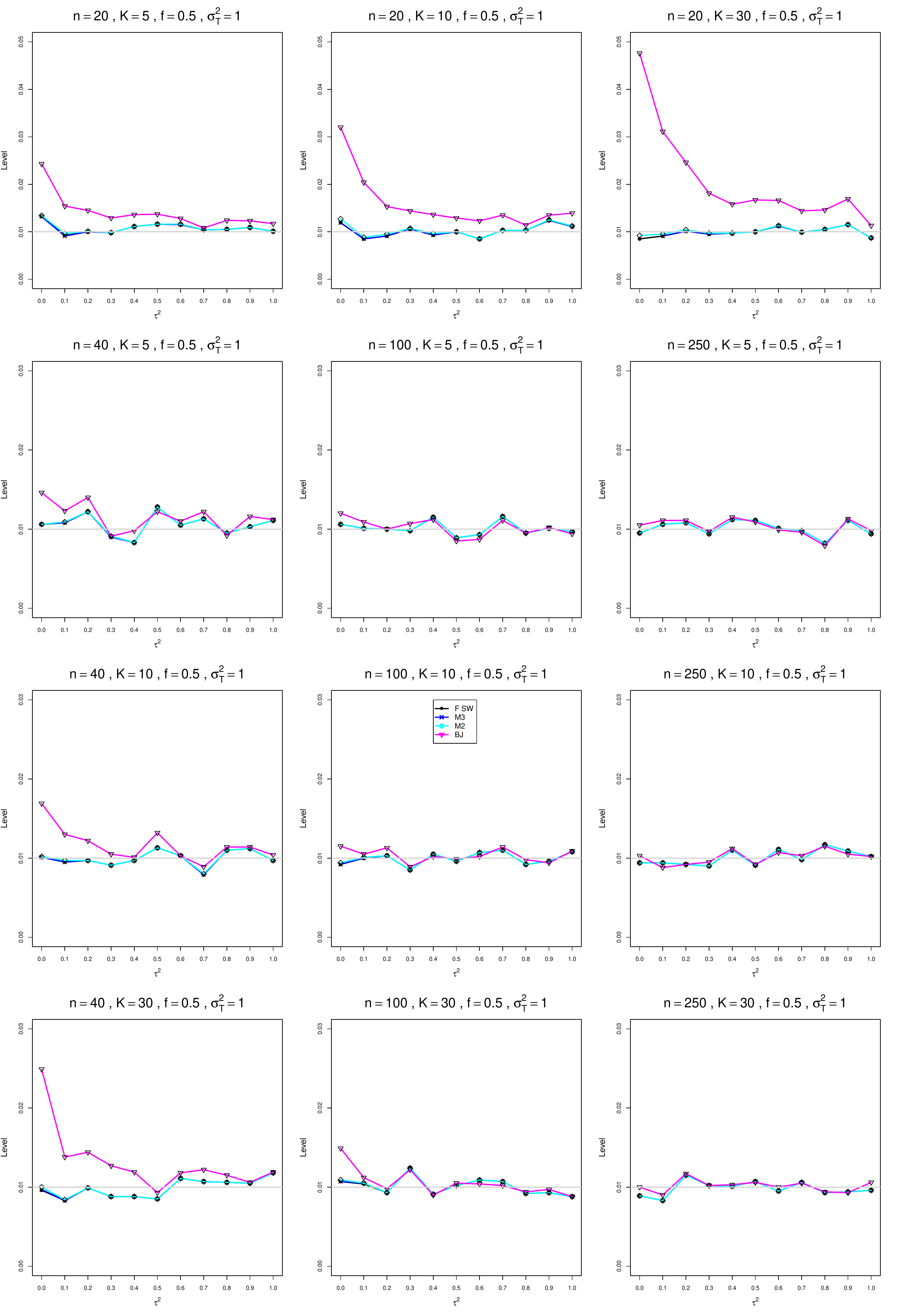}
	\caption{Empirical p-values when $\alpha = .01$ for $\sigma_T^2 = 1$, $f = .5$, and equal sample sizes $n$ = 20, 40, 100, 250
		\label{PlotOfPhatAt001Sigma2T1andq05MD_underH1}}
\end{figure}

\begin{figure}[t]
	\centering
	\includegraphics[scale=0.33]{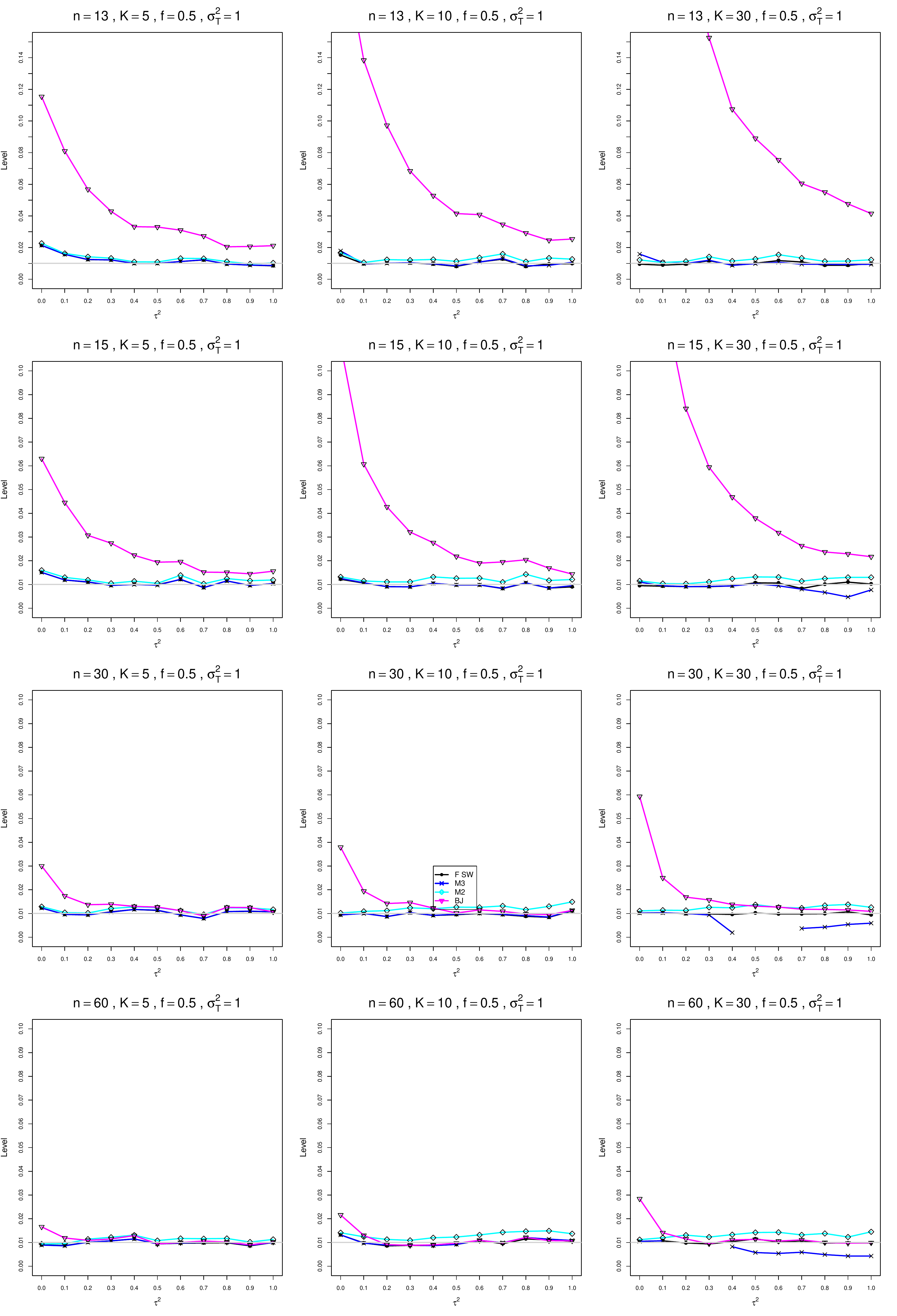}
	\caption{Empirical p-values when $\alpha = .01$ for $\sigma_T^2 = 1$, $f = .5$, and unequal sample sizes $\bar{n}$ = 13, 15, 30, 60
		\label{PlotOfPhatAt001Sigma2T1andq05MD_underH1_unequal}}
\end{figure}

\begin{figure}[t]
	\centering
	\includegraphics[scale=0.33]{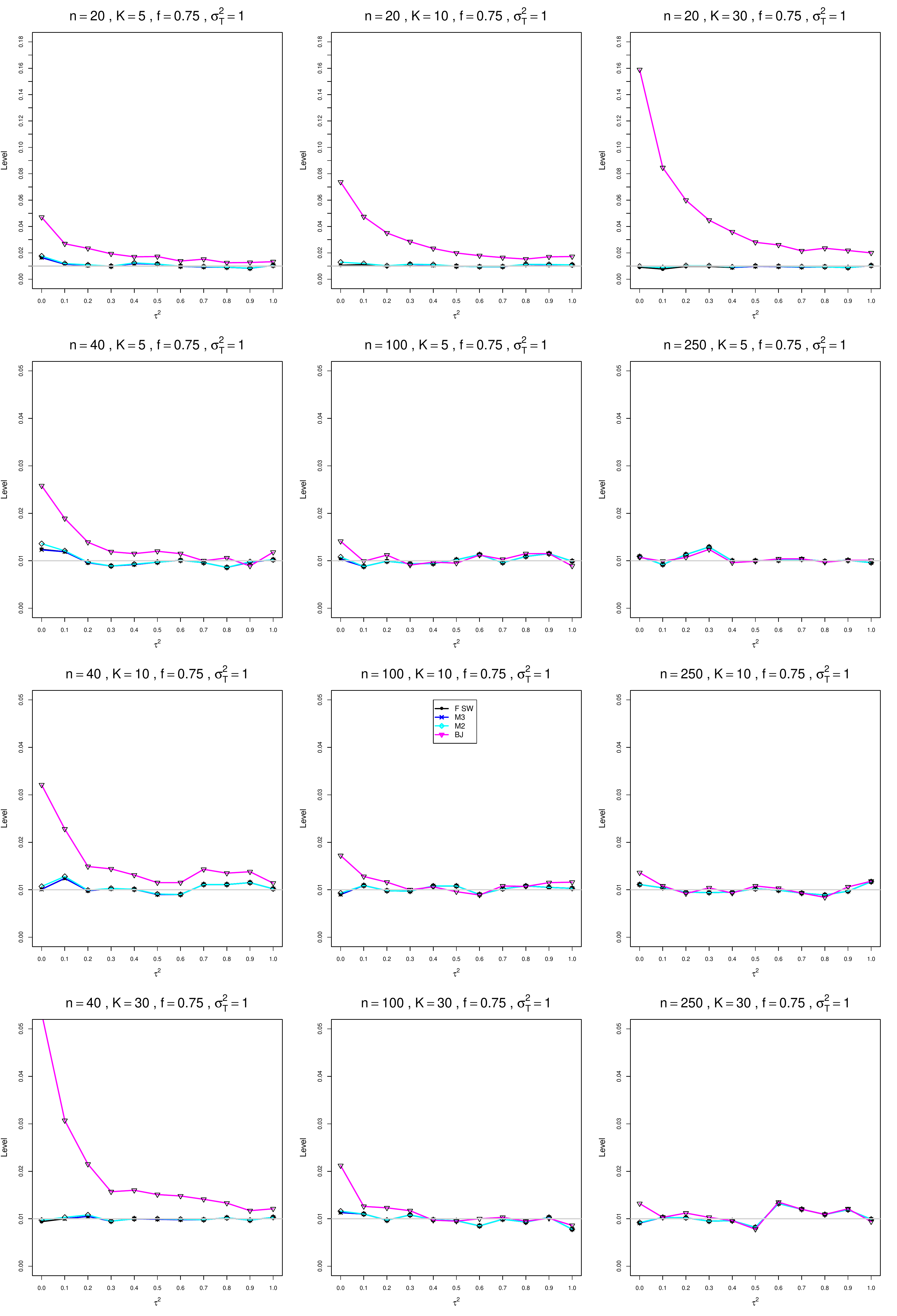}
	\caption{Empirical p-values when $\alpha = .01$ for $\sigma_T^2 = 1$, $f = .75$, and equal sample sizes $n$ = 20, 40, 100, 250
		\label{PlotOfPhatAt001Sigma2T1andq075MD_underH1}}
\end{figure}

\begin{figure}[t]
	\centering
	\includegraphics[scale=0.33]{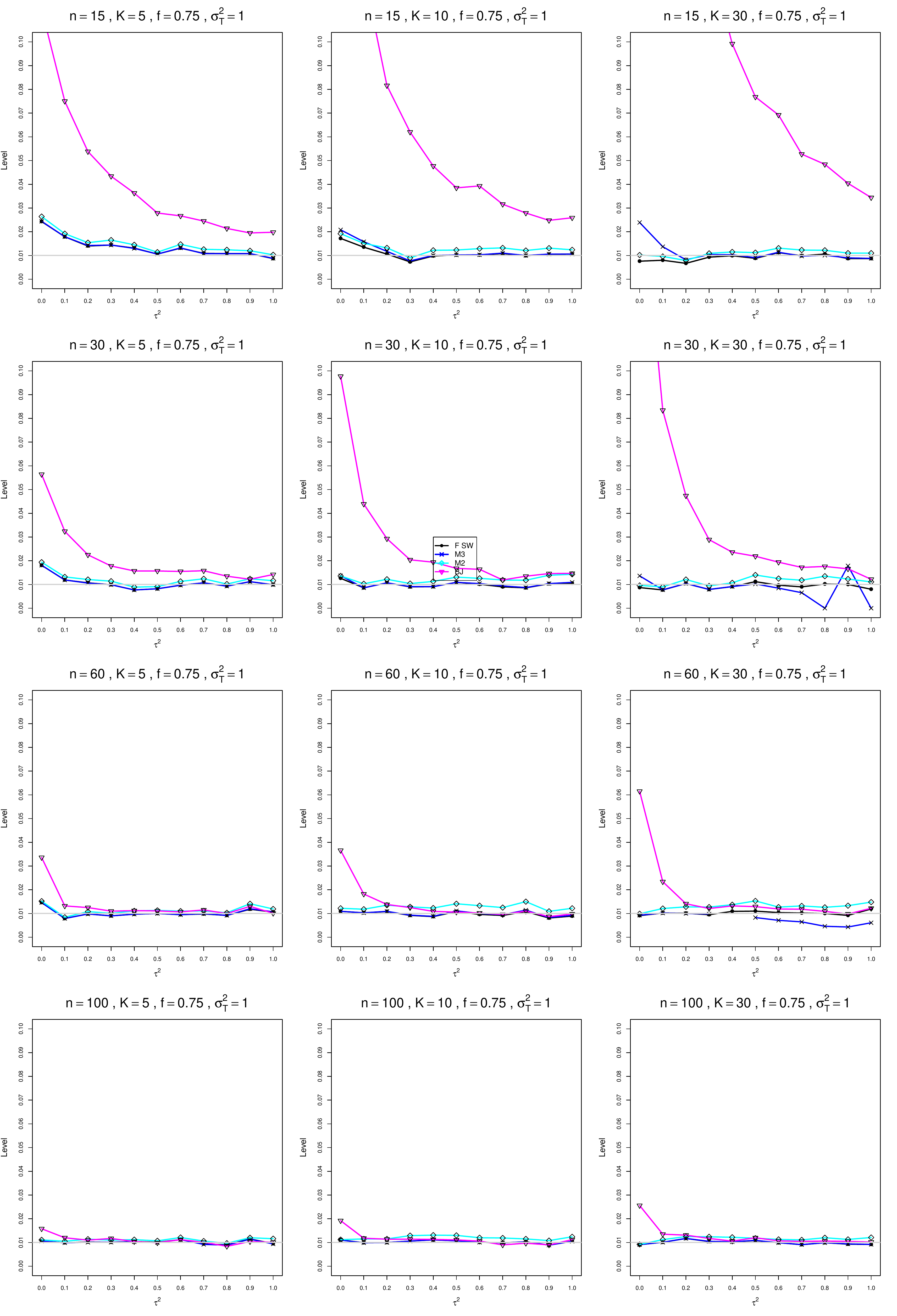}
	\caption{Empirical p-values when $\alpha = .01$ for $\sigma_T^2 = 1$, $f = .75$, and unequal sample sizes $\bar{n}$ = 15, 30, 60, 100
		\label{PlotOfPhatAt001Sigma2T1andq075MD_underH1_unequal}}
\end{figure}

\begin{figure}[t]
	\centering
	\includegraphics[scale=0.33]{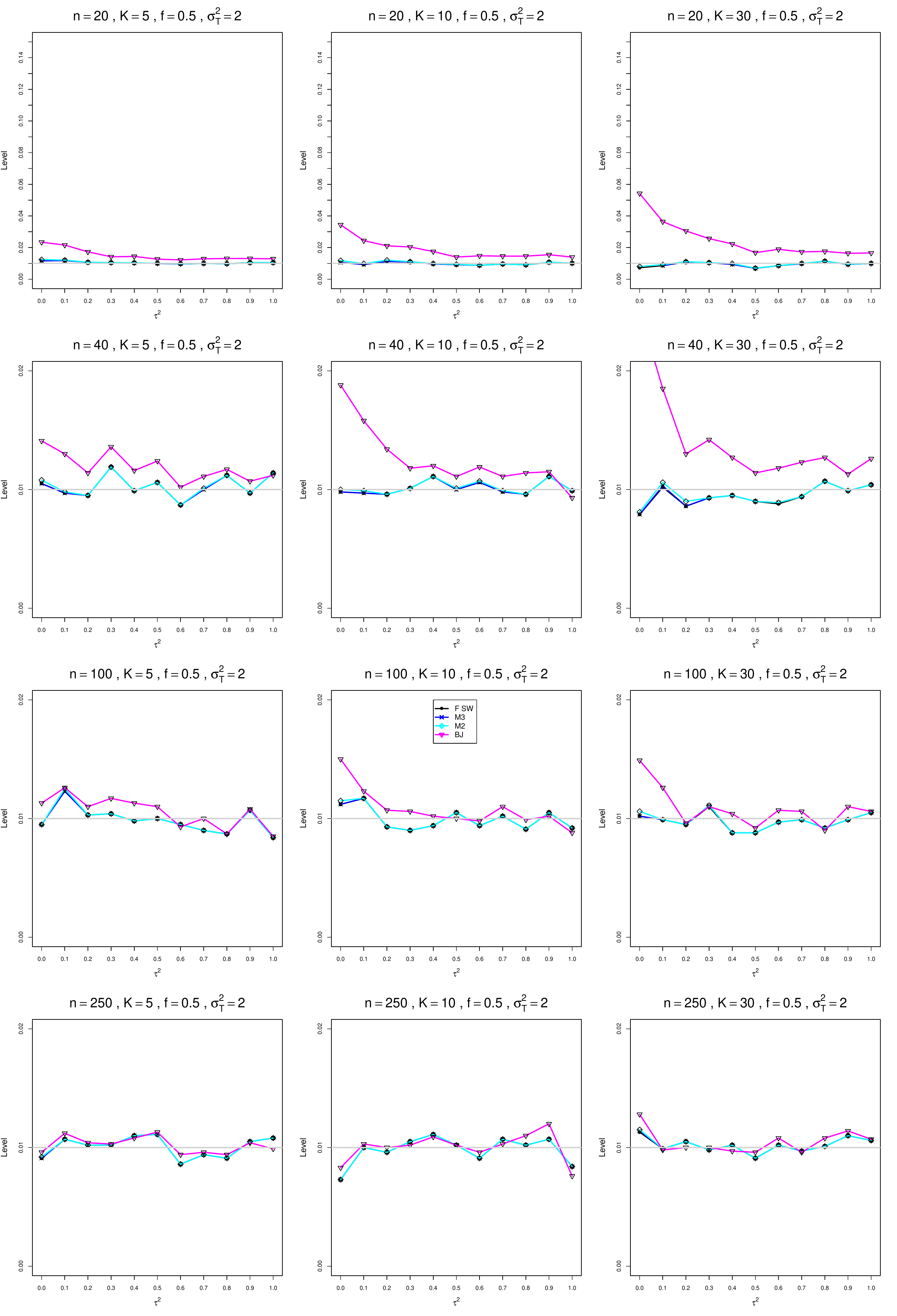}
	\caption{Empirical p-values when $\alpha = .01$ for $\sigma_T^2 = 2$, $f = .5$, and equal sample sizes $n$ = 20, 40, 100, 250
		\label{PlotOfPhatAt001Sigma2T2andq05MD_underH1}}
\end{figure}

\begin{figure}[t]
	\centering
	\includegraphics[scale=0.33]{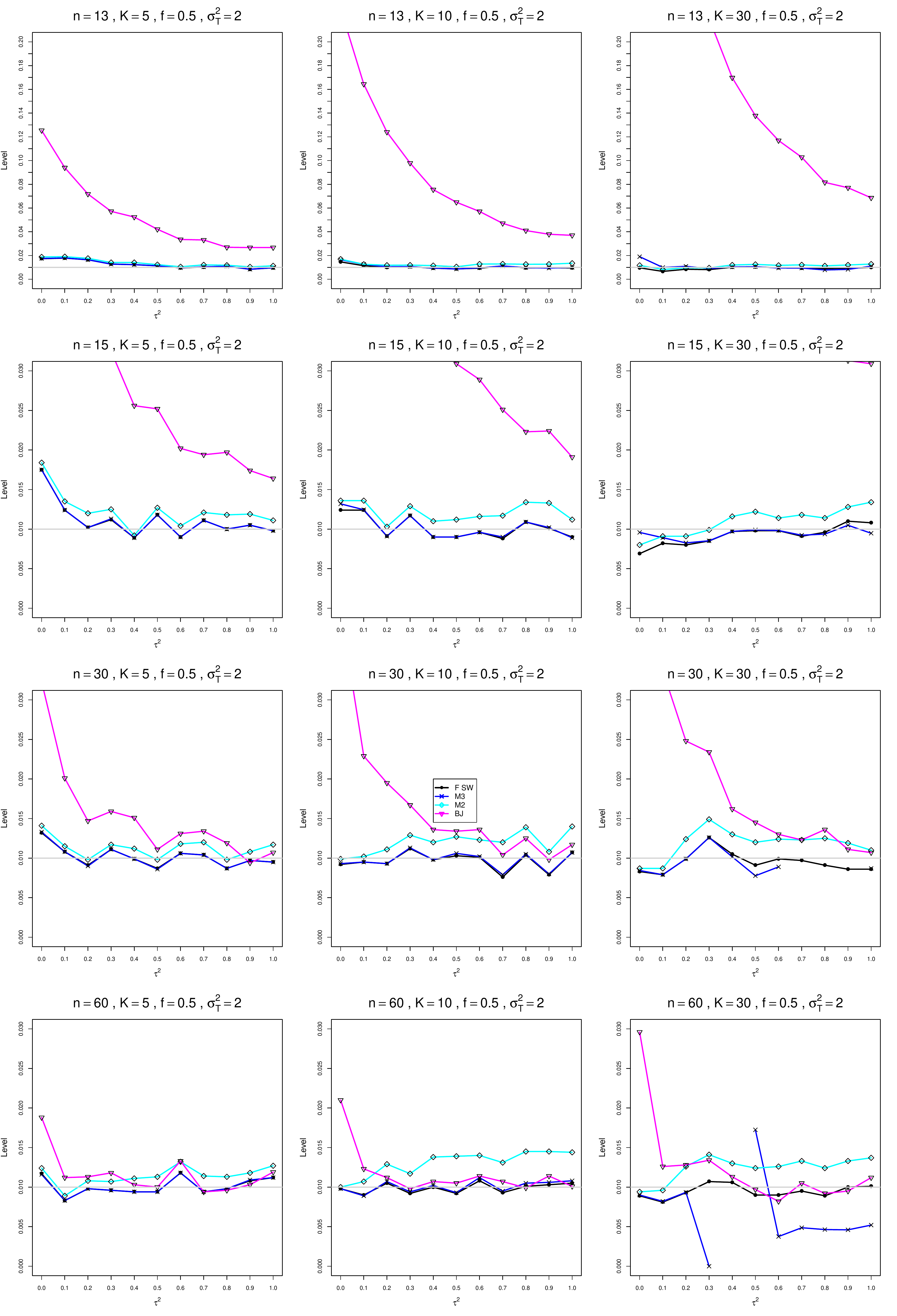}
	\caption{Empirical p-values when $\alpha = .01$ for $\sigma_T^2 = 2$, $f = .5$, and unequal sample sizes $\bar{n}$ = 13, 15, 30, 60
		\label{PlotOfPhatAt001Sigma2T2andq05MD_underH1_unequal}}
\end{figure}

\begin{figure}[t]
	\centering
	\includegraphics[scale=0.33]{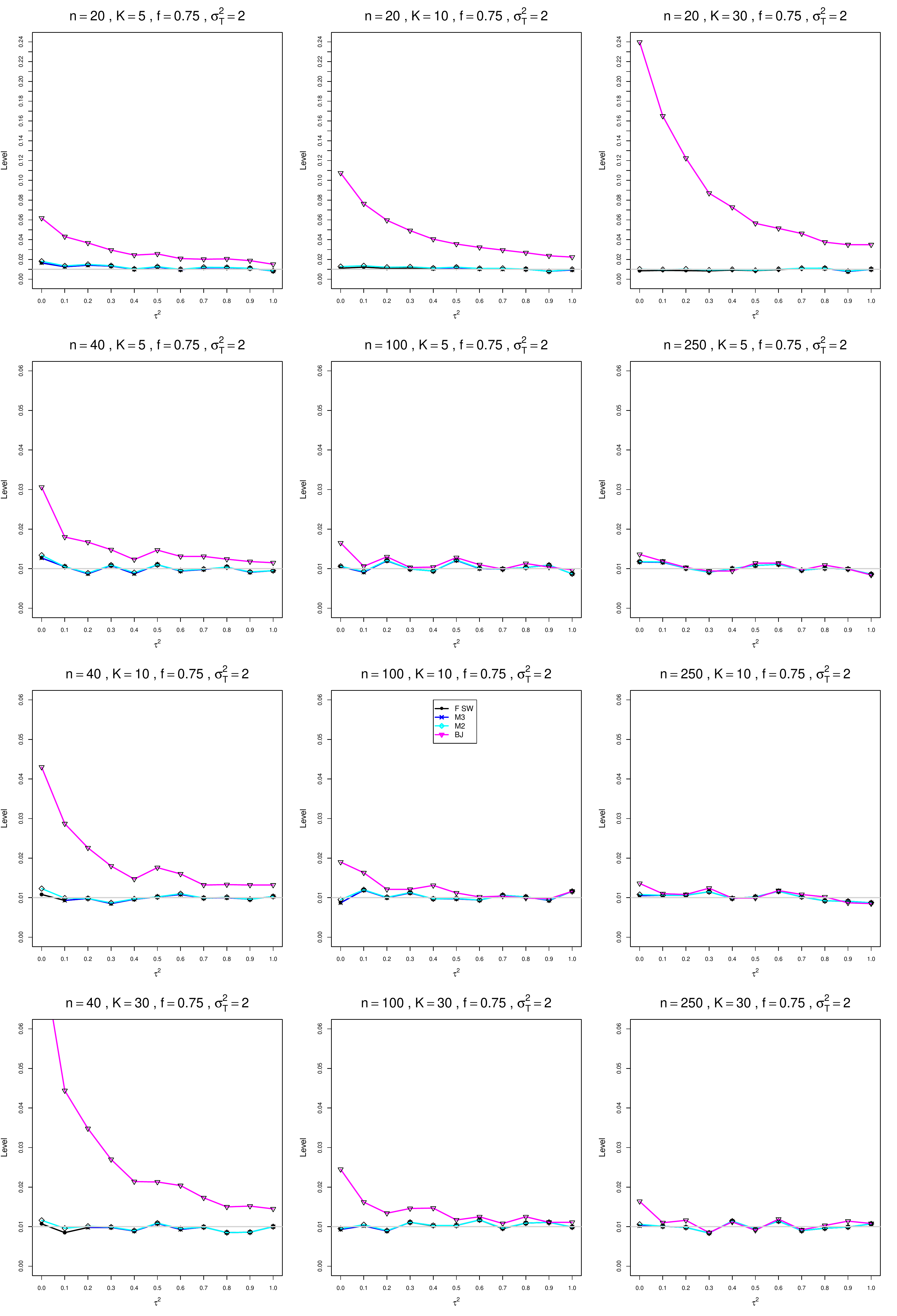}
	\caption{Empirical p-values when $\alpha = .01$ for $\sigma_T^2 = 2$, $f = .75$, and equal sample sizes $n$ = 20, 40, 100, 250
		\label{PlotOfPhatAt001Sigma2T2andq075MD_underH1}}
\end{figure}

\begin{figure}[t]
	\centering
	\includegraphics[scale=0.33]{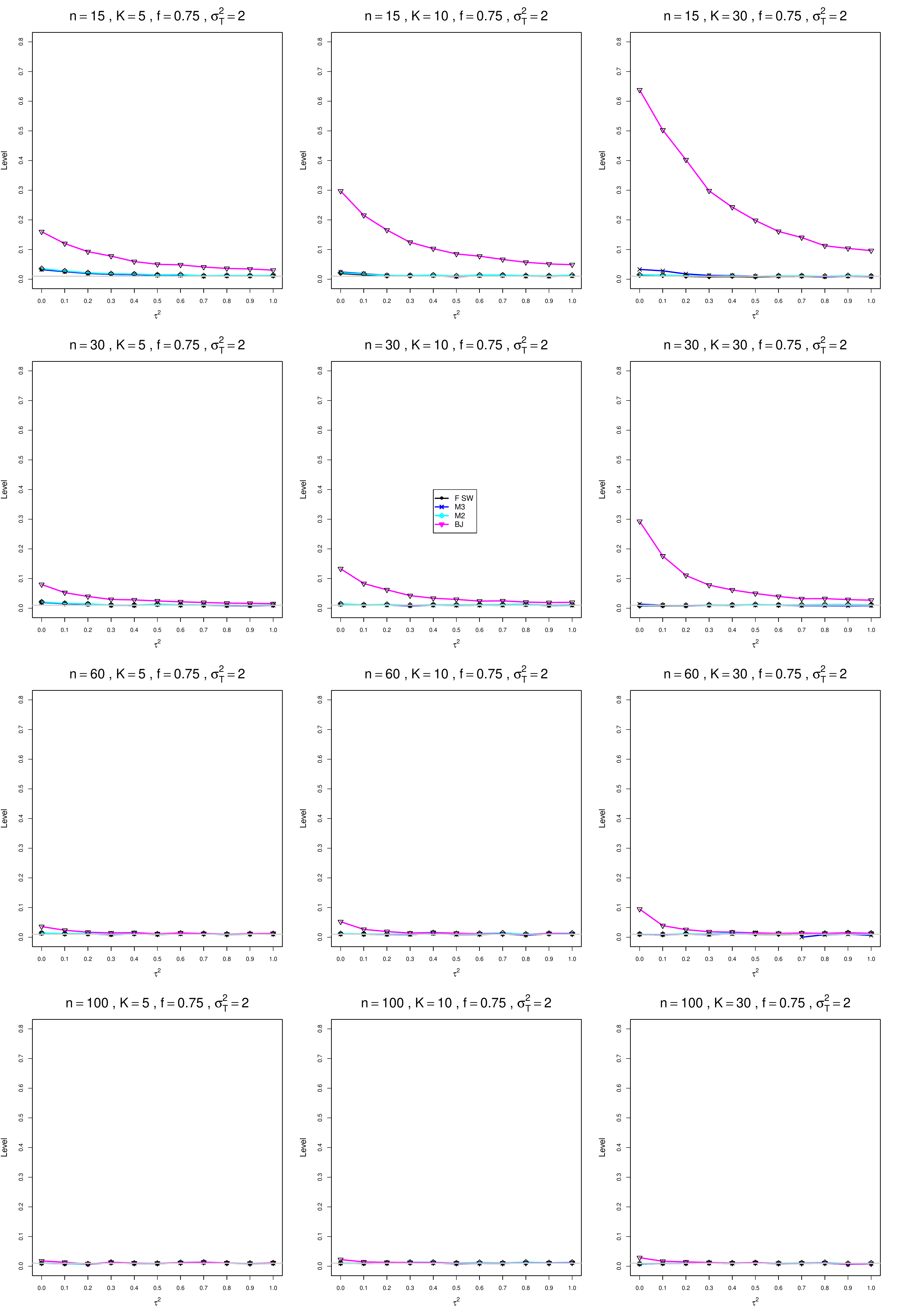}
	\caption{Empirical p-values when $\alpha = .01$ for $\sigma_T^2 = 2$, $f = .75$, and unequal sample sizes $\bar{n}$ = 15, 30, 60, 100
		\label{PlotOfPhatAt001Sigma2T2andq075MD_underH1_unequal}}
\end{figure}

\begin{figure}[t]
	\centering
	\includegraphics[scale=0.33]{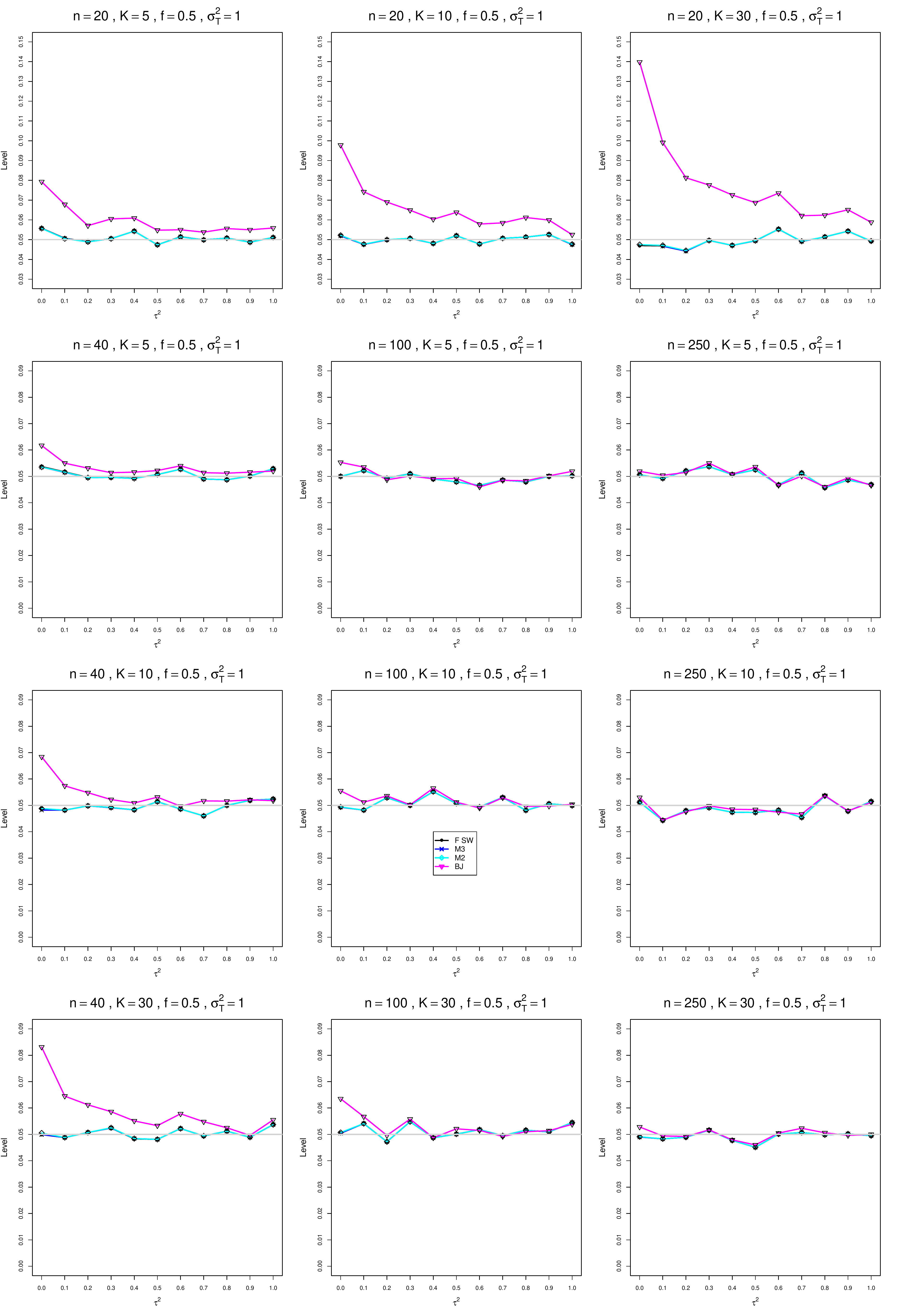}
	\caption{Empirical p-values when $\alpha = .05$ for $\sigma_T^2 = 1$, $f = .5$, and equal sample sizes $n$ = 20, 40, 100, 250
		\label{PlotOfPhatAt005Sigma2T1andq05MD_underH1}}
\end{figure}

\begin{figure}[t]
	\centering
	\includegraphics[scale=0.33]{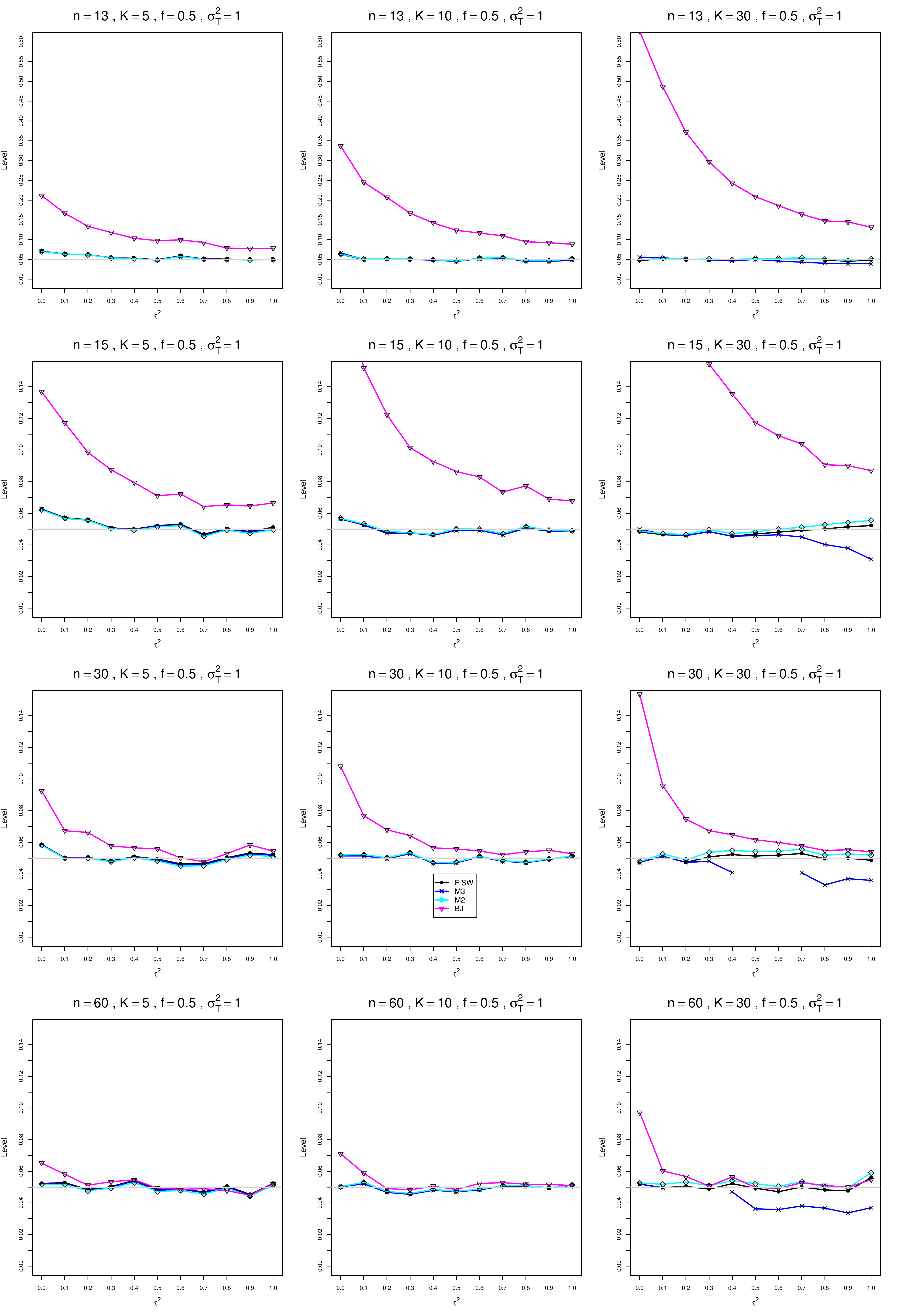}
	\caption{Empirical p-values when $\alpha = .05$ for $\sigma_T^2 = 1$, $f = .5$, and unequal sample sizes $\bar{n}$ = 13, 15, 30, 60
		\label{PlotOfPhatAt005Sigma2T1andq05MD_underH1_unequal}}
\end{figure}

\begin{figure}[t]
	\centering
	\includegraphics[scale=0.33]{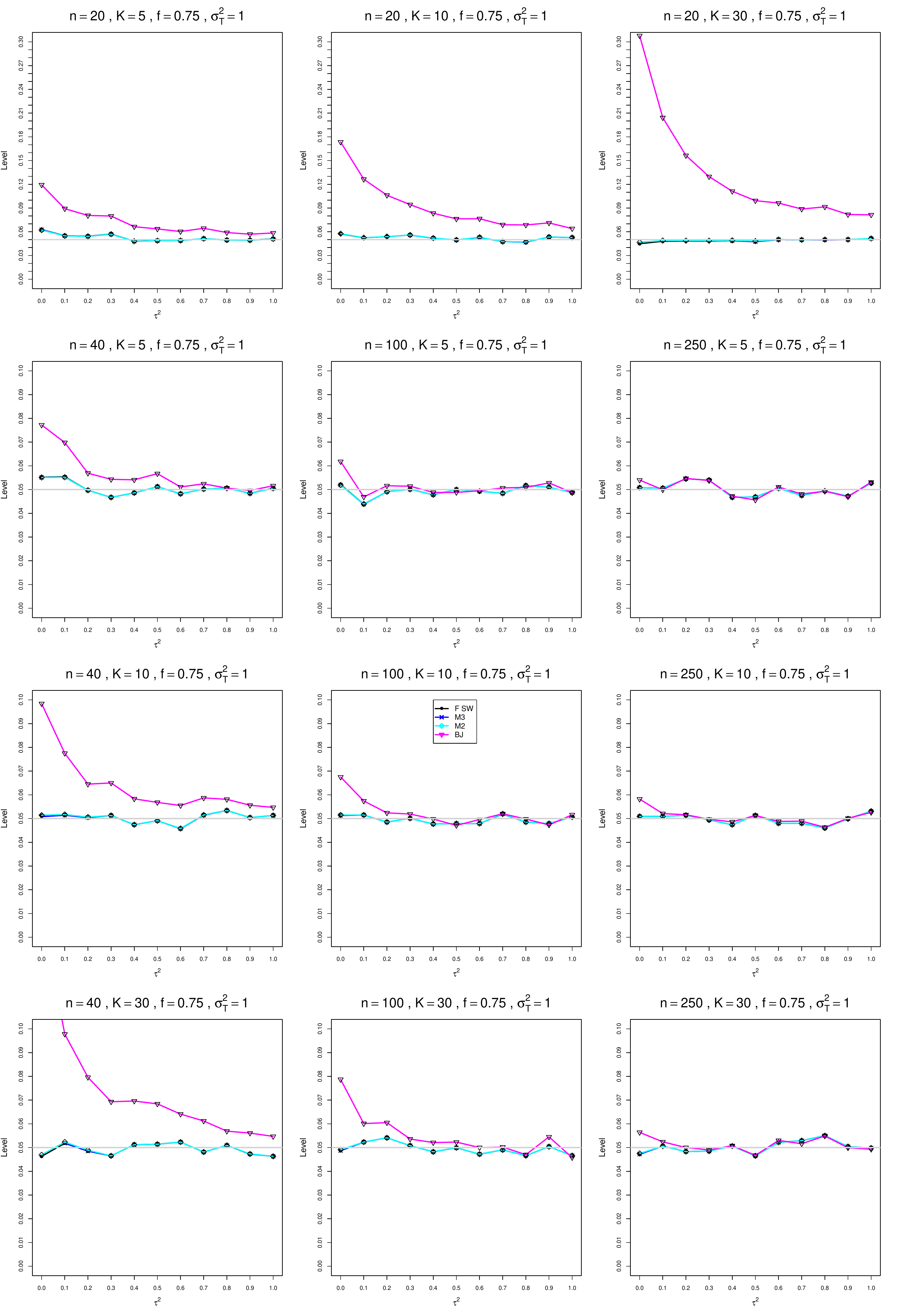}
	\caption{Empirical p-values when $\alpha = .05$ for $\sigma_T^2 = 1$, $f = .75$, and equal sample sizes $n$ = 20, 40, 100, 250
		\label{PlotOfPhatAt005Sigma2T1andq075MD_underH1}}
\end{figure}

\begin{figure}[t]
	\centering
	\includegraphics[scale=0.33]{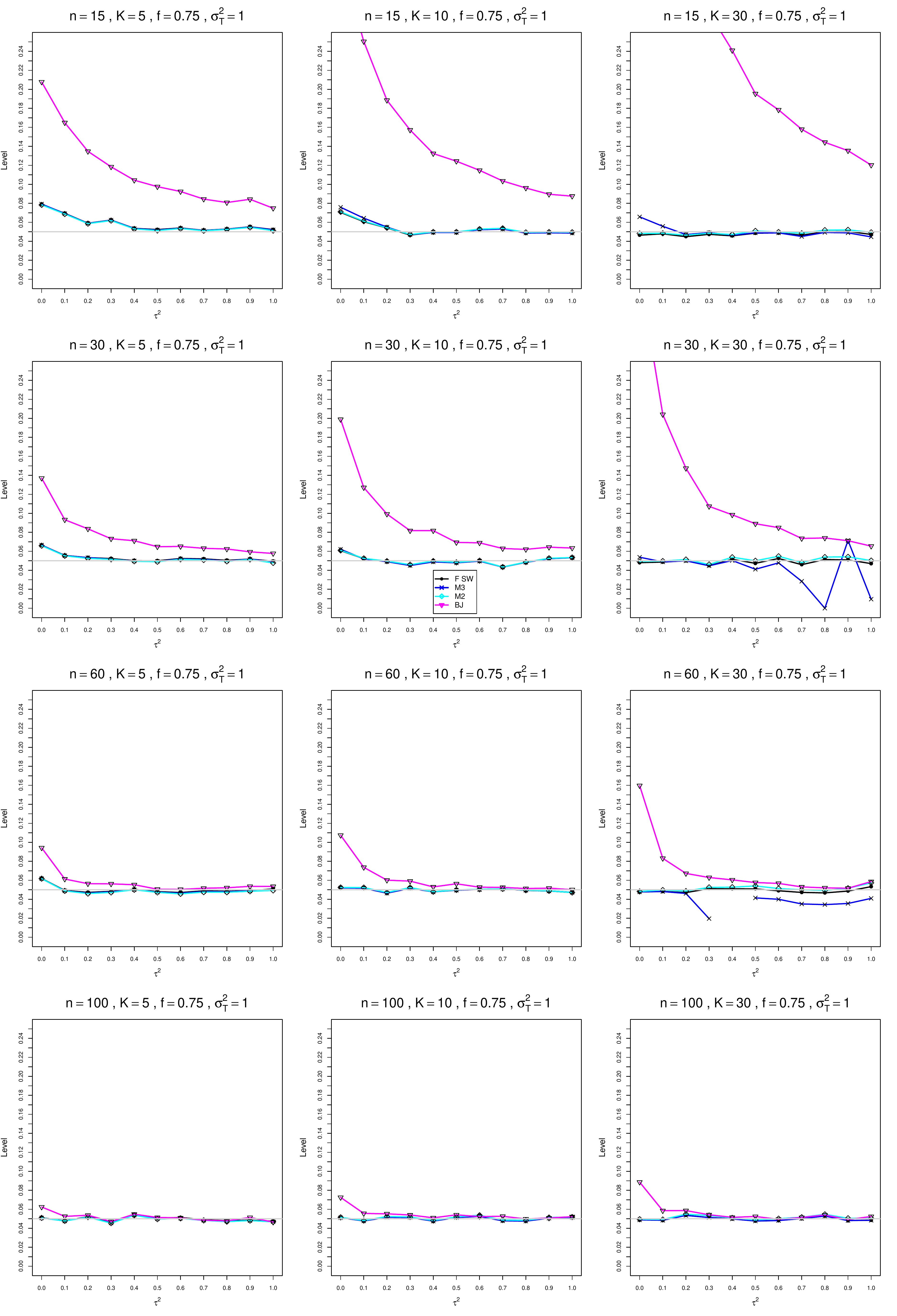}
	\caption{Empirical p-values when $\alpha = .05$ for $\sigma_T^2 = 1$, $f = .75$, and unequal sample sizes $\bar{n}$ = 15, 30, 60, 100
		\label{PlotOfPhatAt005Sigma2T1andq075MD_underH1_unequal}}
\end{figure}

\begin{figure}[t]
	\centering
	\includegraphics[scale=0.33]{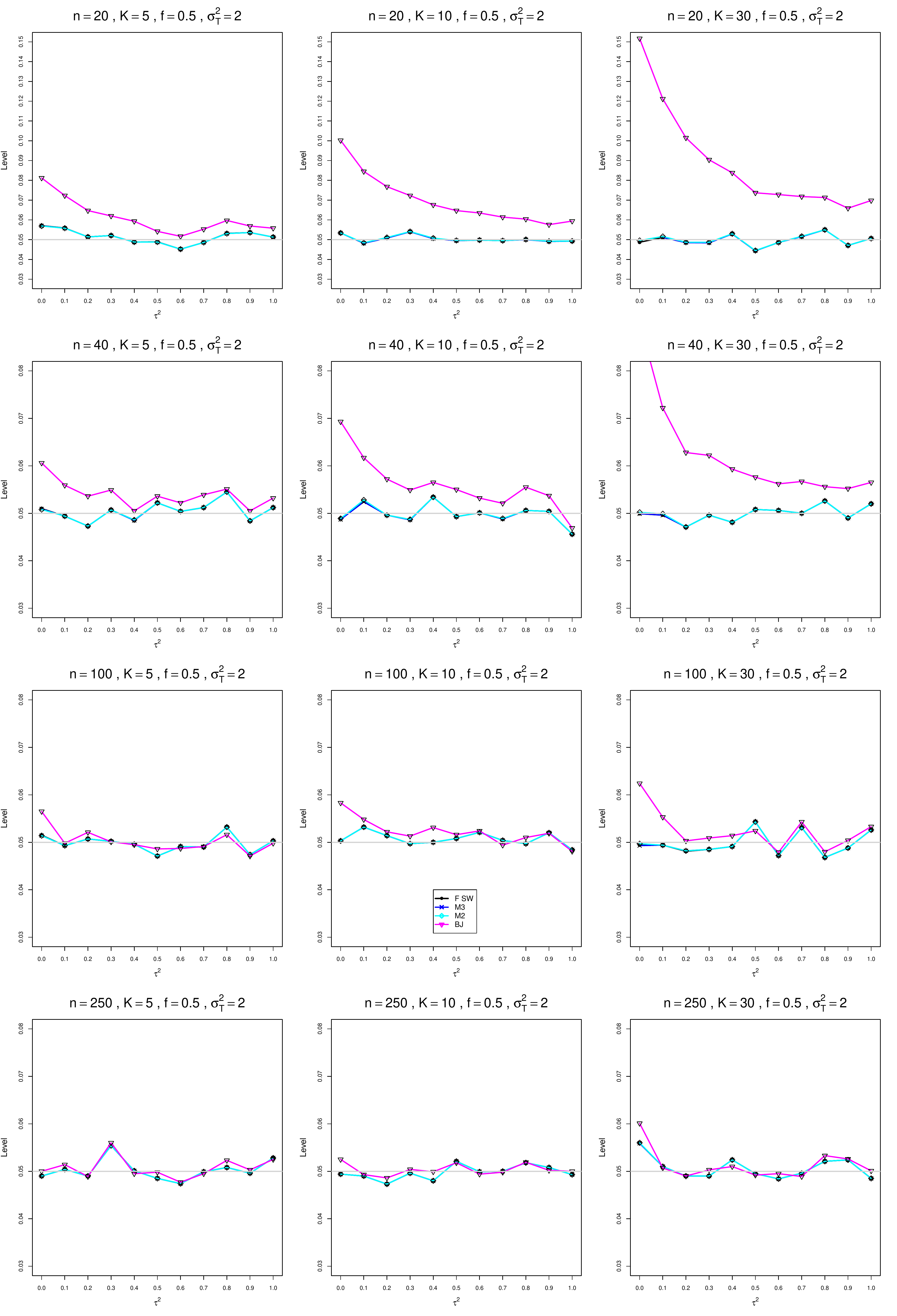}
	\caption{Empirical p-values when $\alpha = .05$ for $\sigma_T^2 = 2$, $f = .5$, and equal sample sizes $n$ = 20, 40, 100, 250
		\label{PlotOfPhatAt005Sigma2T2andq05MD_underH1}}
\end{figure}

\begin{figure}[t]
	\centering
	\includegraphics[scale=0.33]{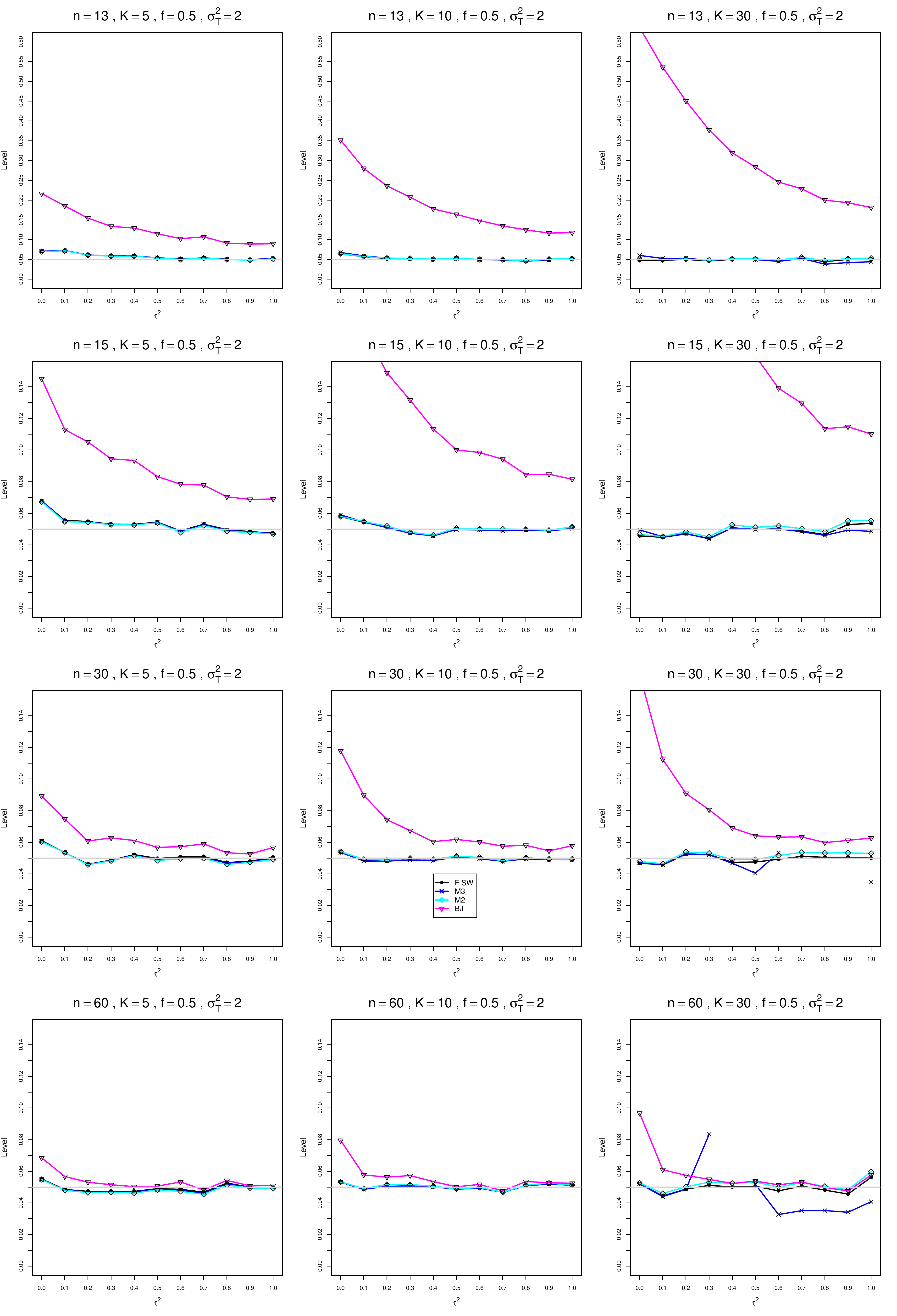}
	\caption{Empirical p-values when $\alpha = .05$ for $\sigma_T^2 = 2$, $f = .5$, and unequal sample sizes $\bar{n}$ = 13, 15, 30, 60
		\label{PlotOfPhatAt005Sigma2T2andq05MD_underH1_unequal}}
\end{figure}

\begin{figure}[t]
	\centering
	\includegraphics[scale=0.33]{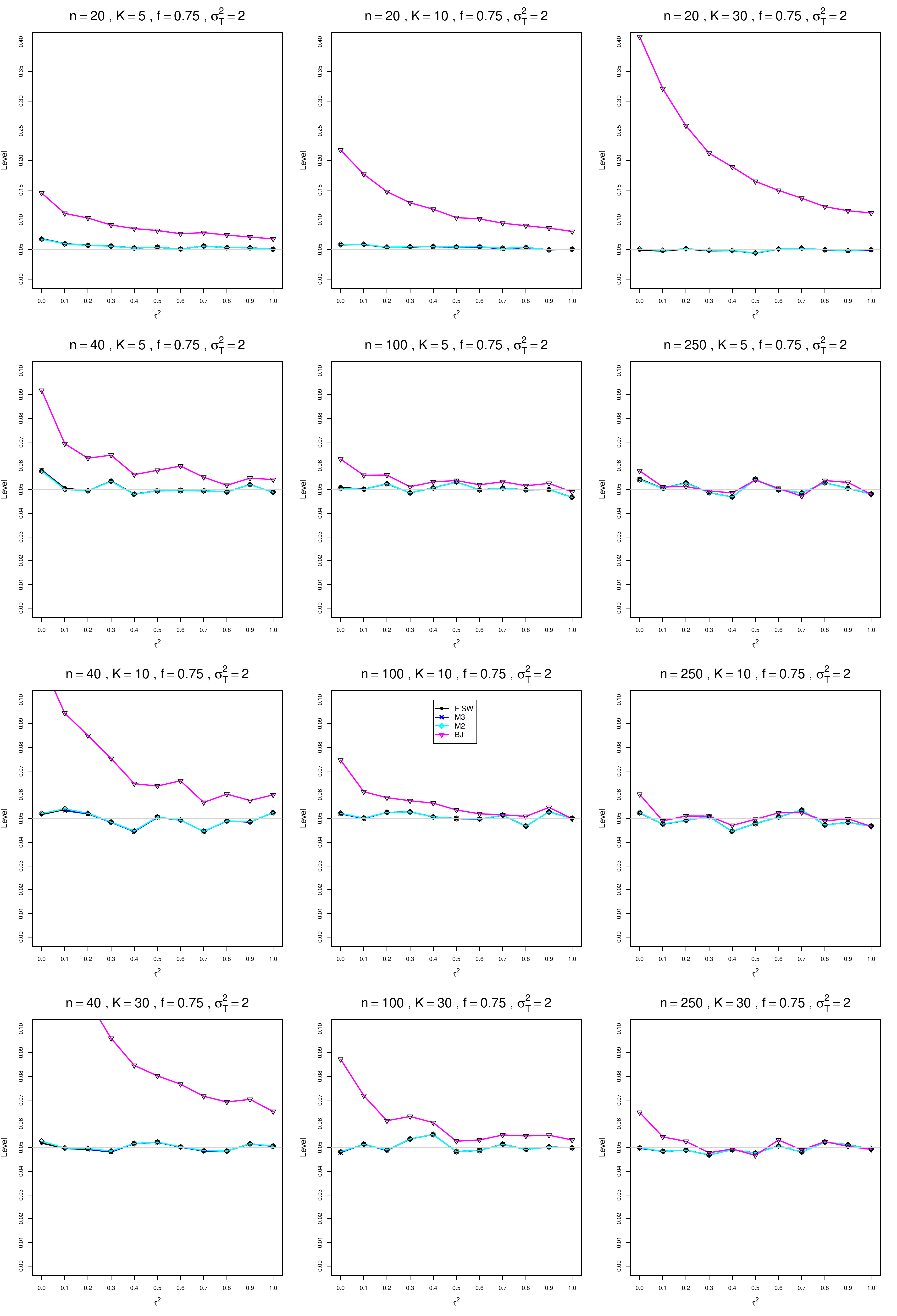}
	\caption{Empirical p-values when $\alpha = .05$ for $\sigma_T^2 = 2$, $f = .75$, and equal sample sizes $n$ = 20, 40, 100, 250
		\label{PlotOfPhatAt005Sigma2T2andq075MD_underH1}}
\end{figure}

\begin{figure}[t]
	\centering
	\includegraphics[scale=0.33]{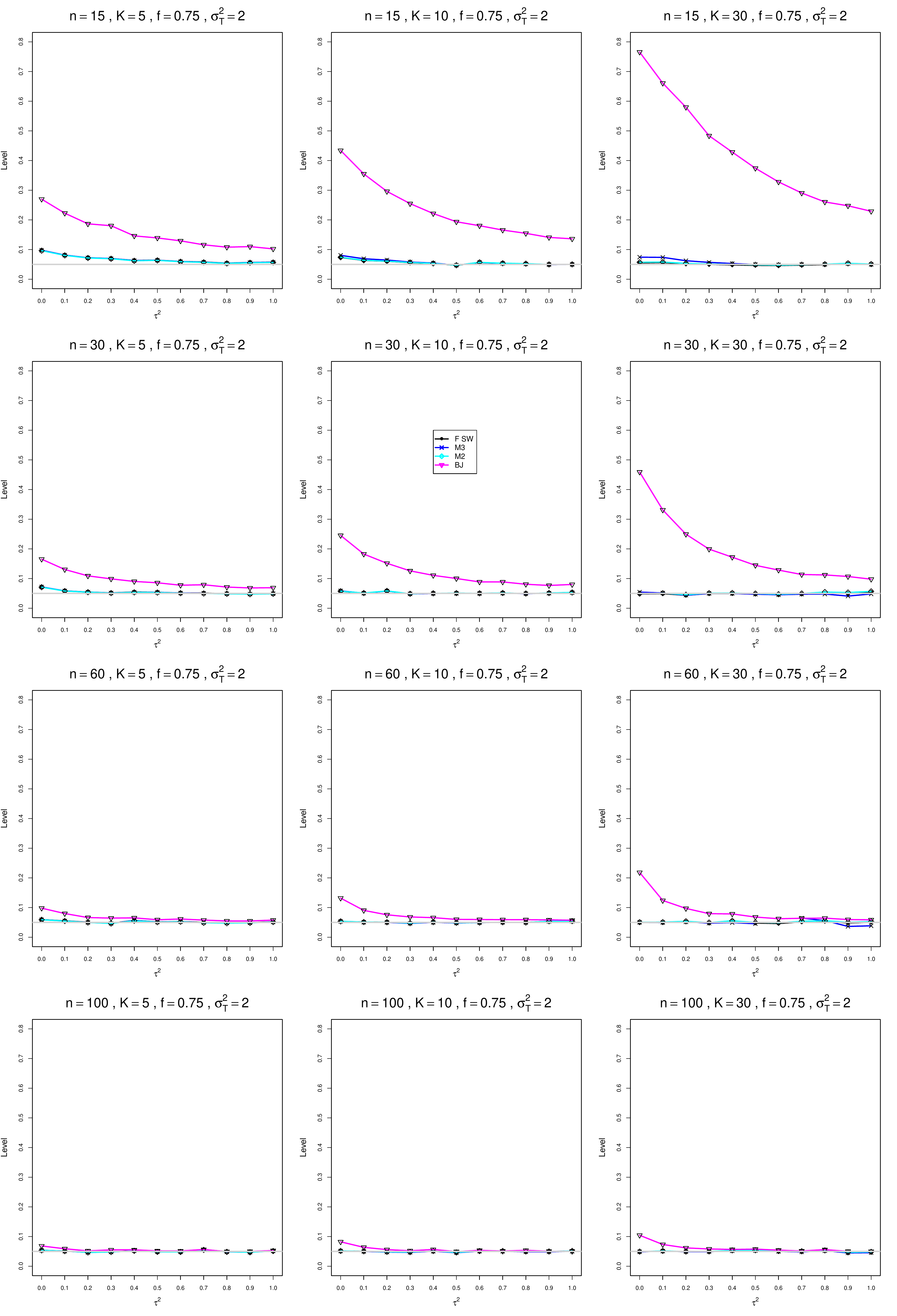}
	\caption{Empirical p-values when $\alpha = .05$ for $\sigma_T^2 = 2$, $f = .75$, and unequal sample sizes $\bar{n}$ = 15, 30, 60, 100
		\label{PlotOfPhatAt005Sigma2T2andq075MD_underH1_unequal}}
\end{figure}

\clearpage
\setcounter{figure}{0}
\setcounter{section}{0}
\renewcommand{\thesection}{B4.\arabic{section}}
\section*{B4. Power of tests for heterogeneity ($\tau^2 = 0$ versus $\tau^2 > 0$) based on approximations to the distribution of $Q$}
Each figure corresponds to a value of $\alpha$ (= .01, .05), a value of $\sigma_T^2$ (= 1, 2), a value of $f$ (= .5, .75), and a pattern of sample sizes (equal or unequal). (For all figures, $\sigma_C^2 = 1$.)\\
For each combination of a value of $n$ (= 20, 40, 100, 250) or $\bar{n}$ (= 13, 15, 30, 60 or 15, 30, 60, 100) and a value of $K$ (= 5, 10, 30), a panel plots power versus $\tau^2$ = 0.0(0.1)1.0. \\
The approximations to the distribution of $Q$ are
\begin{itemize}
	\item F SW (Farebrother approximation, effective-sample-size weights)
	\item M3 (Three-moment approximation, effective-sample-size weights)
	\item M2 (Two-moment approximation, effective-sample-size weights)
	\item $\chi_{K-1}^2$ (Chi-square, IV weights)
	\item Welch (Welch approximation, IV weights)
\end{itemize}

\clearpage

\renewcommand{\thefigure}{B4.\arabic{figure}}
\begin{figure}[t]
	\centering
	\includegraphics[scale=0.33]{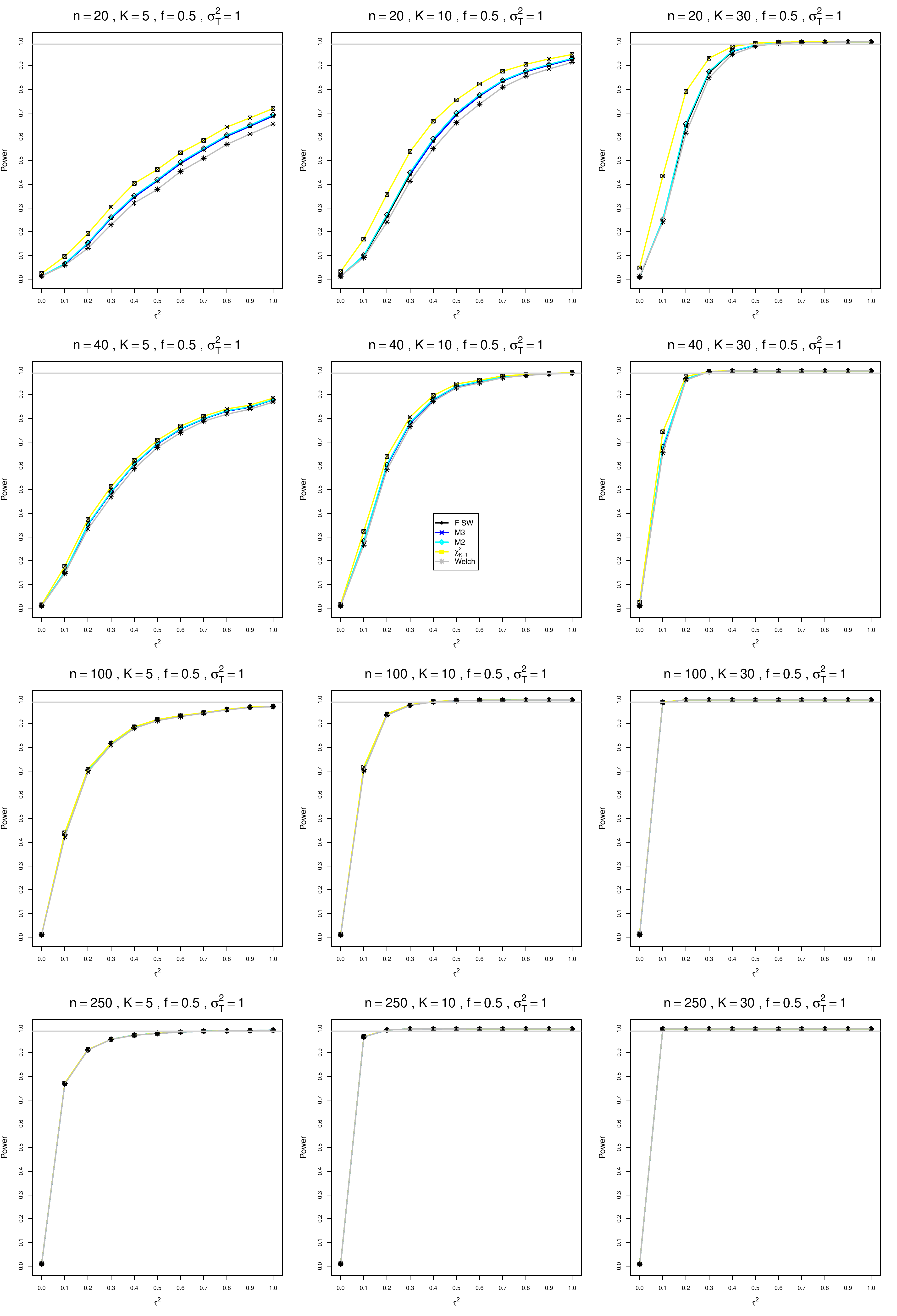}
	\caption{Power when $\alpha = .01$ for $\sigma_T^2 = 1$, $f = .5$, and equal sample sizes $n$ = 20, 40, 100, 250
		\label{PlotOfPhatAt001Sigma2T1andq05MD_underH0}}
\end{figure}

\begin{figure}[t]
	\centering
	\includegraphics[scale=0.33]{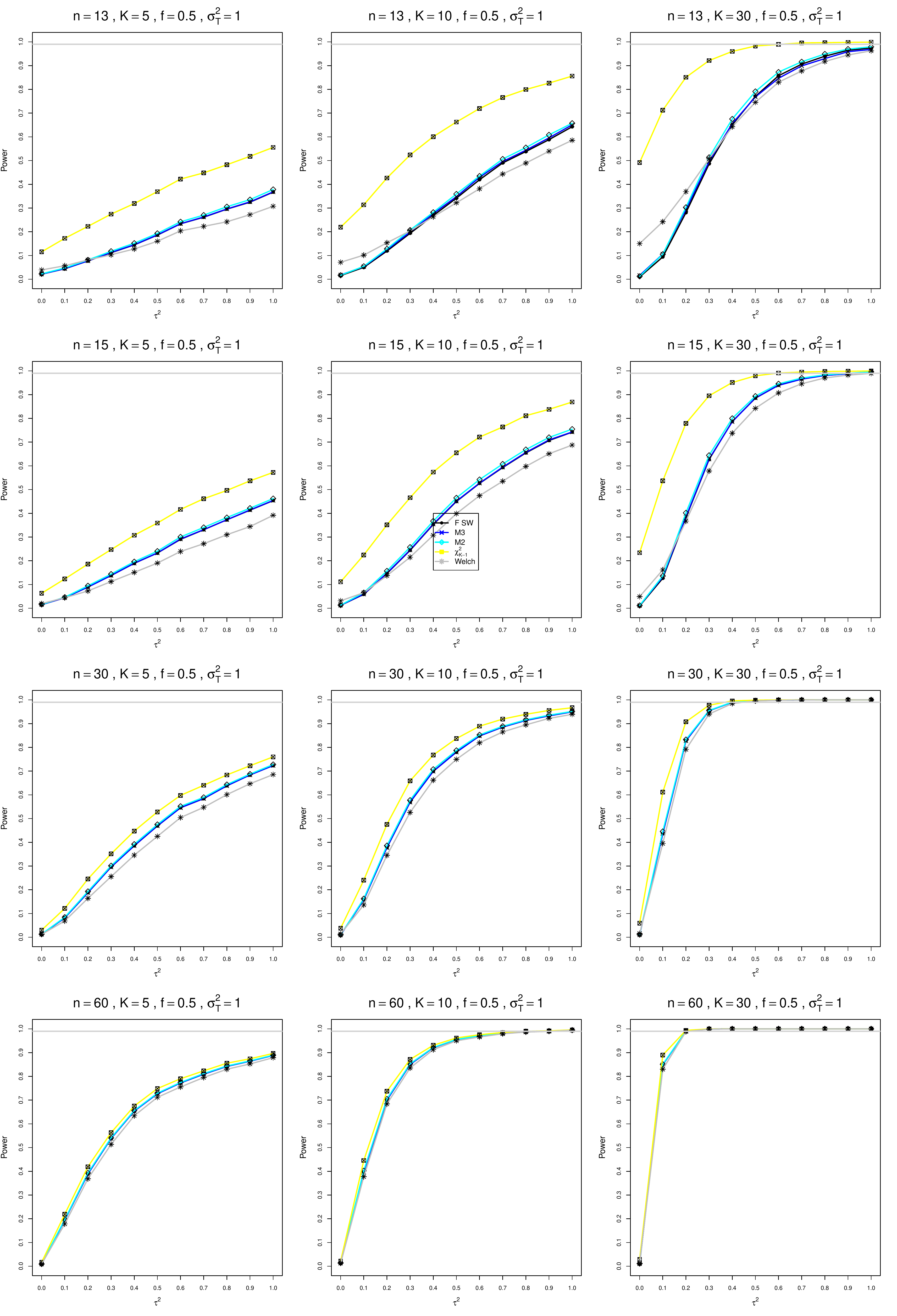}
	\caption{Power when $\alpha = .01$ for $\sigma_T^2 = 1$, $f = .5$, and unequal sample sizes $\bar{n}$ = 13, 15, 30, 60
		\label{PlotOfPhatAt001Sigma2T1andq05MD_underH0_unequal}}
\end{figure}

\begin{figure}[t]
	\centering
	\includegraphics[scale=0.33]{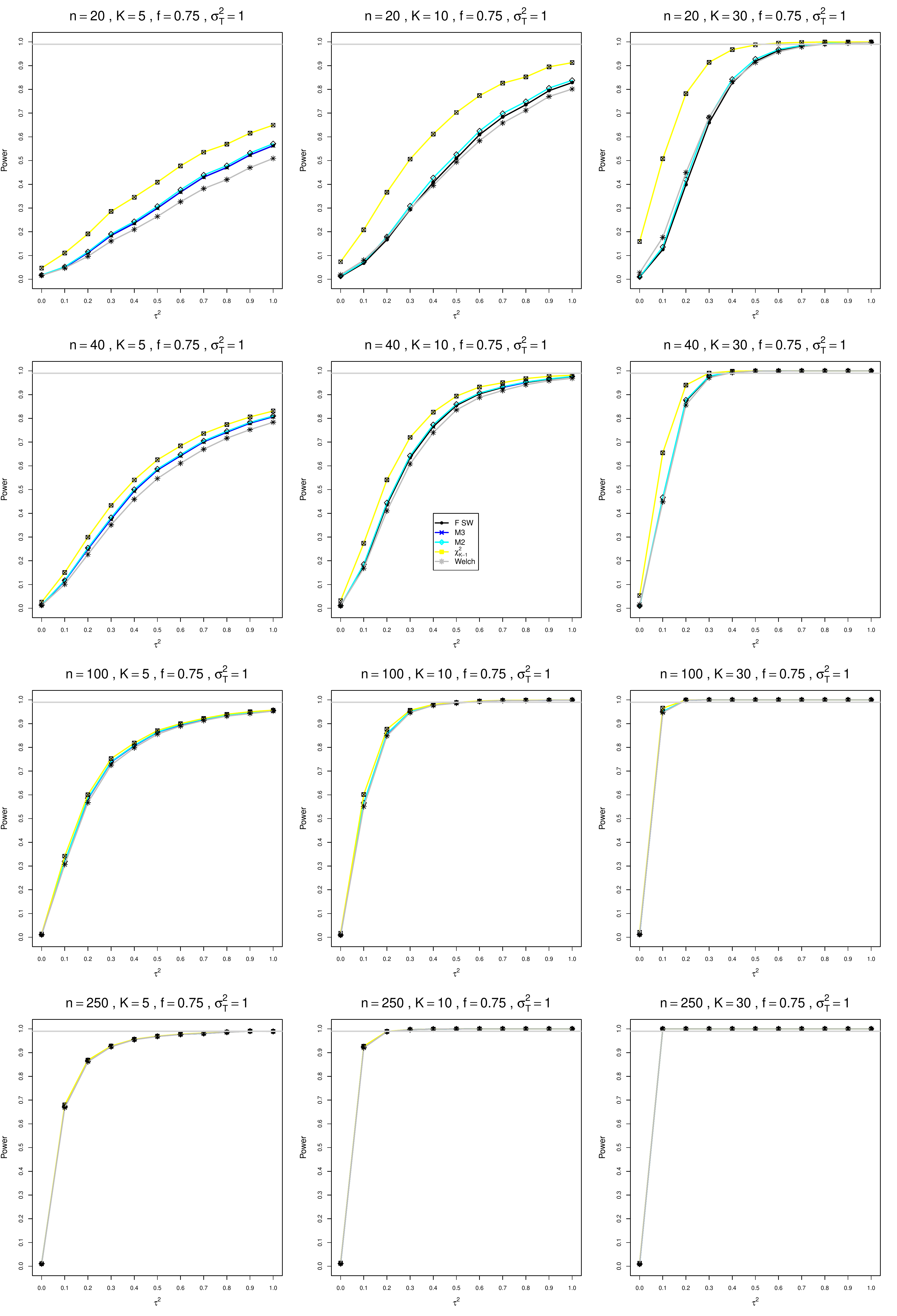}
	\caption{Power when $\alpha = .01$ for $\sigma_T^2 = 1$, $f = .75$, and equal sample sizes $n$ = 20, 40, 100, 250
		\label{PlotOfPhatAt001Sigma2T1andq075MD_underH0}}
\end{figure}

\begin{figure}[t]
	\centering
	\includegraphics[scale=0.33]{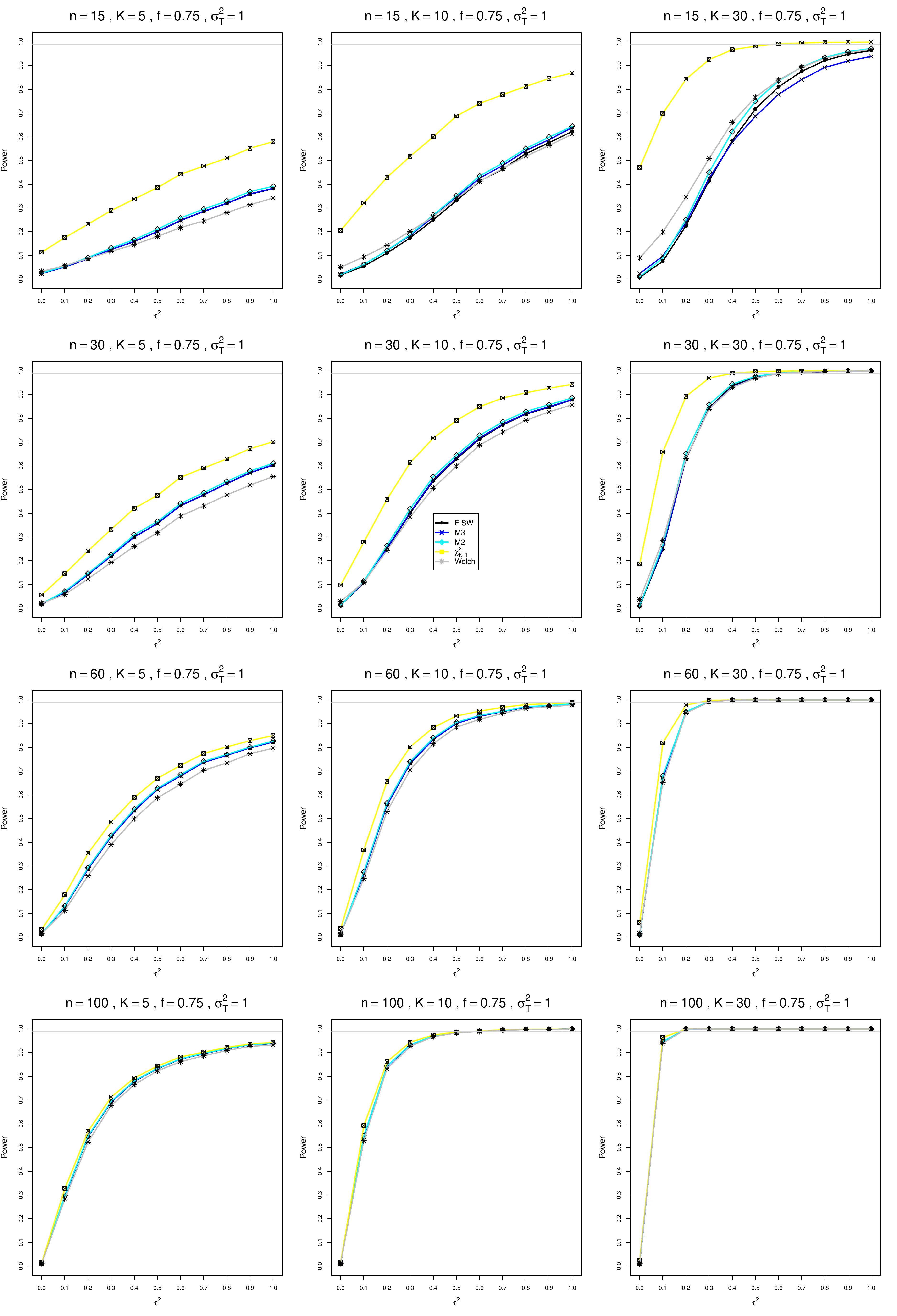}
	\caption{Power when $\alpha = .01$ for $\sigma_T^2 = 1$, $f = .75$, and unequal sample sizes $\bar{n}$ = 15, 30, 60, 100
		\label{PlotOfPhatAt001Sigma2T1andq075MD_underH0_unequal}}
\end{figure}

\begin{figure}[t]
	\centering
	\includegraphics[scale=0.33]{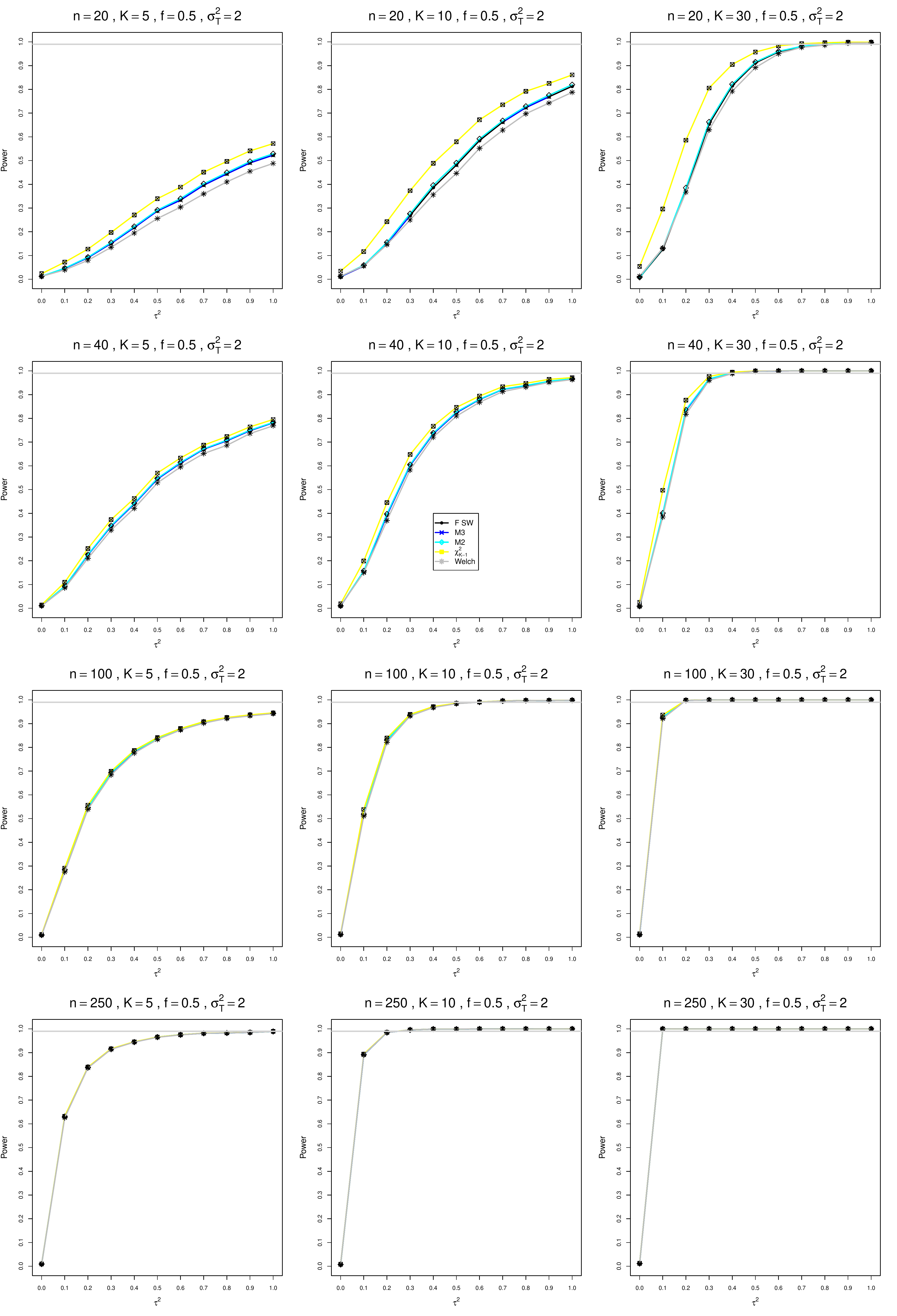}
	\caption{Power when $\alpha = .01$ for $\sigma_T^2 = 2$, $f = .5$, and equal sample sizes $n$ = 20, 40, 100, 250
		\label{PlotOfPhatAt001Sigma2T2andq05MD_underH0}}
\end{figure}

\begin{figure}[t]
	\centering
	\includegraphics[scale=0.33]{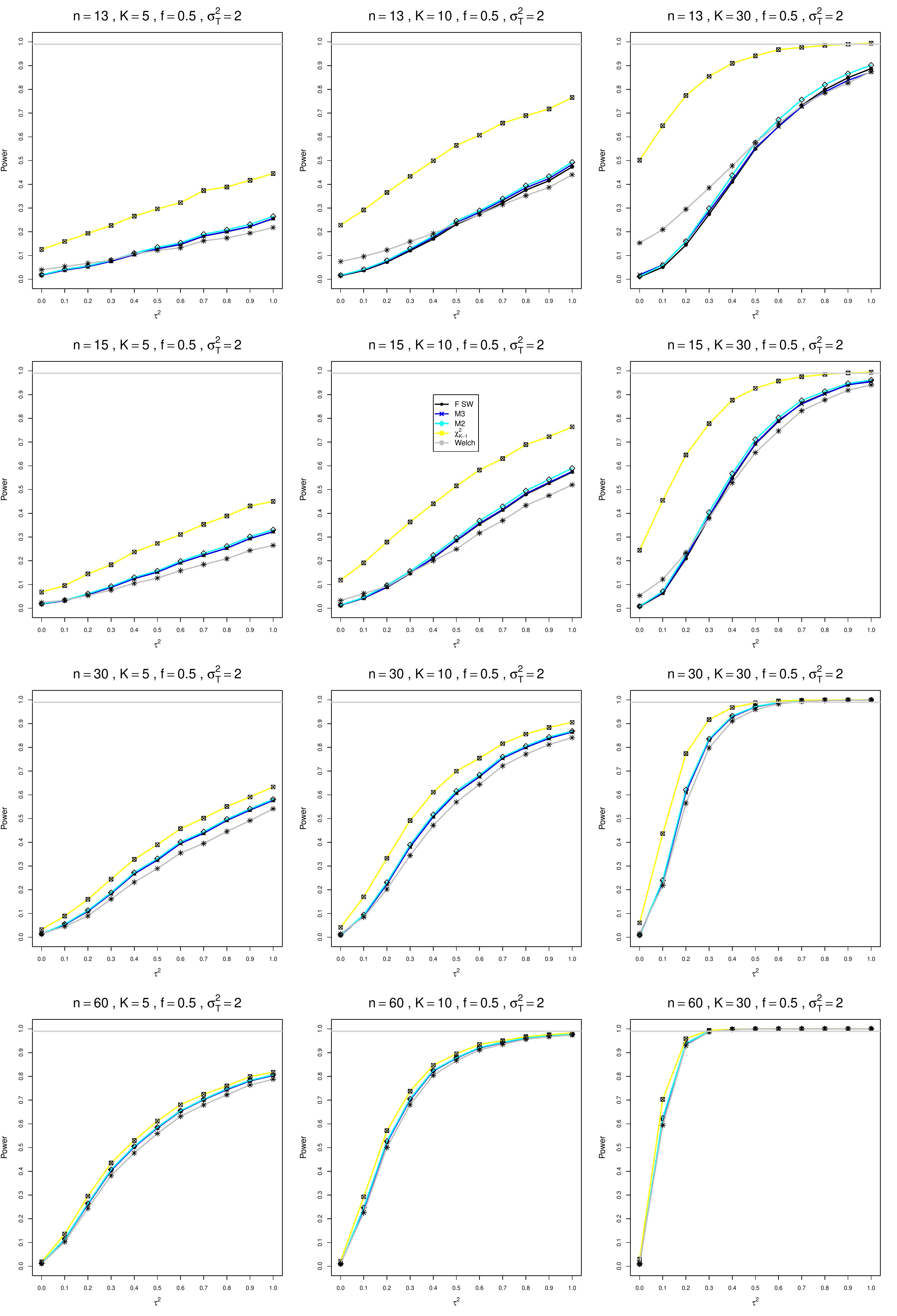}
	\caption{Power when $\alpha = .01$ for $\sigma_T^2 = 2$, $f = .5$, and unequal sample sizes $\bar{n}$ = 13, 15, 30, 60
		\label{PlotOfPhatAt001Sigma2T2andq05MD_underH0_unequal}}
\end{figure}
\begin{figure}[t]
	\centering
	\includegraphics[scale=0.33]{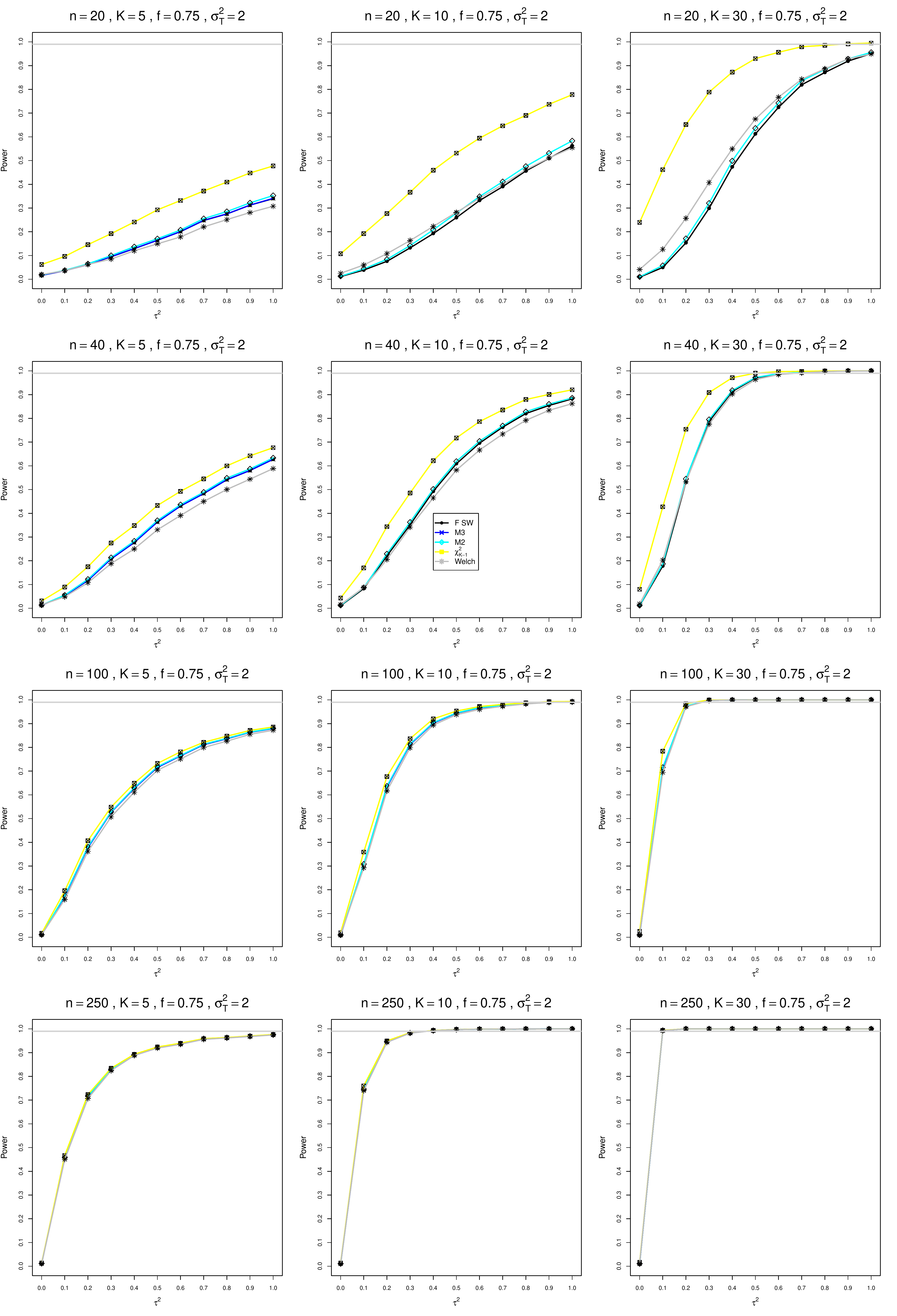}
	\caption{Power when $\alpha = .01$ for $\sigma_T^2 = 2$, $f = .75$, and equal sample sizes $n$ = 20, 40, 100, 250
		\label{PlotOfPhatAt001Sigma2T2andq075MD_underH0}}
\end{figure}

\begin{figure}[t]
	\centering
	\includegraphics[scale=0.33]{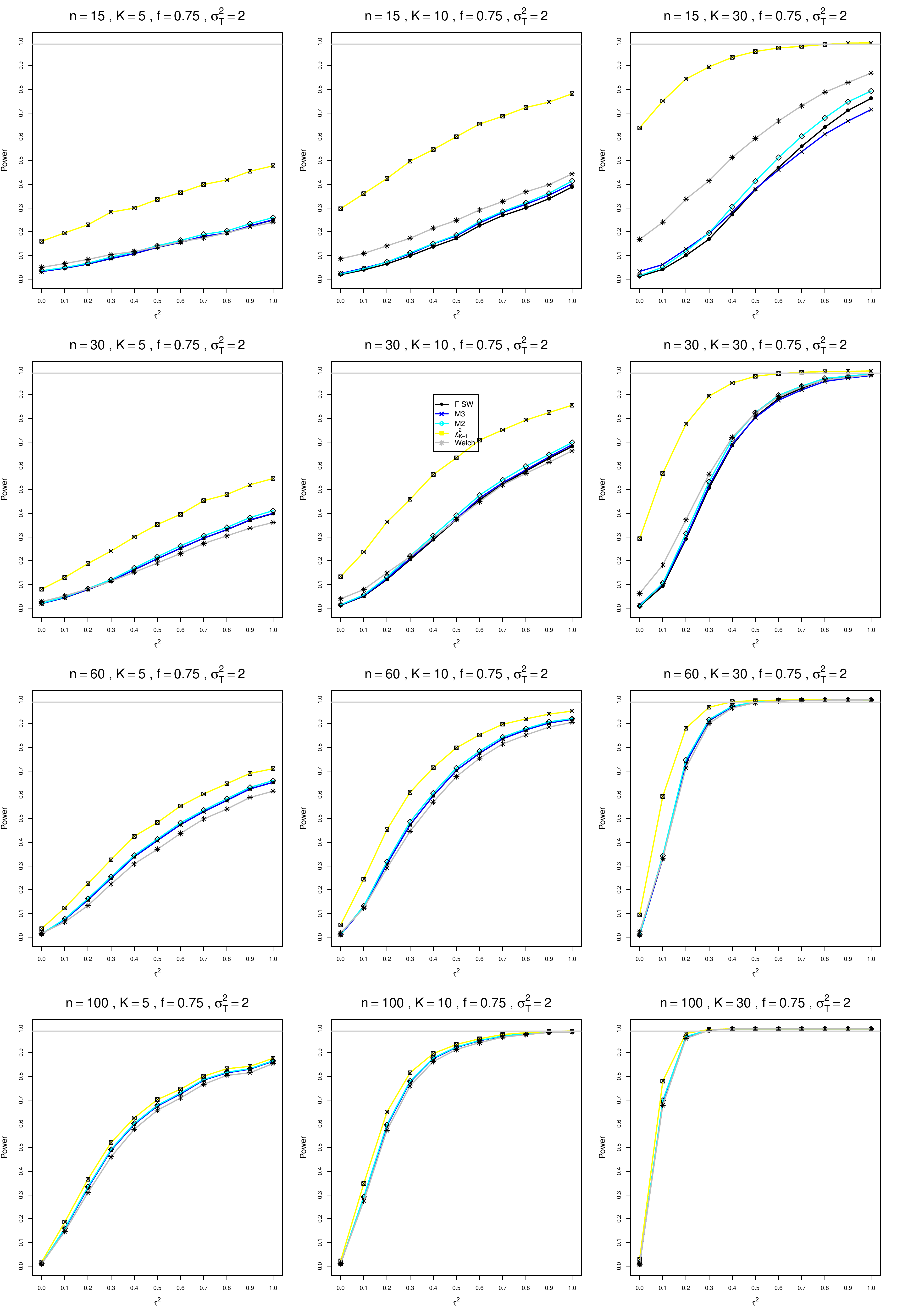}
	\caption{Power when $\alpha = .01$ for $\sigma_T^2 = 2$, $f = .75$, and unequal sample sizes $\bar{n}$ = 15, 30, 60, 100
		\label{PlotOfPhatAt001Sigma2T2andq075MD_underH0_unequal}}
\end{figure}

\begin{figure}[t]
	\centering
	\includegraphics[scale=0.33]{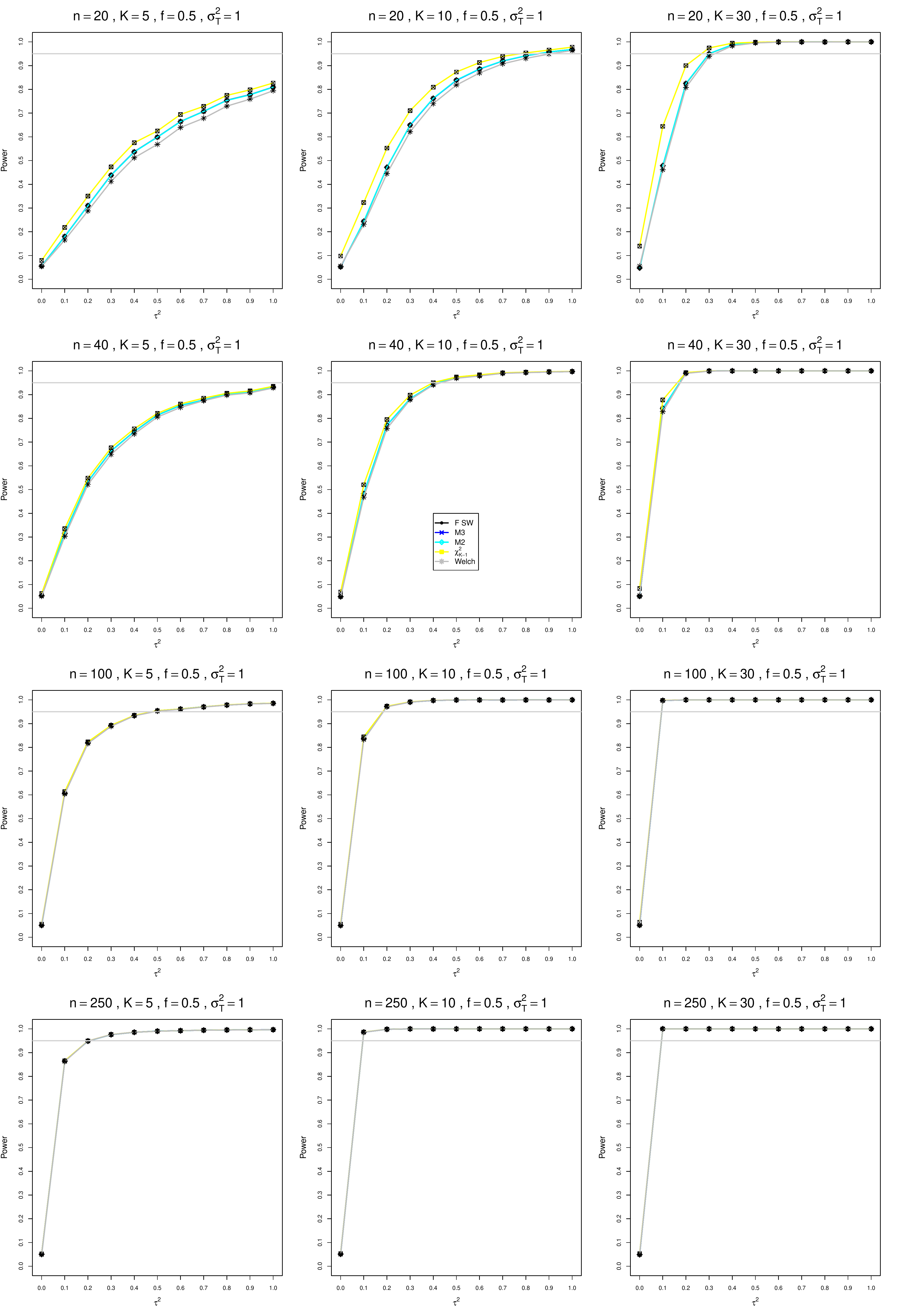}
	\caption{Power when $\alpha = .05$ for $\sigma_T^2 = 1$, $f = .5$, and equal sample sizes $n$ = 20, 40, 100, 250
		\label{PlotOfPhatAt005Sigma2T1andq05MD_underH0}}
\end{figure}

\begin{figure}[t]
	\centering
	\includegraphics[scale=0.33]{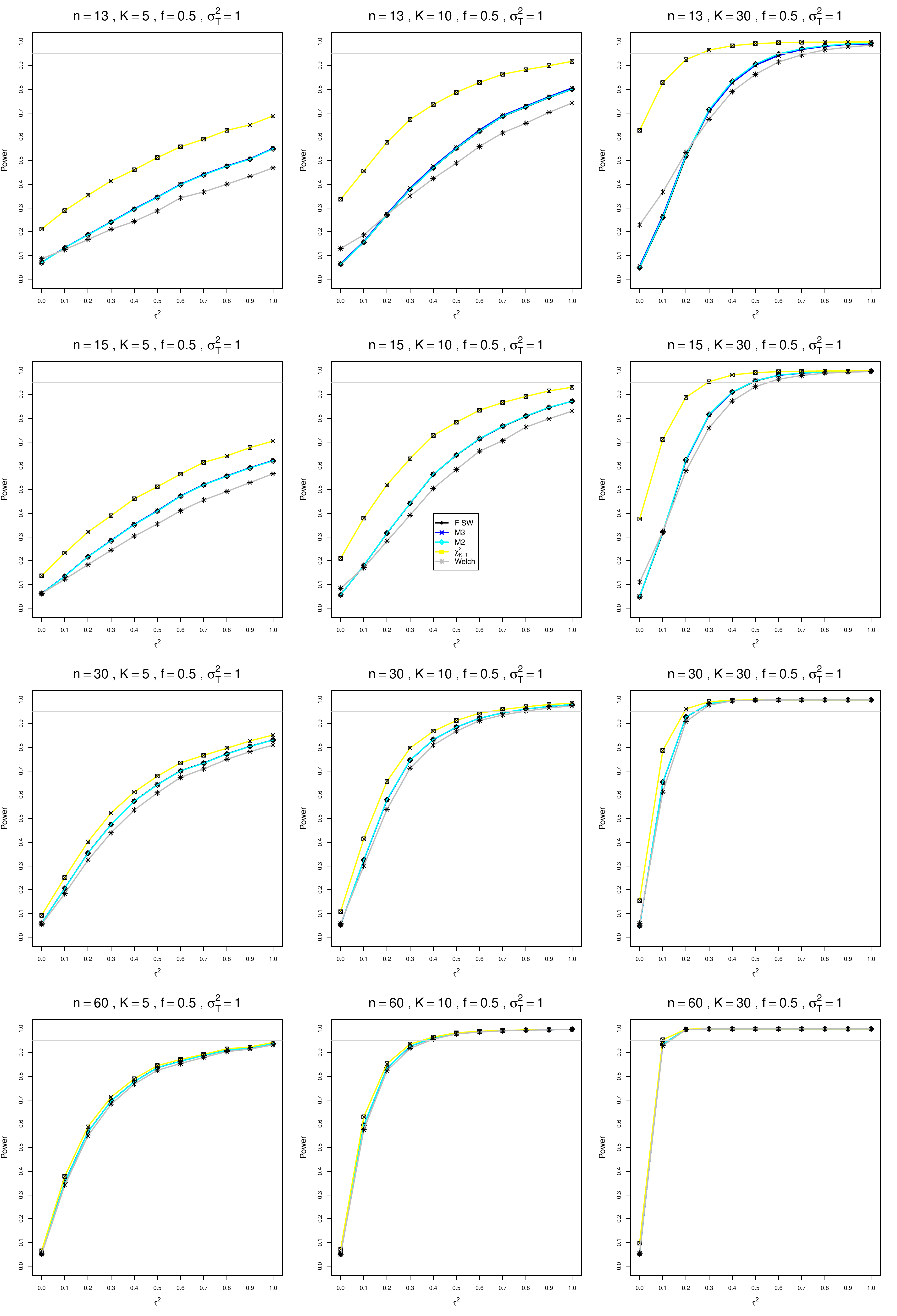}
	\caption{Power when $\alpha = .05$ for $\sigma_T^2 = 1$, $f = .5$, and unequal sample sizes $\bar{n}$ = 13, 15, 30, 60
		\label{PlotOfPhatAt005Sigma2T1andq05MD_underH0_unequal}}
\end{figure}


\begin{figure}[t]
	\centering
	\includegraphics[scale=0.33]{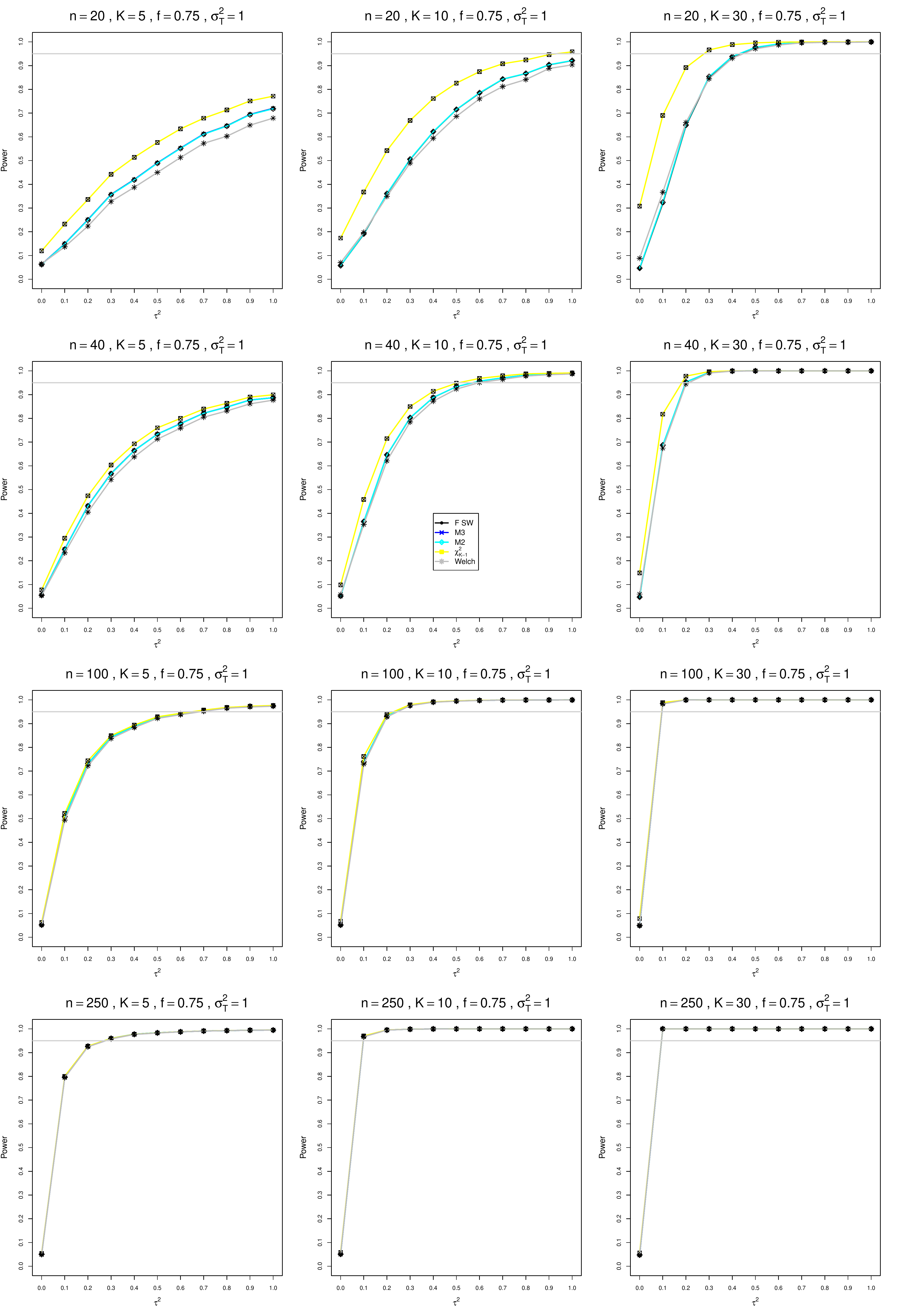}
	\caption{Power when $\alpha = .05$ for $\sigma_T^2 = 1$, $f = .75$, and equal sample sizes $n$ = 20, 40, 100, 250
		\label{PlotOfPhatAt005Sigma2T1andq075MD_underH0}}
\end{figure}

\begin{figure}[t]
	\centering
	\includegraphics[scale=0.33]{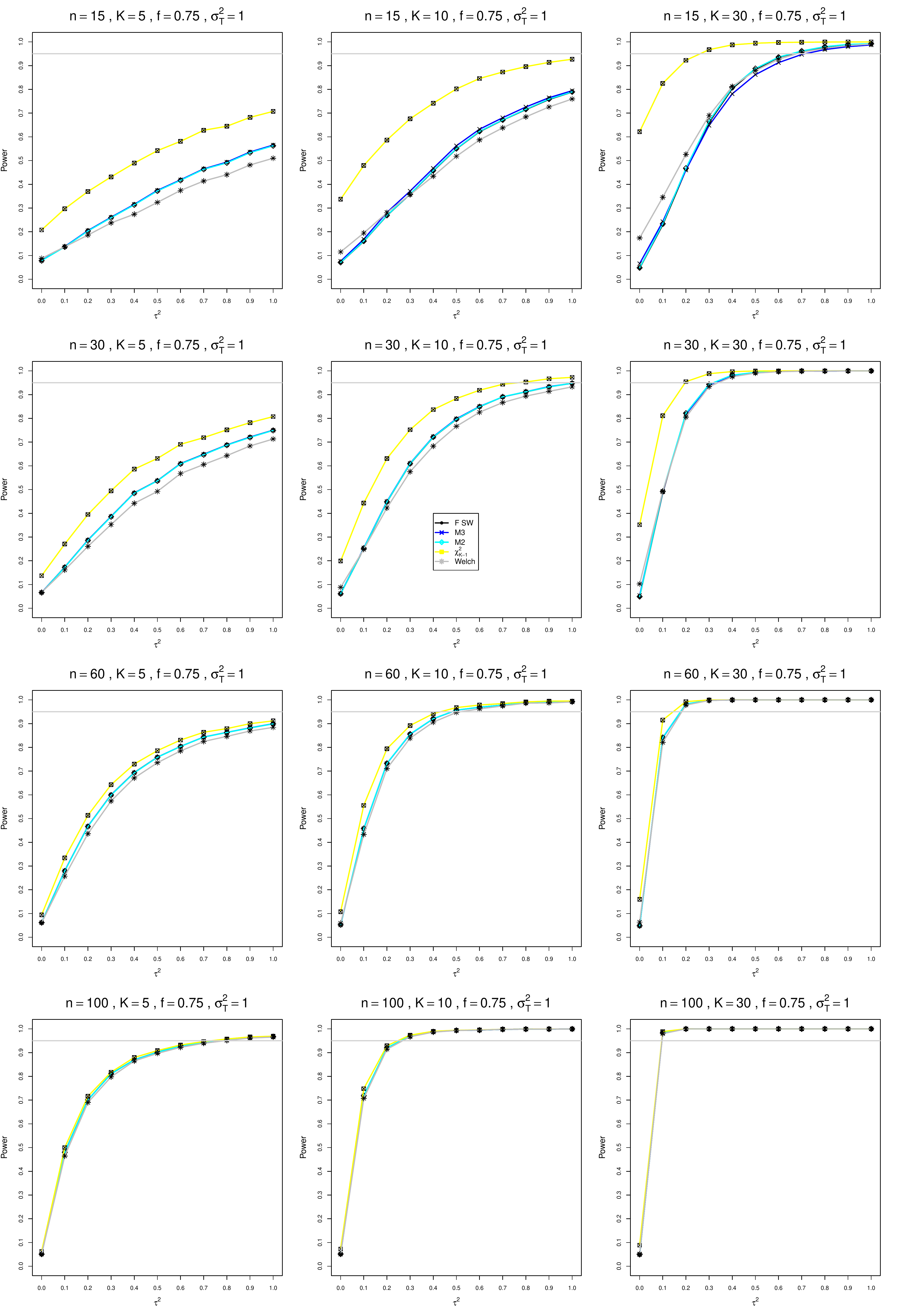}
	\caption{Power when $\alpha = .05$ for $\sigma_T^2 = 1$, $f = .75$, and unequal sample sizes $\bar{n}$ = 15, 30, 60, 100
		\label{PlotOfPhatAt005Sigma2T1andq075MD_underH0_unequal}}
\end{figure}

\begin{figure}[t]
	\centering
	\includegraphics[scale=0.33]{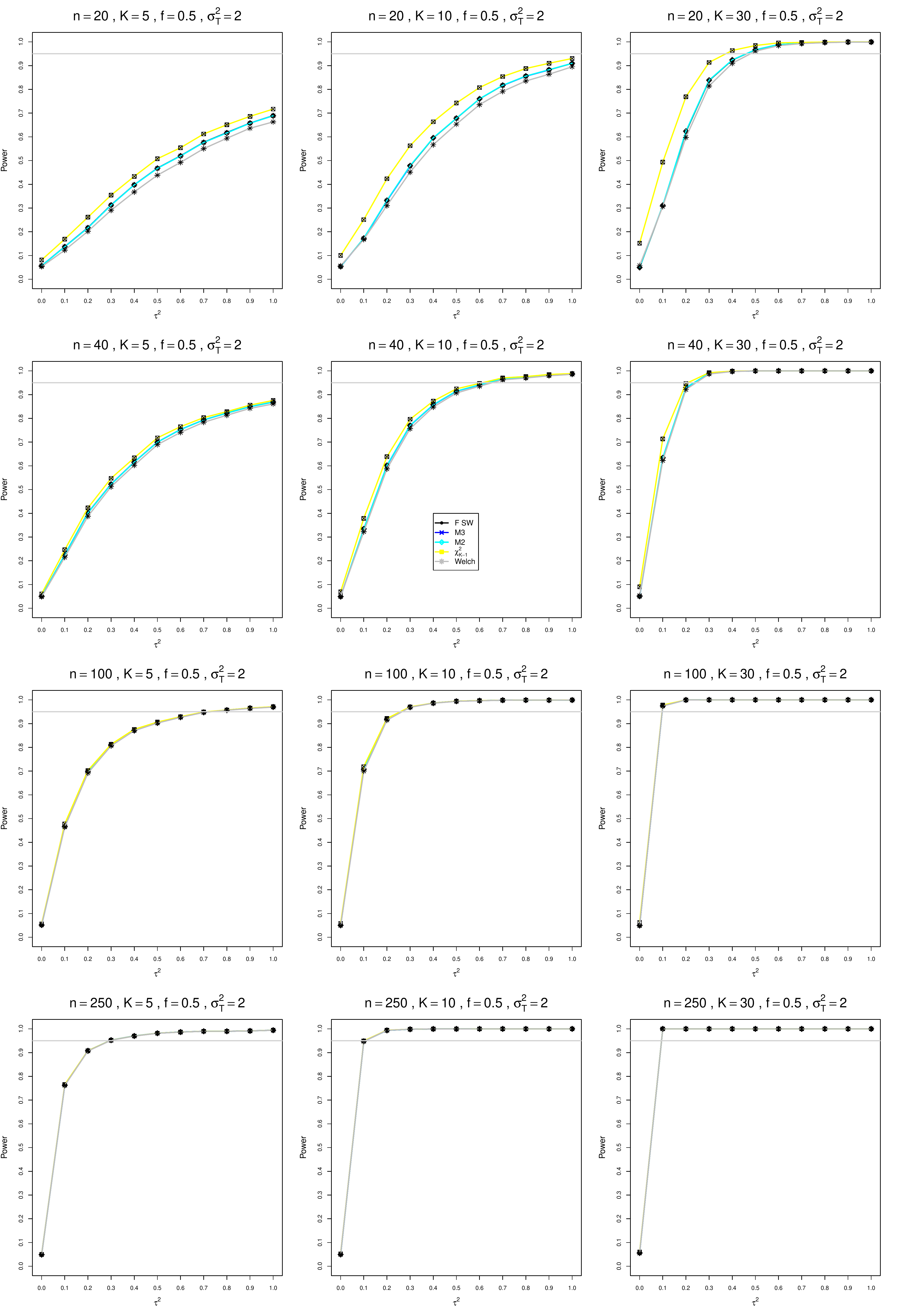}
	\caption{Power when $\alpha = .05$ for $\sigma_T^2 = 2$, $f = .5$, and equal sample sizes $n$ = 20, 40, 100, 250
		\label{PlotOfPhatAt005Sigma2T2andq05MD_underH0}}
\end{figure}

\begin{figure}[t]
	\centering
	\includegraphics[scale=0.33]{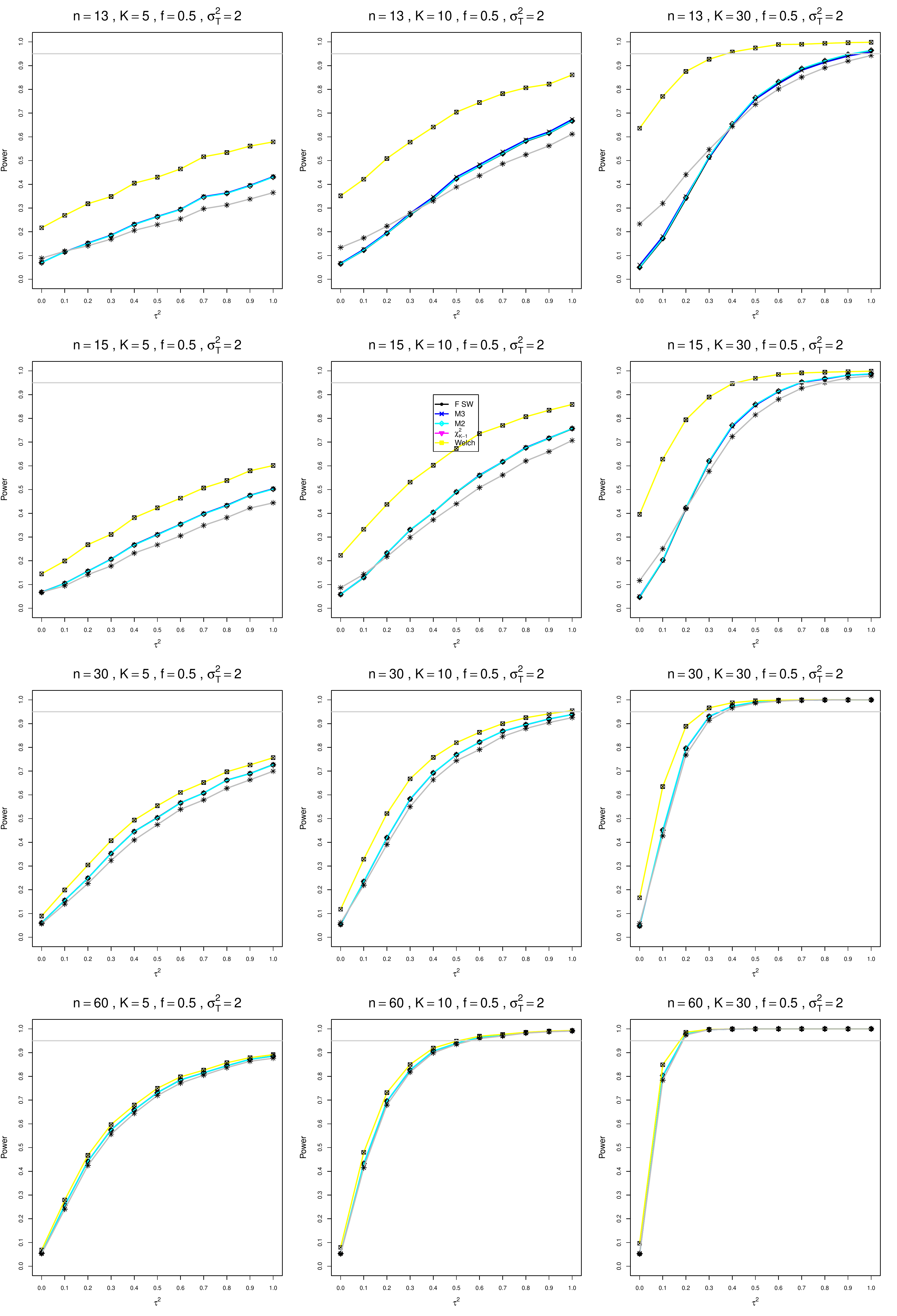}
	\caption{Power when $\alpha = .05$ for $\sigma_T^2 = 2$, $f = .5$, and unequal sample sizes $\bar{n}$ = 13, 15, 30, 60
		\label{PlotOfPhatAt005Sigma2T2andq05MD_underH0_unequal}}
\end{figure}

\begin{figure}[t]
	\centering
	\includegraphics[scale=0.33]{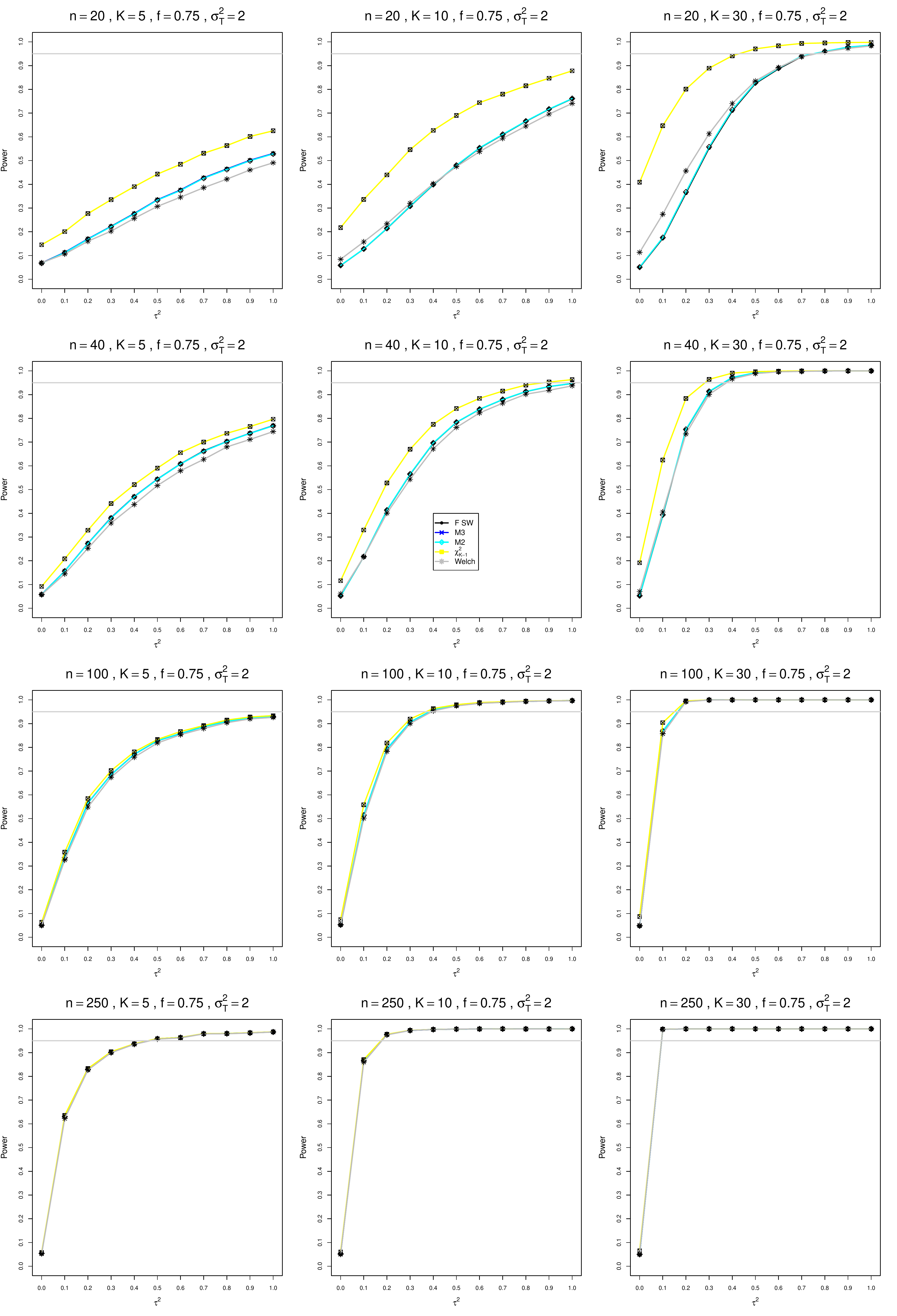}
	\caption{Power when $\alpha = .05$ for $\sigma_T^2 = 2$, $f = .75$, and equal sample sizes $n$ = 20, 40, 100, 250
		\label{PlotOfPhatAt005Sigma2T2andq075MD_underH0}}
\end{figure}
\begin{figure}[t]
	\centering
	\includegraphics[scale=0.33]{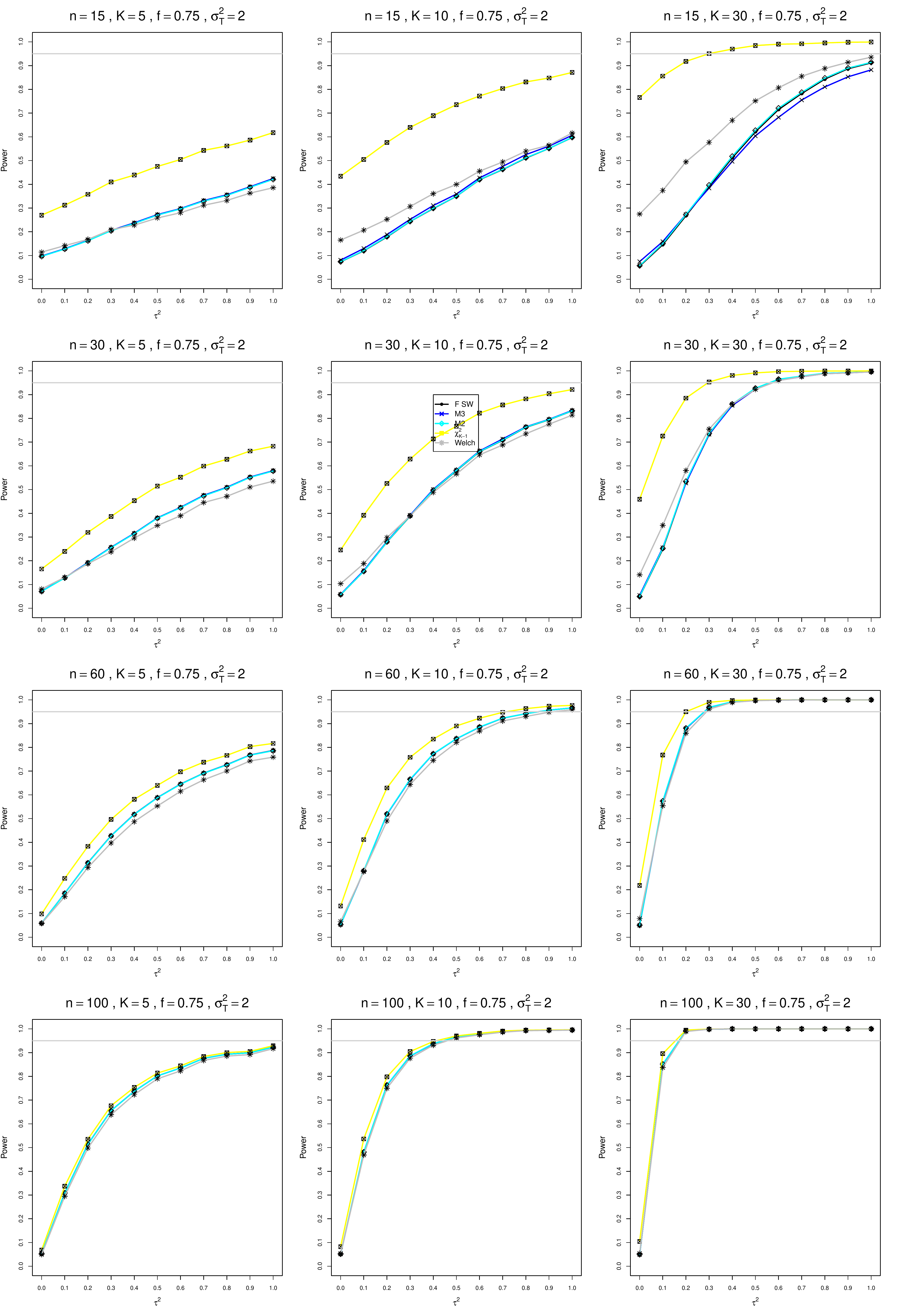}
	\caption{Power when $\alpha = .05$ for $\sigma_T^2 = 2$, $f = .75$, and unequal sample sizes $\bar{n}$ = 15, 30, 60, 100
		\label{PlotOfPhatAt005Sigma2T2andq075MD_underH0_unequal}}
\end{figure}

\clearpage
\setcounter{figure}{0}
\setcounter{section}{0}
\renewcommand{\thesection}{B5.\arabic{section}}
\section*{B5. Bias in estimation of $\tau^2$}
Each figure corresponds to a value of $\sigma_T^2$ (= 1, 2), a value of $f$ (= .5, .75), and a pattern of sample sizes (equal or unequal). (For all figures, $\sigma_C^2 = 1$.) \\
For each combination of a value of $n$ (= 20, 40, 100, 250) or $\bar{n}$ (= 13, 15, 30, 60 or 15, 30, 60, 100) and a value of $K$ (= 5, 10, 30), a panel plots bias of estimators of $\tau^2$ versus $\tau^2$ = 0.0(0.1)1.0.\\
The point estimators of $\tau^2$ are
\begin{itemize}
	\item SDL (DerSimonian-Laird with constant effective-sample-size weights)
	\item DL (standard DerSimonian-Laird: inverse-variance weights)
	\item REML (restricted maximum likelihood)
	\item MP (Mandel-Paule)
	\item CDL (Corrected DerSimonian-Laird)
\end{itemize}

\clearpage
\renewcommand{\thefigure}{B5.\arabic{figure}}
\begin{figure}[t]
	\centering
	\includegraphics[scale=0.33]{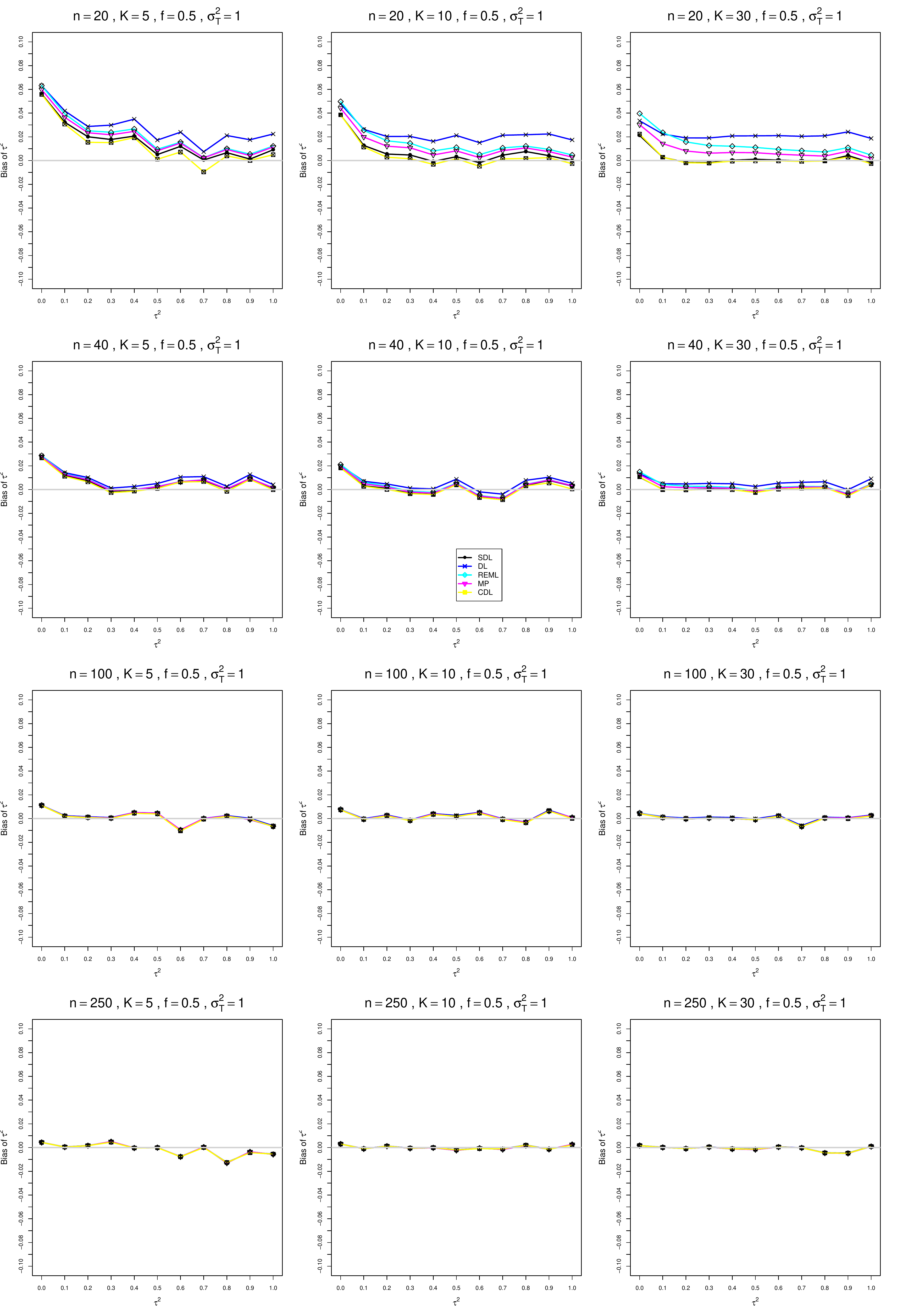}
	\caption{Bias of estimators of the between-studies variance $\tau^2 = 0.0(0.1)1.0$ for $\sigma_T^2 = 1$, $f = .5$, and equal sample sizes $n$ = 20, 40, 100, 250		
		\label{BiasTauMD0_S1_1q052Sigma2T1}}
\end{figure}

\begin{figure}[t]
	\centering
	\includegraphics[scale=0.33]{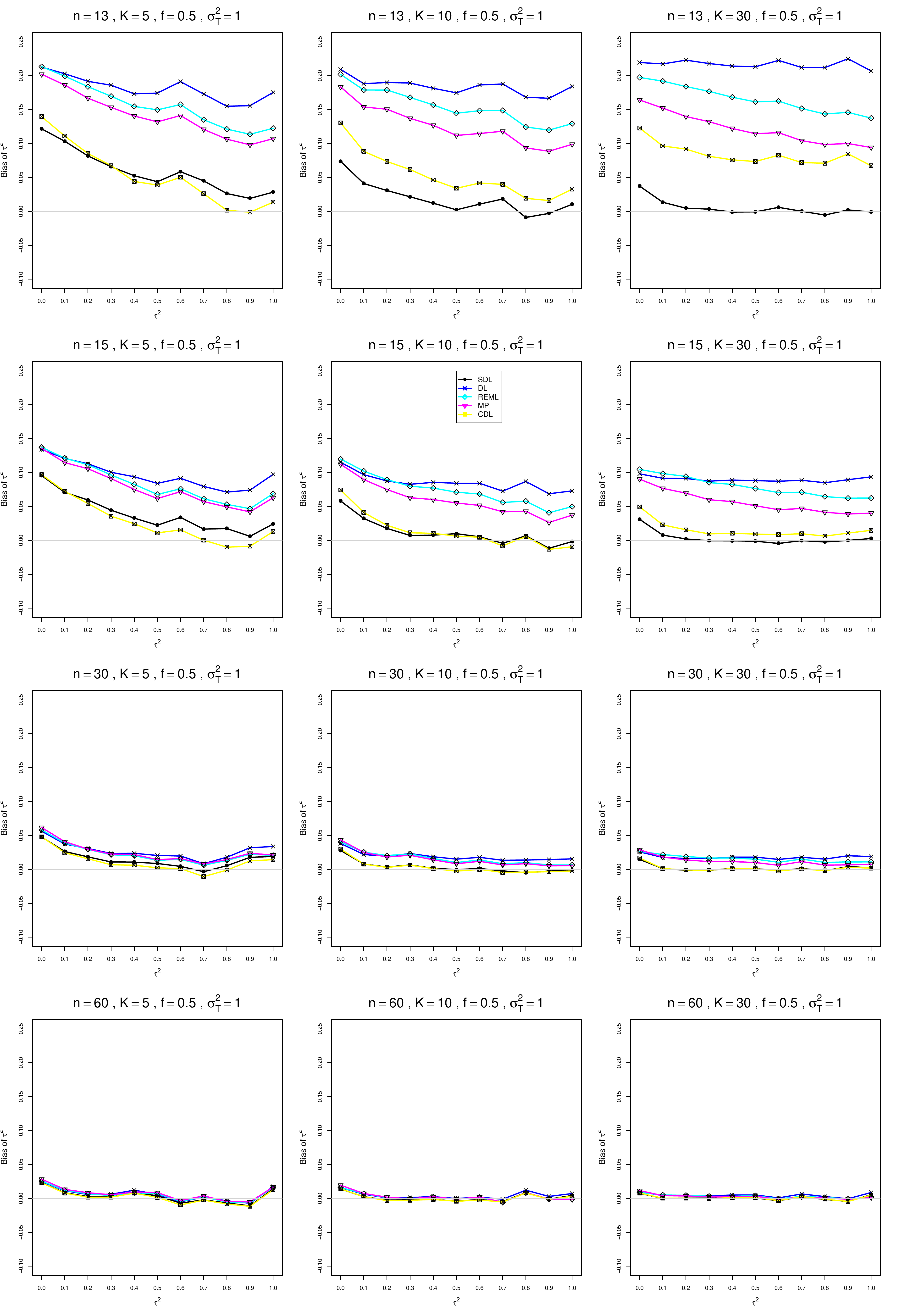}
	\caption{Bias of estimators of the between-studies variance $\tau^2 = 0.0(0.1)1.0$ for $\sigma_T^2 = 1$, $f = .5$, and unequal sample sizes $\bar{n}$ = 13, 15, 30, 60
		\label{BiasTauMD0_S1_1q052Sigma2T1_unequal}}
\end{figure}

\begin{figure}[t]
	\centering
	\includegraphics[scale=0.33]{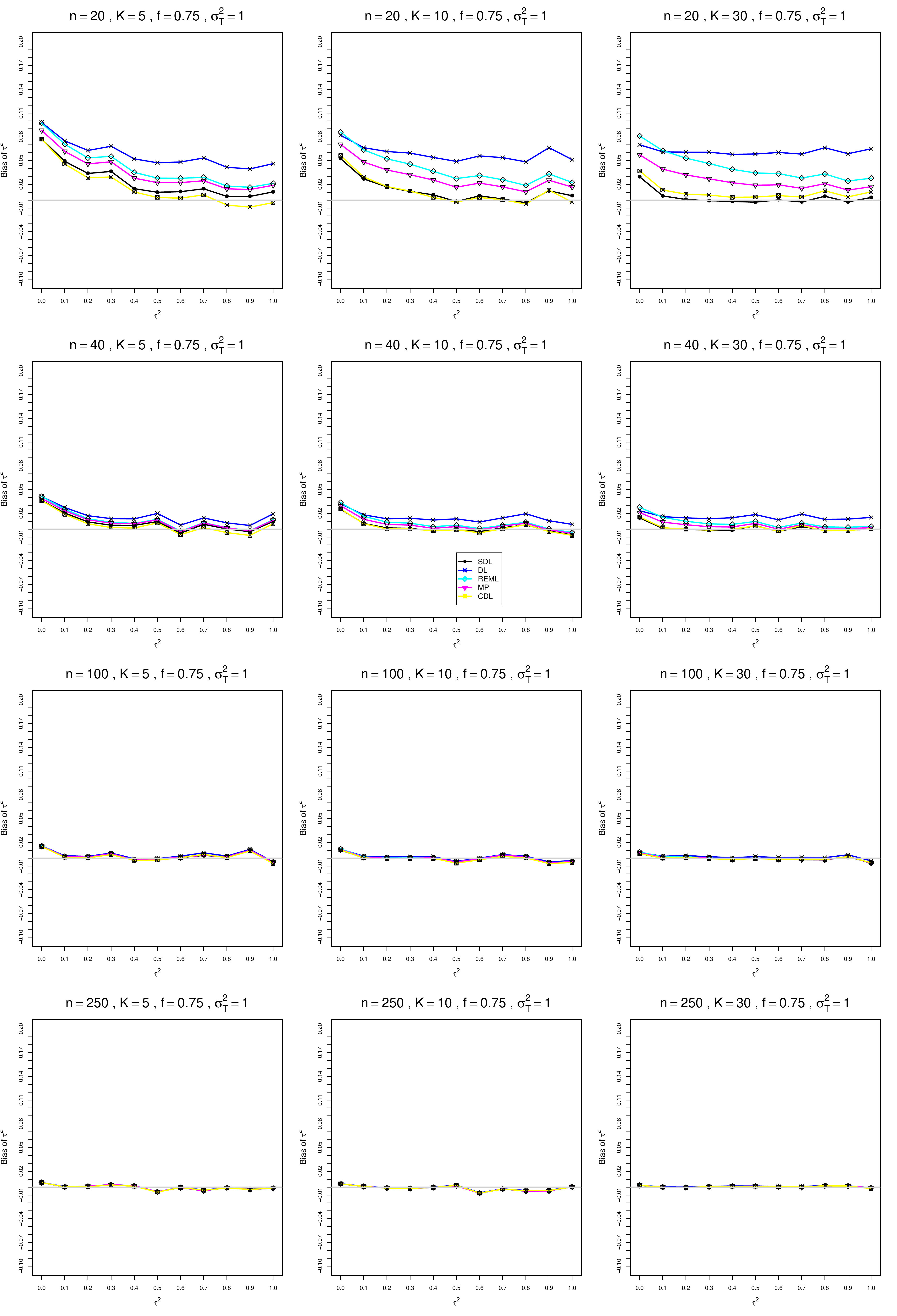}
	\caption{Bias of estimators of the between-studies variance $\tau^2 = 0.0(0.1)1.0$ for $\sigma_T^2 = 1$, $f = .75$, and equal sample sizes $n$ = 20, 40, 100, 250
		\label{BiasTauMD0_S1_1q0752Sigma2T1}}
\end{figure}

\begin{figure}[t]
	\centering
	\includegraphics[scale=0.33]{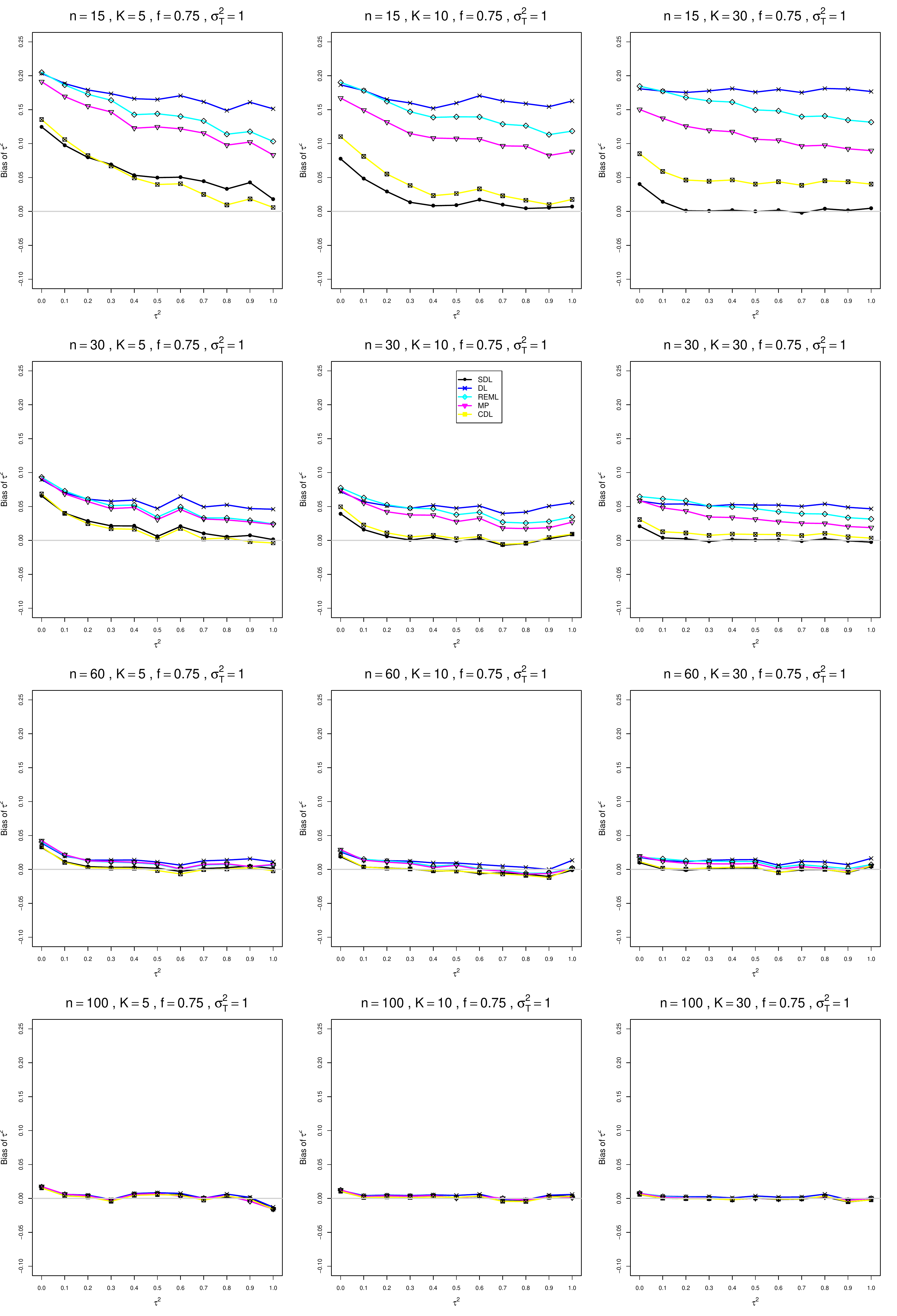}
	\caption{Bias of estimators of the between-studies variance $\tau^2 = 0.0(0.1)1.0$ for $\sigma_T^2 = 1$, $f = .75$, and unequal sample sizes $\bar{n}$ = 15, 30, 60, 100
		\label{BiasTauMD0_S1_1q0752Sigma2T1_unequal}}
\end{figure}

\begin{figure}[t]
	\centering
	\includegraphics[scale=0.33]{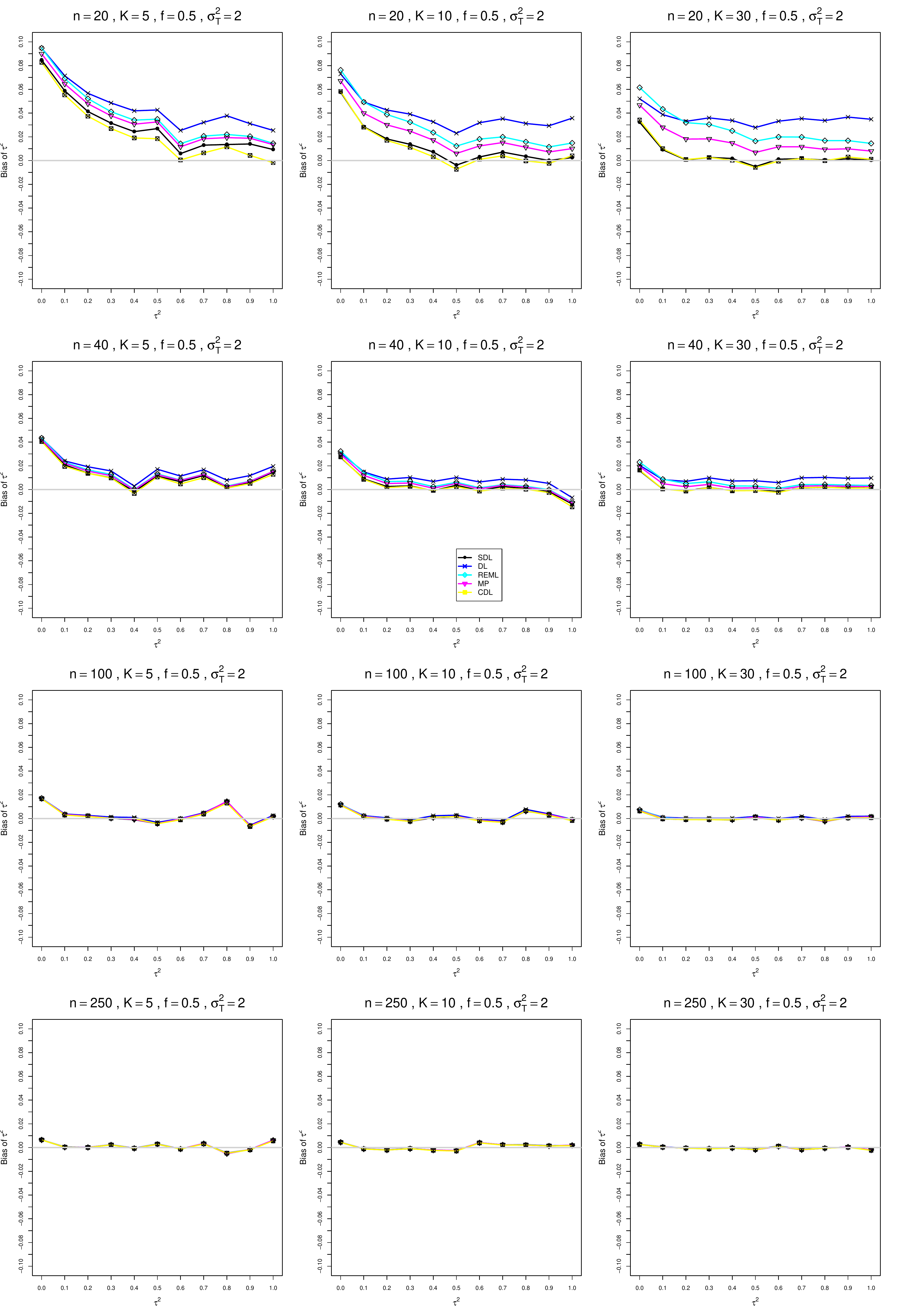}
	\caption{Bias of estimators of the between-studies variance $\tau^2 = 0.0(0.1)1.0$ for $\sigma_T^2 = 2$, $f = .5$, and equal sample sizes $n$ = 20, 40, 100, 250
		\label{BiasTauMD0_S2_1q052Sigma2T1}}
\end{figure}

\begin{figure}[t]
	\centering
	\includegraphics[scale=0.33]{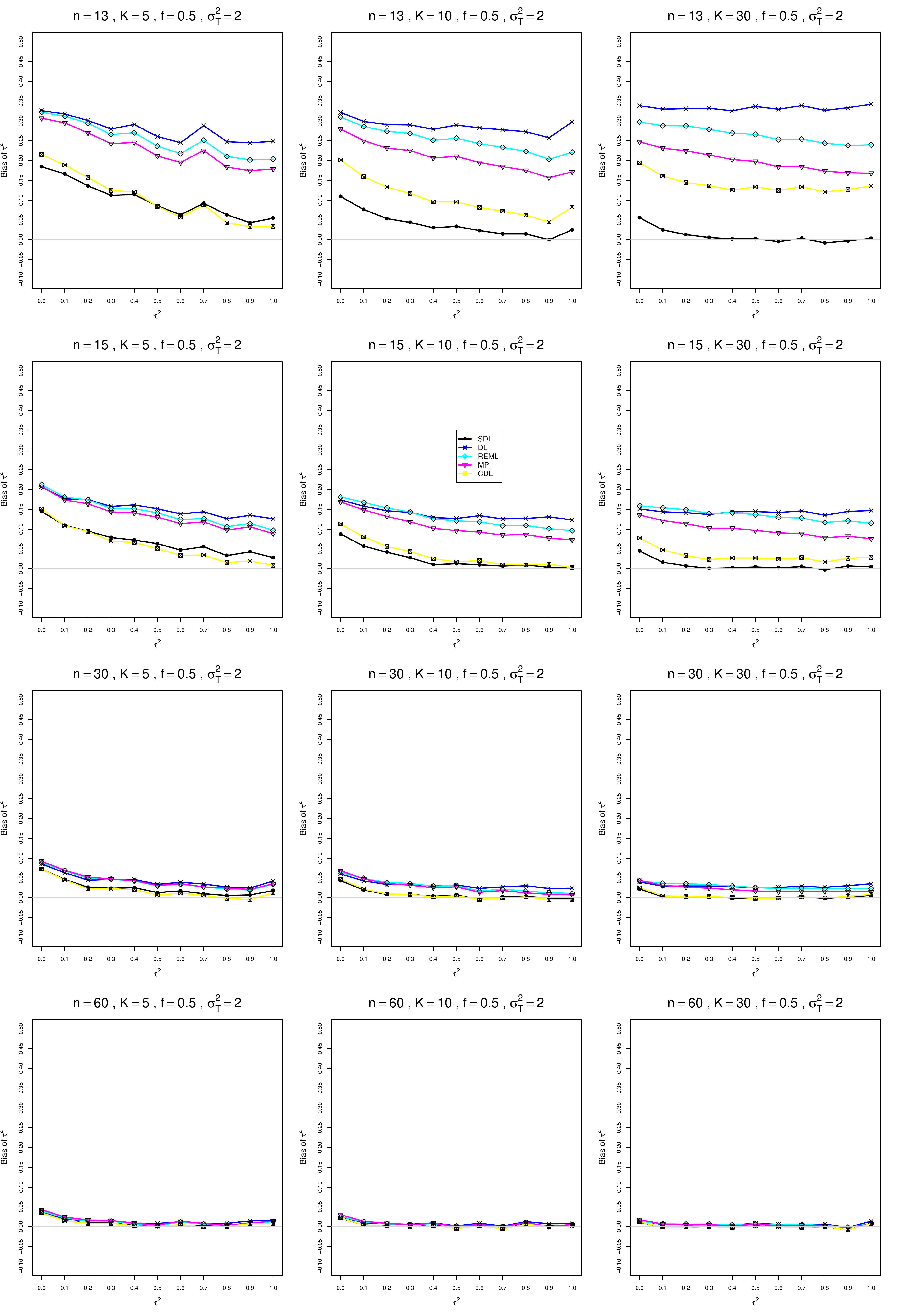}
	\caption{Bias of estimators of the between-studies variance $\tau^2 = 0.0(0.1)1.0$ for $\sigma_T^2 = 2$, $f = .5$, and unequal sample sizes $\bar{n}$ = 13, 15, 30, 60
		\label{BiasTauMD0_S2_1q052Sigma2T1_unequal}}
\end{figure}

\begin{figure}[t]
	\centering
	\includegraphics[scale=0.33]{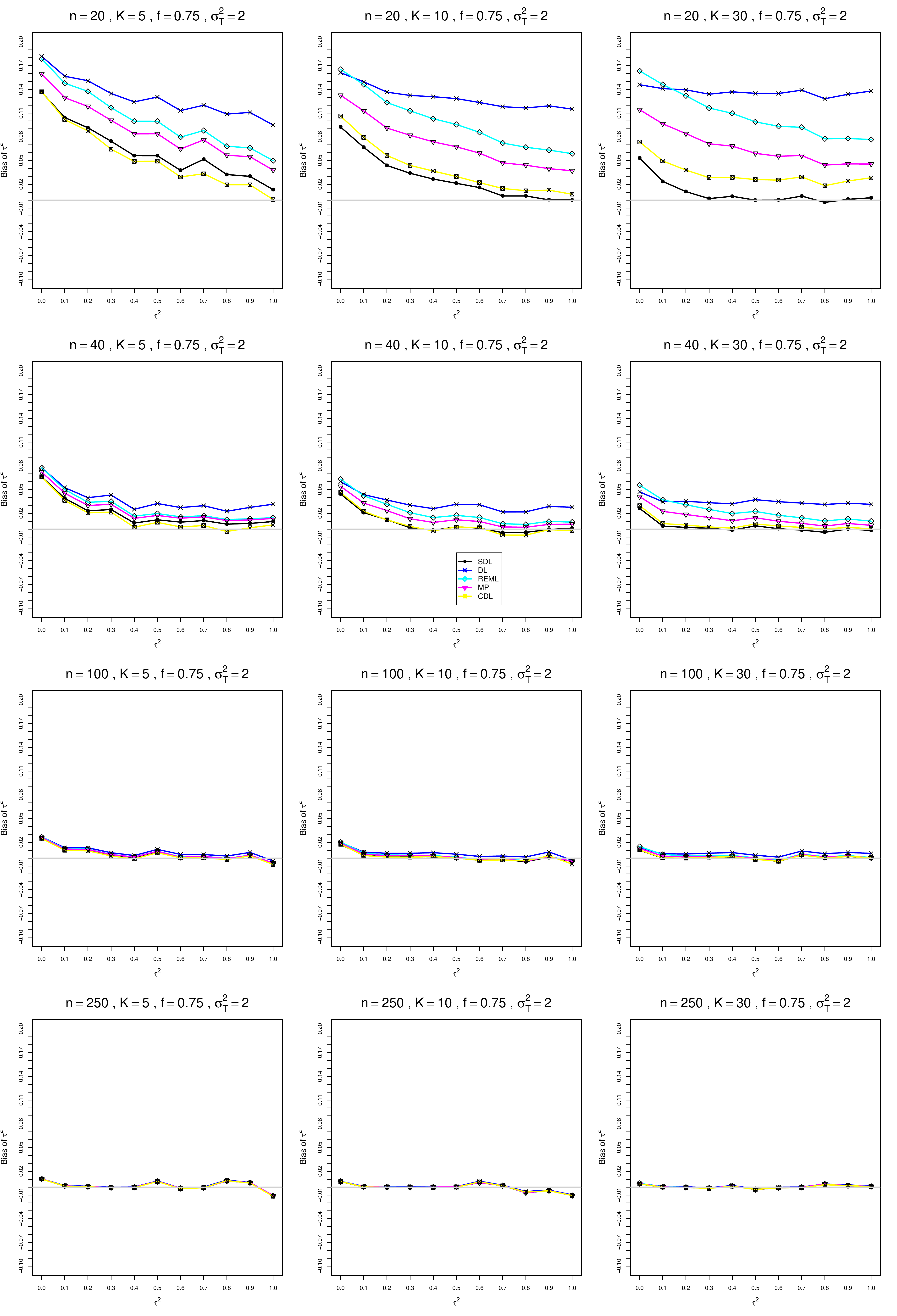}
	\caption{Bias of estimators of the between-studies variance $\tau^2 = 0.0(0.1)1.0$ for $\sigma_T^2 = 2$, $f = .75$, and equal sample sizes $n$ = 20, 40, 100, 250
		\label{BiasTauMD0_S2_1q0752Sigma2T1}}
\end{figure}

\begin{figure}[t]
	\centering
	\includegraphics[scale=0.33]{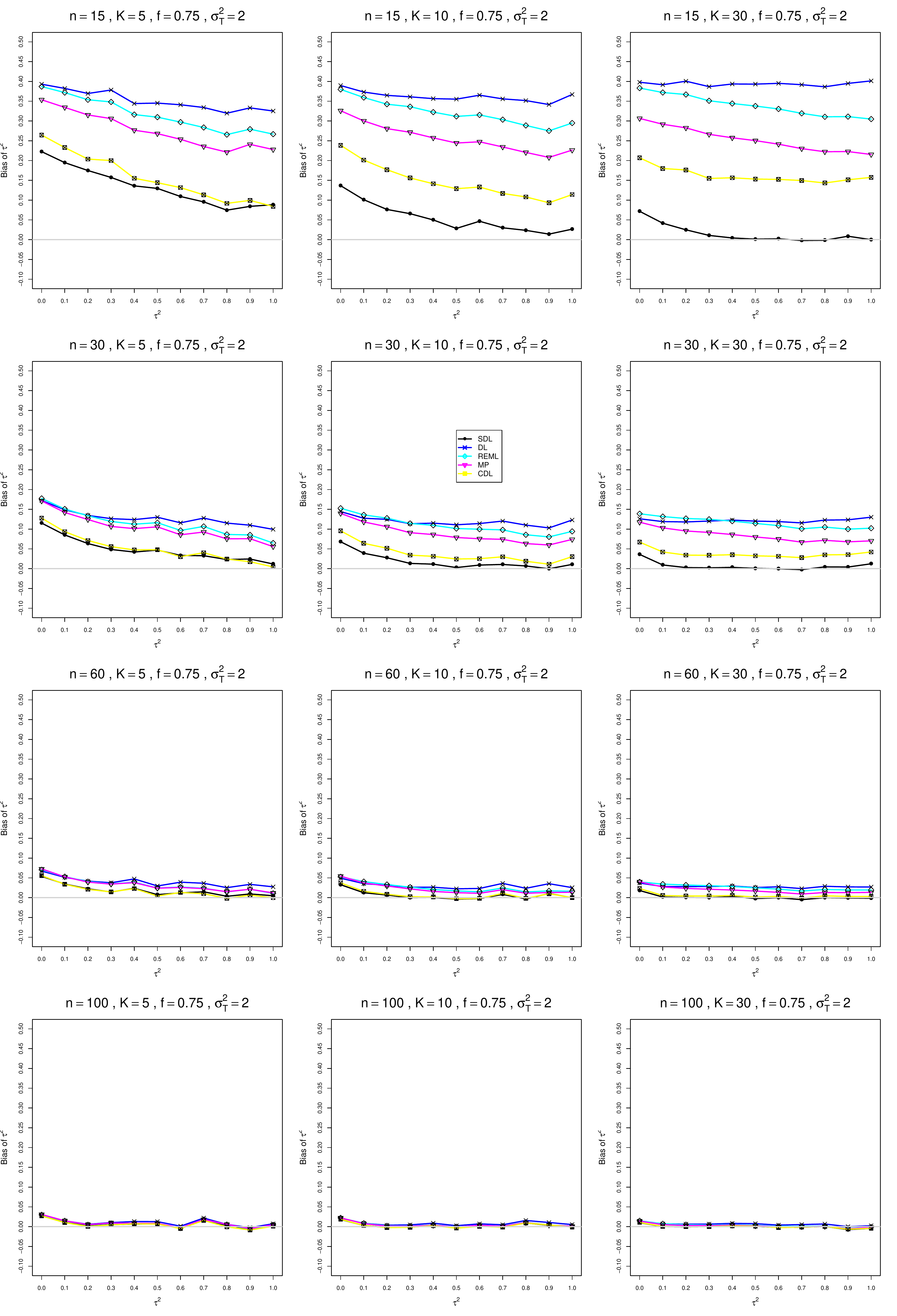}
	\caption{Bias of estimators of the between-studies variance $\tau^2 = 0.0(0.1)1.0$ for $\sigma_T^2 = 2$, $f = .75$, and unequal sample sizes $\bar{n}$ = 15, 30, 60, 100
		\label{BiasTauMD0_S2_1q0752Sigma2T1_unequal}}
\end{figure}

\end{document}